\documentclass[iop,revtex4,12pt,preprint]{emulateapj_recent}
\usepackage{graphics,graphicx}
\usepackage{helvet}
\usepackage{subfigure}
\usepackage[utf8]{inputenc}
\usepackage[T1]{fontenc}
\usepackage{soul}
\usepackage{hyperref}
\usepackage{amsmath, amssymb, gensymb}
\usepackage{morefloats}
\usepackage{float}
\usepackage{ulem}
\usepackage{natbib}
\usepackage{microtype}
\usepackage{multirow}
\usepackage{morefloats}
\usepackage{rotating}
\usepackage{here}
\usepackage{xspace}
\usepackage{color}
\usepackage{lipsum}
\usepackage{xspace}
\def\eg {e.g.,\xspace} 
\def\ie {i.e.,\xspace} 

\def\kms {km\,s$^{-1}$\xspace}

\def\jybmkms {Jy\,beam$^{-1}$\,km\,s$^{-1}$\xspace}
\def\mjybmkms {mJy\,beam$^{-1}$\,km\,s$^{-1}$\xspace}

\def\mjybm {mJy\,beam$^{-1}$\xspace}
\def\mujybm {$\mu$Jy\,beam$^{-1}$\xspace}

\def\jybmkms {Jy\,beam$^{-1}$\,km\,s$^{-1}$\xspace}

\def\etal {\textit{et al.}\xspace}

\def\co {CO\,($J$\,=\,2\,$\rightarrow\,$1)\xspace}

\def\dco {DCO$^+$\,($J$\,=\,3\,$\rightarrow\,$2)\xspace}
\def\ceighteeno {C$^{18}$O\,($J$\,=\,2\,$\rightarrow\,$1)\xspace}
\def\thirteencs {$^{13}$CS\,($J$\,=\,5\,$\rightarrow\,$4)\xspace}

\def\peakISMMcaseI {203}
\def\rmsISMMcaseI {0.57}
\def\peakISMMcaseIII {182}
\def\rmsISMMcaseIII {0.7}

\def\peakPSMMcaseI {6.28}
\def\rmsPSMMcaseI {53}
\def\peakPSMMcaseIII {5.6}
\def\rmsPSMMcaseIII {80}

\def\peakIEmbEightdatA {102}
\def\rmsIEmbEightdatA {60}
\def\peakPEmbEightdatA {0.69}
\def\rmsPEmbEightdatA {25}

\def\peakIEmbEightcaseII {75}
\def\rmsIEmbEightcaseII {0.18}
\def\peakIEmbEightcaseIII {53}
\def\rmsIEmbEightcaseIII {0.2}

\def\peakPEmbEightcaseII {0.48}
\def\rmsPEmbEightcaseII {30}
\def\peakPEmbEightcaseIII {0.32}
\def\rmsPEmbEightcaseIII {30}

\def\peakIEmbEightNdatA {55}
\def\rmsIEmbEightNdatA {66}
\def\peakPEmbEightNdatA {0.51}
\def\rmsPEmbEightNdatA {35}

\def\peakIEmbEightNcaseII {44}
\def\rmsIEmbEightNcaseII {55}
\def\peakIEmbEightNcaseIII {27}
\def\rmsIEmbEightNcaseIII {150}

\def\peakPEmbEightNcaseII {0.38}
\def\rmsPEmbEightNcaseII {35}
\def\peakPEmbEightNcaseIII {0.23}
\def\rmsPEmbEightNcaseIII {43}

\def\peakISMMcaseIUnits {\peakISMMcaseI \xspace\mjybm{}}
\def\rmsISMMcaseIUnits {\rmsISMMcaseI\xspace \mjybm{}}
\def\peakISMMcaseIIIUnits {\peakISMMcaseIII\xspace \mjybm{}}
\def\rmsISMMcaseIIIUnits {\rmsISMMcaseIII\xspace \mjybm{}}

\def\peakPSMMcaseIUnits {\peakPSMMcaseI \xspace\mjybm{}}
\def\rmsPSMMcaseIUnits {\rmsPSMMcaseI\xspace \mujybm{}}
\def\peakPSMMcaseIIIUnits {\peakPSMMcaseIII\xspace \mjybm{}}
\def\rmsPSMMcaseIIIUnits {\rmsPSMMcaseIII\xspace \mujybm{}}

\def\rmsIEmbEightdatAUnits {\rmsIEmbEightdatA \xspace\mujybm{}}
\def\peakPEmbEightdatAUnits {\peakPEmbEightdatA \xspace\mjybm{}}
\def\rmsPEmbEightdatAUnits {\rmsPEmbEightdatA \xspace\mujybm{}}

\def\peakIEmbEightcaseIIUnits {\peakIEmbEightcaseII \xspace\mjybm{}}
\def\rmsIEmbEightcaseIIUnits {\rmsIEmbEightcaseII \xspace\mjybm{}}
\def\peakIEmbEightcaseIIIUnits {\peakIEmbEightcaseIII \xspace\mjybm{}}
\def\rmsIEmbEightcaseIIIUnits {\rmsIEmbEightcaseIII \xspace\mjybm{}}

\def\peakPEmbEightcaseIIUnits {\peakPEmbEightcaseII\xspace \mjybm{}}
\def\rmsPEmbEightcaseIIUnits {\rmsPEmbEightcaseII \xspace\mujybm{}}
\def\peakPEmbEightcaseIIIUnits {\peakPEmbEightcaseIII\xspace \mjybm{}}
\def\rmsPEmbEightcaseIIIUnits {\rmsPEmbEightcaseIII \xspace\mujybm{}}

\def\peakIEmbEightNdatAUnits {\peakIEmbEightNdatA\xspace \mjybm{}}
\def\rmsIEmbEightNdatAUnits {\rmsIEmbEightNdatA\xspace \mujybm{}}
\def\peakPEmbEightNdatAUnits {\peakPEmbEightNdatA\xspace \mjybm{}}
\def\rmsPEmbEightNdatAUnits {\rmsPEmbEightNdatA\xspace\mujybm{}}

\def\peakIEmbEightNcaseIIUnits {\peakIEmbEightNcaseII \xspace\mjybm{}}
\def\rmsIEmbEightNcaseIIUnits {\rmsIEmbEightNcaseII\xspace \mujybm{}}
\def\peakIEmbEightNcaseIIIUnits {\peakIEmbEightNcaseIII\xspace \mjybm{}}
\def\rmsIEmbEightNcaseIIIUnits {\rmsIEmbEightNcaseIII\xspace \mujybm{}}

\def\peakPEmbEightNcaseIIUnits {\peakPEmbEightNcaseII\xspace \mjybm{}}
\def\rmsPEmbEightNcaseIIUnits {\rmsPEmbEightNcaseII\xspace \mujybm{}}
\def\peakPEmbEightNcaseIIIUnits {\peakPEmbEightNcaseIII\xspace\mjybm{}}
\def\rmsPEmbEightNcaseIIIUnits {\rmsPEmbEightNcaseIII\xspace \mujybm{}}

\begin{document}

\title{Characterizing magnetic field morphologies in three \\ Serpens protostellar cores with ALMA}

\shorttitle{High-resolution magnetic fields in Serpens protostars}


\author{Valentin J. M. Le Gouellec\altaffilmark{1,2}}
\author{Charles L. H. Hull\altaffilmark{3,4}\footnote{NAOJ Fellow}}
\author{Anaëlle J. Maury\altaffilmark{2,5}}
\author{Josep M. Girart\altaffilmark{6,7}}
\author{\L ukasz Tychoniec\altaffilmark{8}}
\author{Lars E. Kristensen\altaffilmark{9}}
\author{Zhi-Yun Li\altaffilmark{10}}
\author{Fabien Louvet\altaffilmark{11}}
\author{Paulo C. Cortes\altaffilmark{4,12}}
\author{Ramprasad Rao\altaffilmark{13}}

\affiliation{\altaffilmark{1}European Southern Observatory, Alonso de C\'ordova 3107, Vitacura, Santiago, Chile}
\affiliation{\altaffilmark{2}AIM, CEA, CNRS, Université Paris-Saclay, Université Paris Diderot, Sorbonne Paris Cité, F-91191 Gif-sur-Yvette, France}
\affiliation{\altaffilmark{3}National Astronomical Observatory of Japan, NAOJ Chile, Alonso de C\'ordova 3788, Office 61B, 7630422, Vitacura, Santiago, Chile}
\affiliation{\altaffilmark{4}Joint ALMA Observatory, Alonso de C\'ordova 3107, Vitacura, Santiago, Chile}
\affiliation{\altaffilmark{5}Harvard-Smithsonian Center for Astrophysics, Cambridge, MA 02138, USA}
\affiliation{\altaffilmark{6}Institut de Ci\`encies de l'Espai (ICE-CSIC), Campus UAB, Carrer de Can Magrans S/N, E-08193 Cerdanyola del Vall\`es, Catalonia}
\affiliation{\altaffilmark{7}Institut d'Estudis Espacials de Catalunya, E-08030 Barcelona, Catalonia}
\affiliation{\altaffilmark{8}Leiden Observatory, Leiden University, PO Box 9513, 23000RA, Leiden, The Netherlands}
\affiliation{\altaffilmark{9}Centre for Star and Planet Formation, Niels Bohr Institute and Natural History Museum of Denmark, University of Copenhagen, Øster Voldgade 5-7, DK-1350 Copenhagen K, Denmark}
\affiliation{\altaffilmark{10}Department of Astronomy, University of Virginia, 530 McCormick Road, Charlottesville, VA 22904, USA}
\affiliation{\altaffilmark{11}Departamento de Astronom\'ia, Universidad de Chile, Camino el Observatorio 1515, Las Condes, Santiago, Chile}
\affiliation{\altaffilmark{12}National Radio Astronomy Observatory, 520 Edgemont Road, Charlottesville, VA 22903, USA}
\affiliation{\altaffilmark{13}Submillimeter Array, Academia Sinica Institute of Astronomy and Astrophysics, 645 N. A’ohoku Place, Hilo, HI 96720, USA}

\shortauthors{Le Gouellec \etal}
\email{Valentin.LeGouellec@eso.org}

\slugcomment{Accepted for publication in ApJ on 29 August 2019}

\begin{abstract}

With the aim of characterizing the dynamical processes involved in the formation of young protostars, we present high angular resolution ALMA dust polarization observations of the Class 0 protostellar cores Serpens~SMM1, Emb~8(N), and Emb~8. With spatial resolutions ranging from 150 to 40\,au at 870 $\mu$m, we find unexpectedly high values of the polarization fraction along the outflow cavity walls in Serpens Emb8(N).  We use 3\,mm and 1\,mm molecular tracers to investigate outflow and dense gas properties and their correlation with the polarization. These observations allow us to investigate the physical processes involved in the Radiative Alignment Torques (RATs) acting on dust grains along the outflow cavity walls, which experience irradiation from accretion processes and outflow shocks. The inner core of SMM1-a presents a polarization pattern with a poloidal magnetic field at the bases of the two lobes of the bipolar outflow. To the south of SMM1-a we see two polarized filaments, one of which seems to trace the redshifted outflow cavity wall. The other may be an accretion streamer of material infalling onto the central protostar. We propose that the polarized emission we see at millimeter wavelengths along the irradiated cavity walls can be reconciled with the expectations of RAT theory if the aligned grains present at <\,500\,au scales in Class 0 envelopes have grown larger than the 0.1\,$\mu$m size of ISM dust grains.
Our observations allow us to constrain the star-forming sources’ magnetic field morphologies within the central cores, along the outflow cavity walls, and in possible accretion streamers.
\\
\end{abstract}

\vspace{0.5cm}

\maketitle

\section{Introduction}
\label{sec:intro}

Protostellar cores are forming within the densest parts of molecular clouds, where star formation mostly occurs along organized filamentary structures \citep{Andre2000,Andre2014}. Within these dense regions, prestellar cores, which are stellar precursors, are collapsing under their own gravitational field, and form Class 0 protostellar cores. At this evolutionary stage, the protostar is accreting material from the surrounding envelope, where most of the source’s mass is still located. The accretion is known to be ruled by a variety of physical processes, of which the main observational signature is the vigorous ejection of material in the form of a bipolar outflow. The evolution of these young accreting objects is well known to be strongly regulated by magnetic fields, which impact protostellar disk formation \citep{Wurster2018}, accretion and ejection processes, effects of turbulence \citep{Offner2017}, and core fragmentation \citep{Machida2005a}. 

We can better understand these phenomena by observing the polarization of thermal dust emission, which is the most commonly used tracer of magnetic fields in the ISM, from the scales of molecular clouds down to the $\sim$\,100\,au spatial scales of Class 0 disks.  Dust grains are assumed to produce polarized thermal emission thanks to the alignment between their angular momentum (aligned along their minor axis) with respect to the ambient magnetic field, via the actions of Radiative Alignment Torques \citep[RATs][]{Lazarian2007,Andersson2015}. Thus, the emission we detect is polarized orthogonal to the magnetic field component projected in the plane of the sky.

Single-dish observations of the magnetized ISM have revealed organised magnetic field lines toward dense star-forming filamentary structures, unveiling the role of the magnetic field on 0.1 to 10 pc scales \citep{Alves2008,PlanckCollaborationXXXIII2016,Pattle2017}. Interferometric observatories such as the Submillimeter Array (SMA) and the Combined Array for Research in Millimeter-wave Astronomy (CARMA) probed magnetic field morphologies at the protostellar envelope scales ($\sim$ 1000 au). More recently, observations with the Atacama Large Millimeter/submillimeter Array (ALMA) are able to resolve the tiniest features of protostellar cores (see \citealt{HBLi2014,HullZhang2019} for reviews). 

At the core-envelope scale, dust polarization observations have unveiled interesting results about the relative orientation of the magnetic field with respect to the bipolar outflow. The outflow of a protostellar core is considered to be closely linked with the core rotation axis, as the outflow launching mechanisms can consist of magnetohydrodynamic (MHD) disk winds triggered within the rotating circumstellar disk \citep{Pudritz2006,Frank2014,Bally2016}. Consequently, the study of the magnetic field orientation with respect to the bipolar outflow axis is an important proxy to understand the role played by the magnetic field in the regulation of the angular momentum of a protostellar core. At $\sim$ 1000 au envelope scales, \citet{Hull2014} and \citet{HullZhang2019} (with twice the sample), found the magnetic field is randomly aligned compared with the outflow axis, suggesting that the magnetic field at envelope scales is not affecting the magnetically driven winds at disk scales. However, a more recent work by \citet{Galametz2018} using a smaller sample, suggested a bi-modal distribution, exhibiting magnetic fields that are preferentially  aligned either parallel or perpendicular to the outflow orientation. In addition, they noticed that there is more large scale rotation and multiple systems in cores where there is a large angle between the main core-scale field and the outflow axis. Simulations have shown that these results could depend strongly on the relative strengths of the magnetic field, turbulence, and rotation \citep[e.g.,][]{Machida2006,Offner2016,Hull2017a,JLee2017}.

The orientation of the magnetic field in protostellar cores has been the focus of studies investigating the formation of disks, which is strongly impacted by the phenomenon of magnetic braking. This is the case because the magnetic field can remove enough angular momentum from the envelope material to impede the formation of large protostellar disks at early times \citep{Hennebelle2016}. In this respect, \citet{Maury2019} characterized the disk size distribution in a sample of Class 0 protostars and suggested that indeed, the magnetic field may play an important role in the formation of disk structure at the youngest protostellar evolutionary stage. 

The magnetic field morphologies seen in small-scale  (\ie a few $\times$ 100\,au) observations from CARMA, SMA, and ALMA have unveiled a variety of scenarios.  These include a few results showing that the magnetic field seem to follow the edges of the outflow cavity \citep{Hull2017b,Maury2018,Hull2019}, as well as magnetic field morphologies in young embedded disk structures that seem to exhibit both poloidal and toroidally wrapped field components \citep{Stephens2013,Rao2014,SeguraCox2015,Alves2018,Ohashi2018,Harris2018,Sadavoy2018a}. 

The Serpens region exhibits a filamentary structure with two compact star-forming clumps, Serpens Main and Serpens South, which are located at a distance of 436 $\pm$ 9 pc \citep{OrtizLeon2017b}. The recent star-formation episode observed in this region has been tentatively interpreted as resulting from a collision of two molecular clouds \citep{DuarteCabral2010,DuarteCabral2011}. Serpens SMM1, Emb 8(N), and  Emb 8 are three Class 0 protostars in the NW sub-cluster of Serpens Main. The position of the peak dust continuum emission, associated with these three protostellar cores and their surrounding core fragments (possibly containing protostars), can be found in Table \ref{t.source}.

\begin{table}[!tbph]
\centering
\small
\caption[]{Serpens source information}
\label{t.source}
\setlength{\tabcolsep}{0.25em} 
\begin{tabular}{p{0.3\linewidth}lcccc}
\hline \hline \noalign{\smallskip}
Name & $\alpha_{\textrm{J2000}}$ & $\delta_{\textrm{J2000}}$ & $M_\textrm{env}$ & $L_\textrm{bol}$
\vspace{0in}\\
&&&$M_\odot$&$L_\odot$\\
\noalign{\smallskip}  \hline
\noalign{\smallskip}
Serpens SMM1a & 18:29:49.81 & +1:15:20.41 &&\\
\noalign{\smallskip}
Serpens SMM1b1 & 18:29:49.68 & +1:15:21.09 && \\
\noalign{\smallskip}
Serpens SMM1b2 & 18:29:49.66 & +1:15:21.20 & 20 & 100 \\
\noalign{\smallskip}
Serpens SMM1c & 18:29:49.93 & +1:15:22.00 &&\\
\noalign{\smallskip}
Serpens SMM1d & 18:29:49.99 & +1:15:22.98 &&\\
\noalign{\smallskip}
\hline
\noalign{\smallskip}
Serpens Emb 8 & 18:29:48.09 & +1:16:43.30 && \\
\noalign{\smallskip}
Serpens Emb 8-b & 18:29:48.13 & +1:16:44.57 & \multirow{2}{*}{9.4}&\multirow{2}{*}{5.4}\\
\noalign{\smallskip}
Serpens Emb 8-c & 18:29:48.03 & +1:16:42.70 && \\
\noalign{\smallskip}
Serpens Emb 8(N) & 18:29:48.73 & +1:16:55.61 && \\
\noalign{\smallskip}
\hline
\smallskip
\end{tabular}
\vspace*{-0.1in}
\tablecomments{\footnotesize Envelope mass and bolometric luminosity values are from observations that encompass the whole core of SMM1 \citep{Enoch2011,Kristensen2012}, as well as Emb 8 and Emb 8(N) together \citep{Enoch2009b,Enoch2011}.}
\end{table}

The intermediate-mass protostellar source Serpens SMM1\footnote{Serpens SMM1 has been called by many other names, such as Serpens-Emb6, Serpens FIRS1, Serpens-FIR1, IRAS 18273+0113, S68 FIR, S68 FIRS1, and S68-1b.} is the most luminous source in the cloud \citep{Lee2014}, with a luminosity of $L_\textrm{bol}=100\,L_\odot$ \citep{Kristensen2012}. The protostellar envelope was found to have a mass of about $M_\textrm{env}\,\sim\,20\,M_\odot$ \citep{Enoch2011} and is surrounded by a disk-like structure with $M_\textrm{disk}\,\sim\,1.0\,M_\odot$ and $R_\textrm{disk}\,\sim\,300$ au \citep{Enoch2009b}. ALMA observations from \citet{Hull2016a} show a one-sided, high velocity, highly collimated molecular jet ($\sim$ 80 \kms) from the central source SMM1-a. The base of the narrow jet is surrounded by a wide-angle outflow cavity, whose walls were observed in free-free emission by the Karl G. Jansky Very Large Array (VLA) \citep{RodriguezKamenetzky2016}. \citet{Hull2016a} showed an extremely high-velocity (EHV), one-sided redshifted molecular jet from the protobinary system SMM1-b located to the NW of the central source. These three sources were observed in full polarization by CARMA in the TADPOL survey \citep{Hull2014}. \citet{Hull2016a} attributed the ionization of the outflow cavity walls to UV radiation escaping from the accreting central protostar or to the precession of the high-velocity jet, which would impact the surrounding envelope. \citet{Goicoechea2012} proposed an alternative scenario, where the ionizing radiation is caused by distributed shocks throughout the outflow. Interferometric dust polarization observations of this source have suggested that the SE redshifted lobe of the bipolar outflow from SMM1-a is shaping the magnetic field \citep{Hull2017b}. SMM1-a is also known to host a hot corino-like central region, as a few complex organic molecules (COMs) have previously been detected, including methanol, methyl formate, dimethyl ether, vinyl cyanide, and ethyle glycol \citep{Kristensen2010,Oberg2011,Tychoniec2018}. Hot corinos are thought to correspond to the central region of the protostar, where the temperature is high enough to sublimate icy grain mantles, which release COMs into the gas phase \citep{Maury2014,Walsh2014}.

Serpens Emb 8 and 8(N)\footnote{Serpens Emb 8 has been also called S68N, and Serpens Emb 8(N) has also the name of S68Nb.} are two low-mass protostellar sources separated by $15.7^{\prime\prime}$, \ie $\,\sim\,7000$ au. These sources were observed in \citet{Enoch2009b, Enoch2011} with Bolocam and have a combined envelope mass of $M_\textrm{env}\,\sim\,9.4\,M_\odot$ and a bolometric luminosity of $L_\textrm{bol}=5.4\,L_\odot$. These observations have a spatial resolution of $\sim\,13,500$ au, thus encompassing the two protostellar sources. ALMA dust polarization observations of Serpens Emb 8 exhibited a chaotic magnetic field morphology; the authors concluded that the magnetic field is most likely weak with respect to the cloud scale turbulence \citep{Hull2017a}.

Regarding the relative age of Emb 8 and 8(N), the differences between the two bipolar outflows of both sources offer a clue.  Unlike Emb 8, Emb 8(N) exhibits a  pristine EHV jet on both sides, which has not propagated as far as the outflow from Emb 8 \citep{Dionatos2010b,Tychoniec2019}. As molecular jets are generally an indication of the young age of a protostar \citep{Bally2016}, we propose that Emb 8(N) may be younger. Moreover the outflow opening angles of these sources are quite different, which can be related with age  \citep{ArceSargent2006,Velusamy2014,Hsieh2017}.  The opening angle of Emb 8(N) is smaller, again suggesting a younger age for Emb 8(N).  


In this paper we present ALMA 870 $\mu$m polarization observations toward the three Class 0 protostars Emb 8(N), Serpens SMM1, and Serpens Emb 8. We describe in Section \ref{sec:obs} the different observational data and the data reduction. In Section \ref{sec:res} we present the dust polarization and total intensity maps, as well as a few molecular line observations. Finally, we discuss in Section \ref{sec:dis} the different polarization patterns and the potential grain alignment mechanisms implied, as well as the relations between the bipolar outflow and the magnetic field morphology. Finally, we draw our conclusions in Section \ref{sec:con}.

\section{ALMA Observations and Data Reduction}
\label{sec:obs}

We present three 870\xspace$\mu$m ALMA dust polarization observations of our three sources in Serpens.  Each of the datasets A, B, and C, targeted all three sources, and were taken on 2015 June $3\,\&\,7$, 2016 September $12\,\&\,13$, and 2017 July $31^{st}$ (ALMA projects: 2013.1.00726.S, 2015.1.00768.S, 2016.1.00710.S; PI: C. Hull). The synthesized beam of our observations varies from 0$\farcs$33 to 0$\farcs$11, corresponding to a spatial resolutions varying from $\sim\,$144 au in the dataset A, up to $\sim\,$48 au from the dataset C, at a distance of 436 pc. Each dataset consists of four spectral windows of 2 GHz each, ranging in frequency from 336.5 GHz to 350.5 GHz. The details of the observations can be found in Table \ref{t.obs}. In the datasets A, B, and C, the polarization calibrators were respectively J1751+0939, J1751+0939, and J1924-2914, chosen for their high polarization fraction.
The ALMA flux calibration accuracy in Band 7 (870\xspace$\mu$m) is 10$\%$. See \citet{Nagai2016} for a complete description of the ALMA polarization system.

We faced some issues when imaging the datasets B and C, as they were ``semi-pass,'' because the requested resolution and sensitivity were not reached.
To improve our image quality, the datasets were combined together following three different schemes during the production of the Stokes images (see Table \ref{t.imaging}). The choices of which datasets to merge depended on which of them produced the best images at our multiple desired spatial resolutions.

\begin{table}[tbh!]
\centering
\small
\caption[]{ALMA Observation details}
\label{t.obs}
         \setlength{\tabcolsep}{0.61em} 
         \begin{tabular}{p{0.13\linewidth}c|ccc}
            \hline \hline
           \noalign{\smallskip}
            Dataset & Baselines &&Calibrators\\
            & (m) & \\
            \noalign{\smallskip}
            \hline
            \noalign{\smallskip}
             &&bandpass&J1751+0939\\
             A & 16.5 - 763 &phase&J1751+0939\\
             && flux&Titan\\
            \noalign{\smallskip}
            \hline
             \noalign{\smallskip}
             &&bandpass&J1751+0939\\
             B & 12.4 - 3042 &phase&J1751+0939\\
             && flux&J1751+0939\\
            \noalign{\smallskip}
            \hline \noalign{\smallskip}
             &&bandpass&\\ 
             C & 11.7 - 3320 &phase& See note \\
             &&flux&\\ 
            \noalign{\smallskip}
            \end{tabular}
            \tablecomments{\footnotesize In dataset C, all three calibrators were J1751+0939 in one execution, and J1924-2914 in the other.}
            \smallskip
            \smallskip
\end{table}

\begin{table*}[tbh!]
\centering
\small
\caption[]{Imaging details}
\label{t.imaging}
         \setlength{\tabcolsep}{0.7em} 
         \begin{tabular}{p{0.15\linewidth}cccccccccc}
            \hline \hline
           \noalign{\smallskip}
            Case & Source & $I$ & $Q$ \& $U$  & $\theta_{\textrm{res}}$ & $I_{\textrm{peak}}$&$I_{\textrm{rms}}$ &$Q_{\textrm{rms}}$&$U_{\textrm{rms}}$ &Figures\\
             &  & & & ('') & $\left(\frac{\textrm{mJy}}{\textrm{beam}}\right)$ &$\left(\frac{\textrm{mJy}}{\textrm{beam}}\right)$&$\left(\frac{\textrm{mJy}}{\textrm{beam}}\right)$&$\left(\frac{\textrm{mJy}}{\textrm{beam}}\right)$&\\
             \noalign{\smallskip}
            \hline
            \noalign{\smallskip}
             \multirow{2}{*}{Dataset A} & Ser-Emb 8 & A & A & 0.35 $\times$ 0.32 & \peakIEmbEightdatA{} & 0.060 &0.024&0.024&\ref{fig:emb8_pol},\ref{fig:emb8_pol_CO},\ref{fig:emb8_pol_pfrac}\\
              & Ser-Emb 8(N) & A & A & 0.35 $\times$ 0.32 &\peakIEmbEightNdatA{} & 0.066&0.024&0.025&\ref{fig:emb8N_pola}\\             
            \noalign{\smallskip}
            \hline
            \noalign{\smallskip}
             Case-1 & Ser-SMM1 & ABC & ABC & 0.15 $\times$ 0.14 & \peakISMMcaseI{}{} & \rmsISMMcaseI{} & 0.033&0.030 & \ref{fig:smm1_pol},\ref{fig:smm1_pol_CO},\ref{fig:smm1_pol_pfrac},\ref{fig:Tb}\\
            \noalign{\smallskip}
            \hline
             \noalign{\smallskip}   
             \multirow{2}{*}{Case-2} & Ser-Emb 8 & AC & AC& 0.20 $\times$ 0.16 & \peakIEmbEightcaseII{}& \rmsIEmbEightcaseII{} &0.024&0.024 & \ref{fig:emb8_pol}\\
             & Ser-Emb 8(N) & AC & AC &  0.26 $\times$ 0.22 & \peakIEmbEightNcaseII{} & \rmsIEmbEightNcaseII{} &0.025&0.025&\ref{fig:emb8N_pola},\ref{fig:emb8N_CO_pol},\ref{fig:emb8N_pol_pfrac},\ref{fig:emb8N_pol_chemistry}\\
            \noalign{\smallskip}
            \hline \noalign{\smallskip}
             & Ser-SMM1 & C& B &  0.13 $\times$ 0.13 & \peakISMMcaseIII{} & \rmsISMMcaseIII{} &0.066&0.058&\ref{fig:smm1_pol},\ref{fig:smm1a_pola_E_vec},\ref{fig:SMM1a_jet_pol}\\
             Case-3 & Ser-Emb 8 & C & BC &  0.12 $\times$ 0.11 & \peakIEmbEightcaseIII{} & \rmsIEmbEightcaseIII{} &0.033& 0.033 & \ref{fig:emb8_pol}\\
             & Ser-Emb 8(N) & C & BC & 0.14 $\times$ 0.11 & \peakIEmbEightcaseIII{}& \rmsIEmbEightNcaseIII{} &0.035&0.035&\ref{fig:emb8N_pola}\\
             \noalign{\smallskip}
         \hline 
         \end{tabular}
         \smallskip
         \smallskip
         \tablecomments{\footnotesize Case-1, 2, and 3 are different combinations of the datasets A, B, and C. $\theta_{\textrm{res}}$ is the angular resolution of the observations. $I_\textrm{peak}$ is the peak total intensity of the Stokes $I$ total intensity map. $I_{\textrm{rms}}$, $Q_{\textrm{rms}}$, and $U_{\textrm{rms}}$ are the noise values in the Stokes $I$, $Q$, and $U$ maps, respectively. The values are calculated as flux density per unit of synthesized beam $\theta_\textrm{res}$. The maps of Serpens SMM1 and Emb 8 from Dataset A were previously published in \citet{Hull2017a, Hull2017b}.}
         \smallskip
         \smallskip
\end{table*}

The polarized dust continuum images were produced by using the CASA task \texttt{clean}, applying four rounds of consecutive phase-only self-calibration, using the total intensity (Stokes $I$) solutions as a model for the Stokes $Q$ and $U$, with a Briggs weighting parameter of robust = 1. The three Stokes parameters $I$, $Q$, and $U$ were cleaned separately after the last round of self-calibration using an appropriate residual threshold and number of iterations. The linear polarization properties of the radiation field from the thermal dust emission are given by the Stokes parameters $Q$ and $U$, whereas the Stokes $I$ parameter gives the total intensity of the dust continuum emission. The quantities derived from the combined use of the three Stokes maps are the polarized intensity $P$, the polarization fraction $P_\textrm{frac}$, and the polarization position angle $\chi$:
\begin{equation}
P = \sqrt{Q^2 + U^2}
\end{equation}
\begin{equation}
P_{\textrm{frac}} = \frac{P}{I}
\end{equation}
\begin{equation}
\chi = \frac{1}{2}\arctan{\left(\frac{U}{Q}\right)}
\end{equation}
Although the Stokes parameter $Q$ and $U$ can be positive or negative, the polarized intensity $P$ is always positive. This introduces a bias in the measurement of the polarized intensity, especially for low signal-to-noise (S/N) ratio emission, where $P$ is below the $3\sigma_P$ threshold, $\sigma_\textrm{P}$ being the noise in the $P$ map. We corrected this bias in our $P$ maps in order to arrive at the corrected polarized intensity, following the method described in \cite{Vaillancourt2006, Hull2015b}. Note that in Case-3, the Stokes parameters $I$ and $Q\,\&\,U$ come from different combinations of datasets (see Table \ref{t.imaging}). Therefore, before debiasing $P$ and making the $P$ and $I$ images, we used the \texttt{imsmooth} CASA task to smooth the three Stokes parameters $I$, $Q$, and $U$ to have the same reconstructed beam (by convolving the map with a 2D-Gaussian kernel). The resulting beam was chosen in such a way that it encompasses perfectly the two beams resulting from the different combinations. In addition, we performed a primary beam correction on all the total intensity and polarized intensity maps presented in this article.

Finally we present 1.3 mm (Band 6) and 3 mm (Band 3) ALMA spectral-line data (ALMA projects: 2013.1.00726.S and 2016.1.00710.S; PI: C. Hull), which were taken on 2014 August 18 and 2016 October 4, respectively, and have angular resolutions of approximately 0$\farcs$45$\times$0$\farcs$55 and 0$\farcs$56$\times$0$\farcs$6.  The data include the following transitions: \co (used to trace the outflow, shown in \citealt{Hull2016a,Hull2017b} for the case of Serpens SMM1), \thirteencs, \ceighteeno, and \dco.

\section{Results}
\label{sec:res}

Below, we discuss our results from the dust polarization, continuum, and the spectral-line observations of the three protostars Serpens Emb 8(N), SMM1, and Emb 8. In Figures \ref{fig:emb8N_pola}, \ref{fig:smm1_pol}, and \ref{fig:emb8_pol}, we describe the magnetic field morphology recovered at several spatial scales probed by the ALMA data (for example, dataset C recovered angular scales from 0$\farcs$11 to $\sim$\,$1\farcs$3). In Figures \ref{fig:emb8N_CO_pol}, \ref{fig:smm1_pol_CO}, and \ref{fig:emb8_pol_CO}, we discuss the spatial correlation of the dust continuum emission, magnetic field orientation, and the molecular outflows. In Figures \ref{fig:emb8N_pol_pfrac}, \ref{fig:smm1_pol_pfrac}, and \ref{fig:emb8_pol_pfrac}, we present polarization fraction and polarized intensity maps. Finally, in Figure \ref{fig:emb8N_pol_chemistry}, we compare the dust polarization with the emission of molecular species tracing the dense gas (\thirteencs{}, \ceighteeno{}, and \dco{}) toward Serpens Emb 8(N).

\subsection{Serpens Emb 8(N)}
\label{subsec:8N}

\begin{figure*}[!tbph]
\centering
\smallskip
\begin{tabular}{cc}
    \hspace{-0.27cm}
    \includegraphics[scale=0.35,clip,trim=2.6cm 2.2cm 4.2cm 2cm]{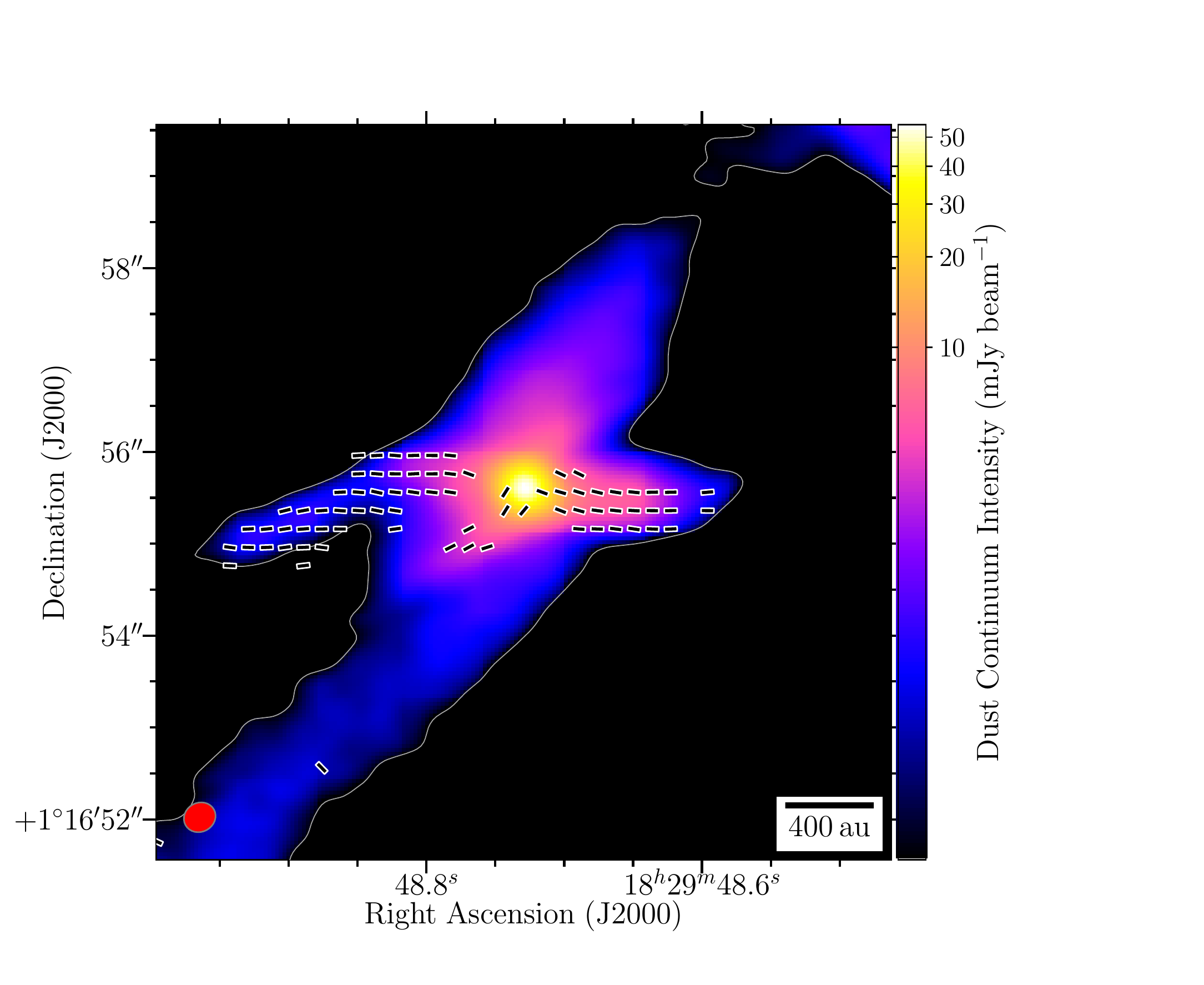}
    &\multirow{2}{*}[4.6cm]{\hspace{-0.4cm}\includegraphics[scale=0.685,clip,trim=0.25cm 0.5cm 2.4cm 2cm]{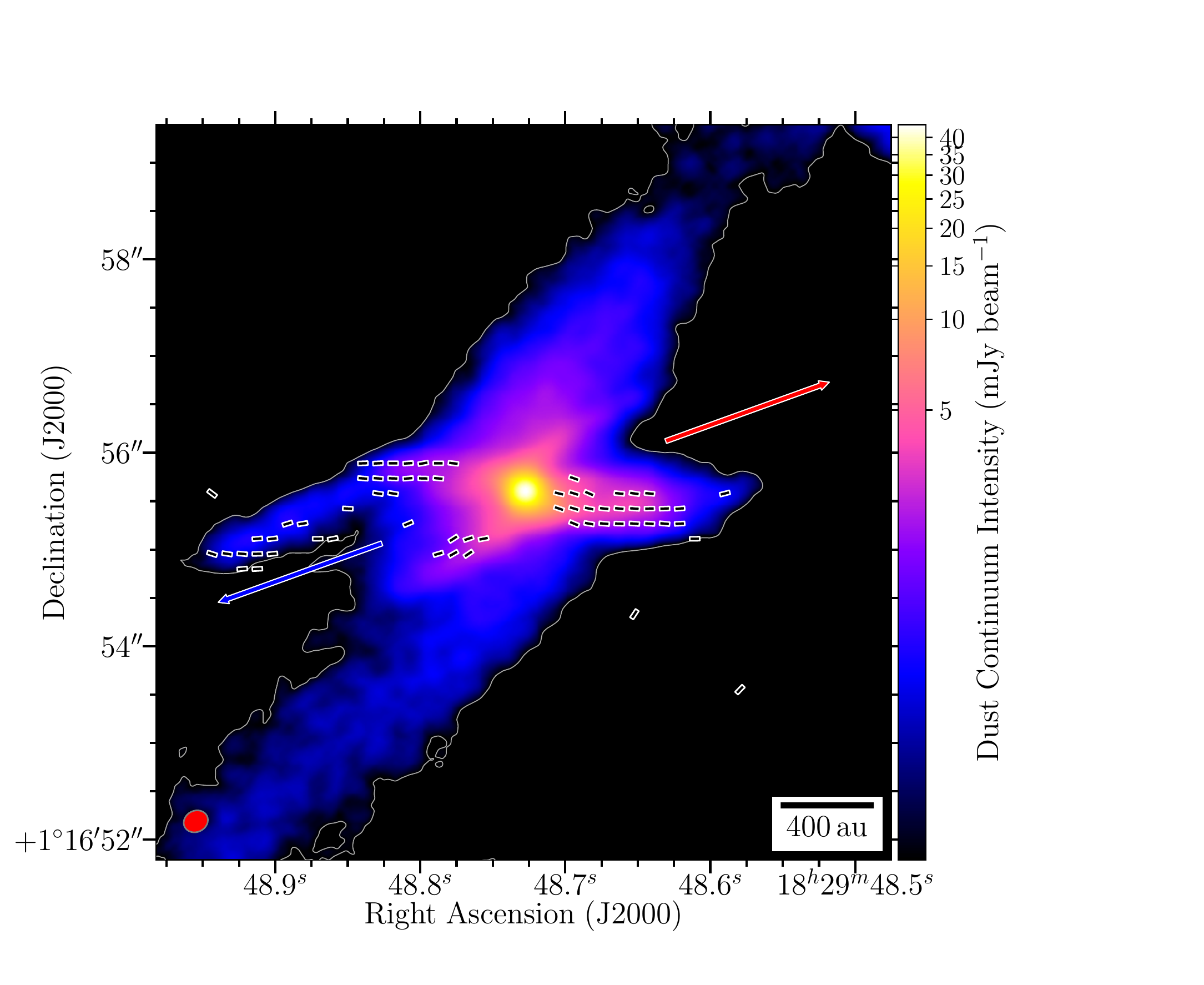}}\\ 
    \hspace{-0.3cm}
    \includegraphics[scale=0.35,clip,trim=3.4cm 2.2cm 4.25cm 2cm]{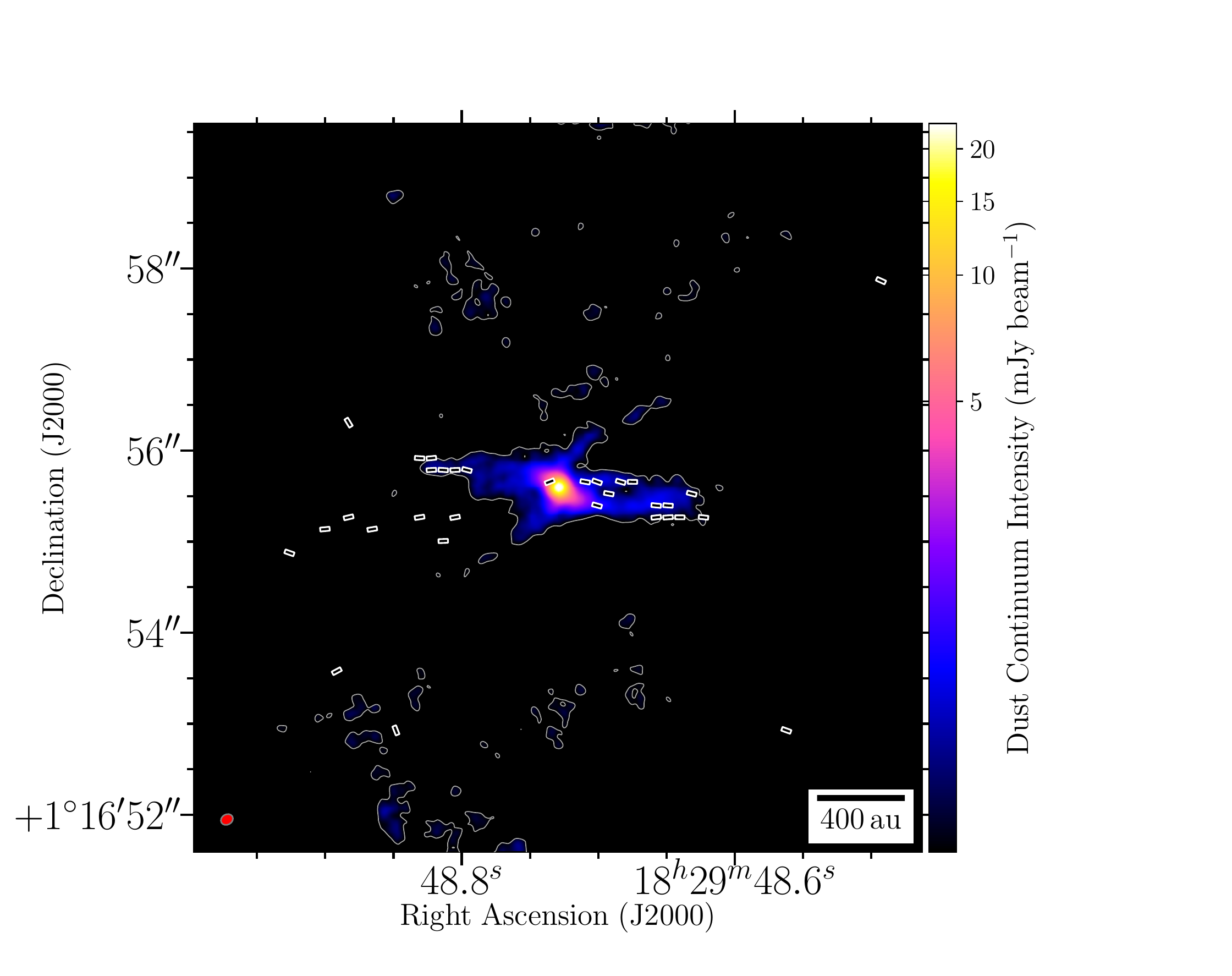} &
\end{tabular}
\smallskip
\smallskip
\smallskip
\caption{\footnotesize
Magnetic field around Serpens Emb 8(N). Line segments represent the magnetic field orientation, rotated by 90$^\circ$ from the dust polarization angle $\chi$ (the length of the segments does not represent any quantity). They are plotted where the polarized intensity $P\,>\,3\sigma_P$. The color scale is the total intensity (Stokes $I$) thermal dust emission, shown from $3\sigma_{I}$. The gray contour indicates the $3\sigma_{I}$ level. \textit{Top Left}: Dataset A, see Table \ref{t.imaging}. $\sigma_P$ = \rmsPEmbEightNdatAUnits{}, $\sigma_I$ = \rmsIEmbEightNdatAUnits{}. The peak polarized and total intensities are \peakPEmbEightNdatAUnits{} and \peakIEmbEightNdatAUnits{}, respectively. The red ellipse in the lower-left corner represents the synthesized beam of ALMA from dataset A only. The beam size is 0$\farcs$35 $\times$ 0$\farcs$32, with a position angle of --61.5$^\circ$. \textit{Right}: Combination of the datasets A and C, see Table \ref{t.imaging} Case-2. $\sigma_P$ = \rmsPEmbEightNcaseIIUnits{}, $\sigma_I$ = \rmsIEmbEightNcaseIIUnits{}. The peak polarized and total intensities are \peakPEmbEightNcaseIIUnits{} and \peakIEmbEightNcaseIIUnits{}, respectively. The blue and red arrows represent the direction of the blueshifted and redshifted lobe of the bipolar outflow, respectively. The beam size is 0$\farcs$26 $\times$ 0$\farcs$22, with a position angle of --64$^\circ$. \textit{Bottom Left}: Combination of the datasets A,B, and C differently for Stokes $I$, $Q$, and $U$, see Table \ref{t.imaging} Case-3. $\sigma_P$ = \rmsPEmbEightNcaseIIIUnits{}, $\sigma_I$ = \rmsIEmbEightNcaseIIIUnits{}. The peak polarized and total intensities are \peakPEmbEightNcaseIIIUnits{} and \peakIEmbEightNcaseIIIUnits{}, respectively. The beam size is 0$\farcs$14 $\times$ 0$\farcs$11, with a position angle of --60.8$^\circ$. \textit{The ALMA data used to make the figures are available in the online version of this publication.}}
\label{fig:emb8N_pola}
\vspace{0.3cm}
\end{figure*}

Serpens Emb 8(N) has never been observed at such high angular resolution. In Figure \ref{fig:emb8N_pola} we show multi-scale observations of the magnetic field and thermal dust continuum emission around the protostar, with spatial resolutions of 146, 105, and 55\,au (from dataset A, Case-2, and Case-3: see Table \ref{t.imaging}). We resolve progressively enhanced dust continuum emission along the outflow cavity walls, which, however becomes faint at the highest angular resolution (Figure \ref{fig:emb8N_pola}, bottom left panel). This type of structure is created by the outflow, which clears the cavity and causes material to accumulate aside the outflow, resulting in a high density, compact feature that is enhanced because of how the emission is spatially filtered by the ALMA interferometer.  The intermediate-resolution map (Figure \ref{fig:emb8N_pola}, right-hand panel) is the one that recovers the highest flux density in both total intensity and polarized dust emission. Most of the features are resolved out in the highest resolution map, which may be because the emission is too faint, resulting in a loss of signal due to a lack of sensitivity in the higher resolution beam. Note that as the polarized intensity is less dynamic-range limited than the total intensity, dust polarization appears where there is no detection of Stokes $I$, especially in the highest resolution maps we present of all three sources.

Apart from the dust emission seen along the outflow walls, a filament is seen in the low- and mid-resolution maps in the dust continuum, orientated NW to SE, which has no obvious relation with the protostellar outflow and appears unpolarized (see Figure \ref{fig:emb8N_pola} top-left and right-hand panels). The mid-resolution map clearly allows us to disentangle the outflow cavity walls (traced by the high resolution polarized emission) from this large, unpolarized filament. We see a clear asymmetry in the polarization of the cavity walls, which may be linked to this filament. On each side of the bipolar outflow, one side of the outflow cavity walls is depolarized: the northern wall on the redshifted side, and the southern wall on the blueshifted side. These depolarized zones overlap with the large scale filament, suggesting that, for example, the polarized emission from the cavity wall and from the large filamentary structure could have blended together, resulting in the depolarization we see. Indeed, the emission from infalling envelope material is known to be polarized (\eg in B335: \citealt{Maury2018}; BHR 71 IRS1: \citealt{Hull2019}; and in SMM1-a, see below). Given the average flux of this filamentary structure, a detection of polarized emission at 3$\sigma_{P}$ would imply a polarization fraction of 20\%. We address the question of the local conditions necessary to enhance the alignment of dust grains in Section \ref{subsec:UV_Cav}; in light of the fact that grain alignment in the filament is not likely to be strongly enhanced, it is possible that the filament appears unpolarized because the recovered continuum is simply too faint to detect polarization. Therefore, the filament may alter the polarized emission at places where it is in the same line of sight with the outflow cavity walls.

This filament does not appear in the highest resolution map (Figure \ref{fig:emb8N_pola} bottom-left panel). If we broadly calculate what would have been its flux at the highest angular resolution given the flux measured in the mid-resolution map (Figure \ref{fig:emb8N_pola} right panel), we obtain a value below the 3$\sigma_I$ threshold, suggesting a lack of sensitivity rather than filtering effect. Moreover, the fact the datasets B and C (Table \ref{t.obs}, \ref{t.imaging}) have significantly less integration time strengthens this hypothesis.

\begin{figure}[!tbh]
\centering
\includegraphics[scale=0.43,clip,trim=0.2cm 0.2cm 1cm 0.2cm]{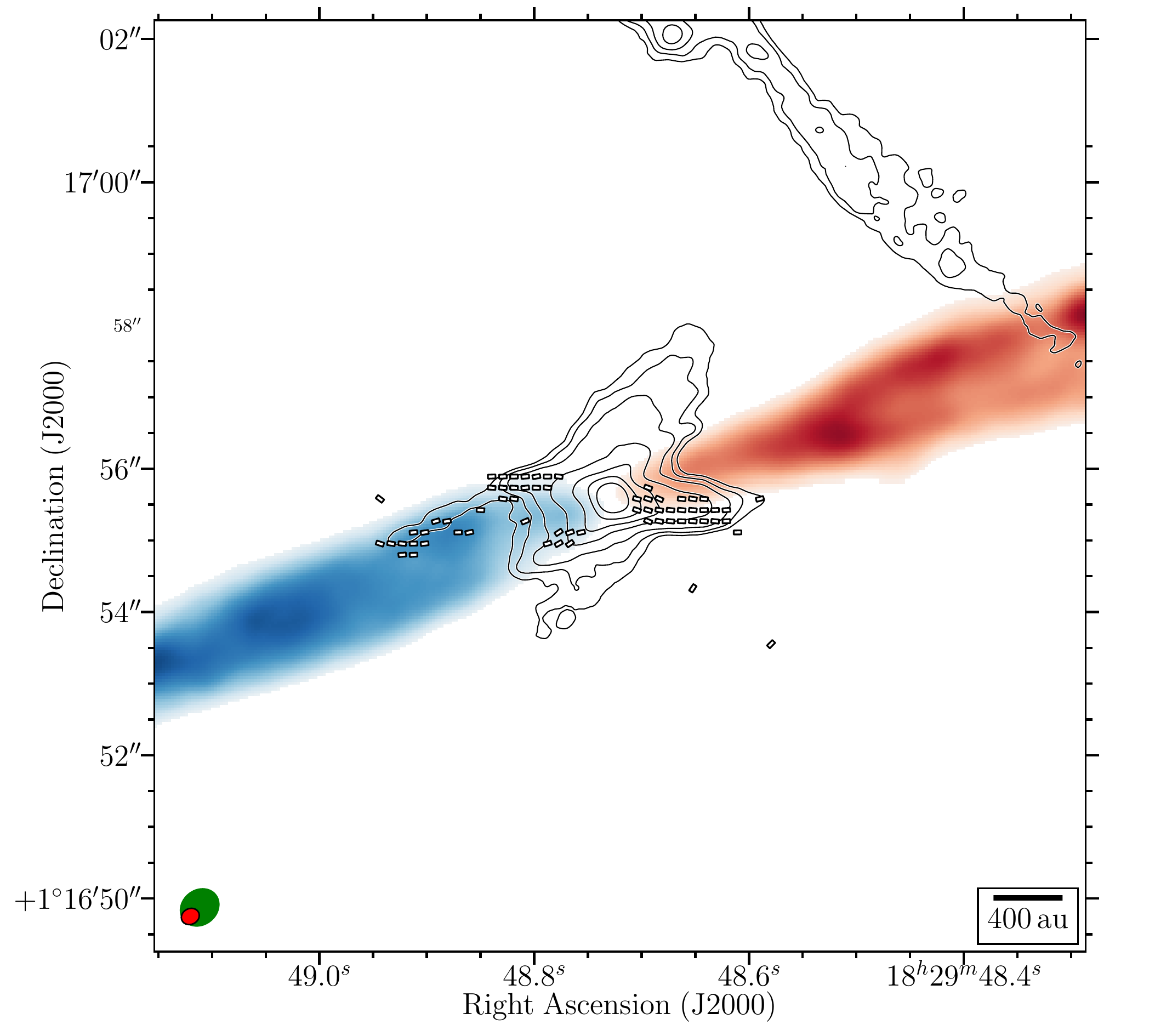}
\caption{\footnotesize Moment 0 map of \co{} in color scale overlaid with the total intensity contours and magnetic field orientations around Serpens Emb 8(N). The moment 0 map is constructed by integrating emission from --53 to 0 \kms (blue) and from 15 to 40 \kms (red). The $v_\textrm{LSR}$ is $\sim$\,8.5\,\kms. The peaks of the red- and blueshifted moment 0 maps are 2.10\,\jybmkms{} and 2.52\,\jybmkms{}, respectively. Same as Figure \ref{fig:emb8N_pola} (right) for the line segments. The black contours trace the dust continuum from the Case-2 (see Table \ref{t.imaging}) at levels of 11, 16, 24, 44, 74, 128, 256 $\times$ $\sigma_I$ , where $\sigma_I$ = \rmsIEmbEightNcaseIIUnits{}. The red ellipse in the lower-left corner represents the synthesized beam of ALMA continuum observations. The beam size is 0$\farcs$26 $\times$ 0$\farcs$22, with a position angle of --64$^\circ$. The green ellipse represents the resolution from the molecular line maps. Its size is 0$\farcs$53 $\times$ 0$\farcs$45. 
}
\label{fig:emb8N_CO_pol}
\vspace{0.3cm}
\end{figure}

\begin{figure}[!tbh]
\centering
\subfigure{\includegraphics[scale=0.42,clip,trim=0.2cm 2.4cm 2.3cm 2cm]{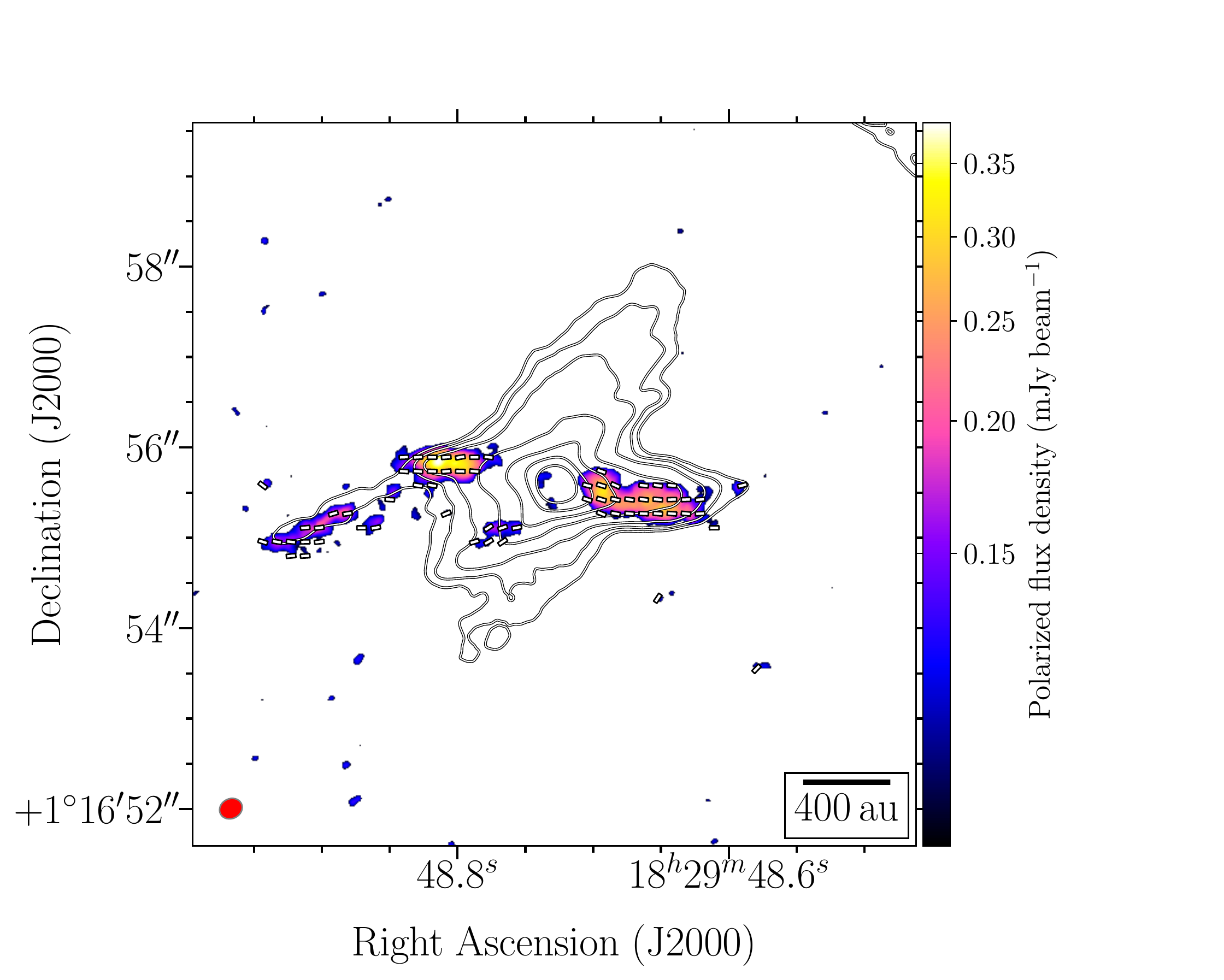}}
\subfigure{\includegraphics[scale=0.42,clip,trim=0.2cm 0.3cm 2.5cm 2cm]{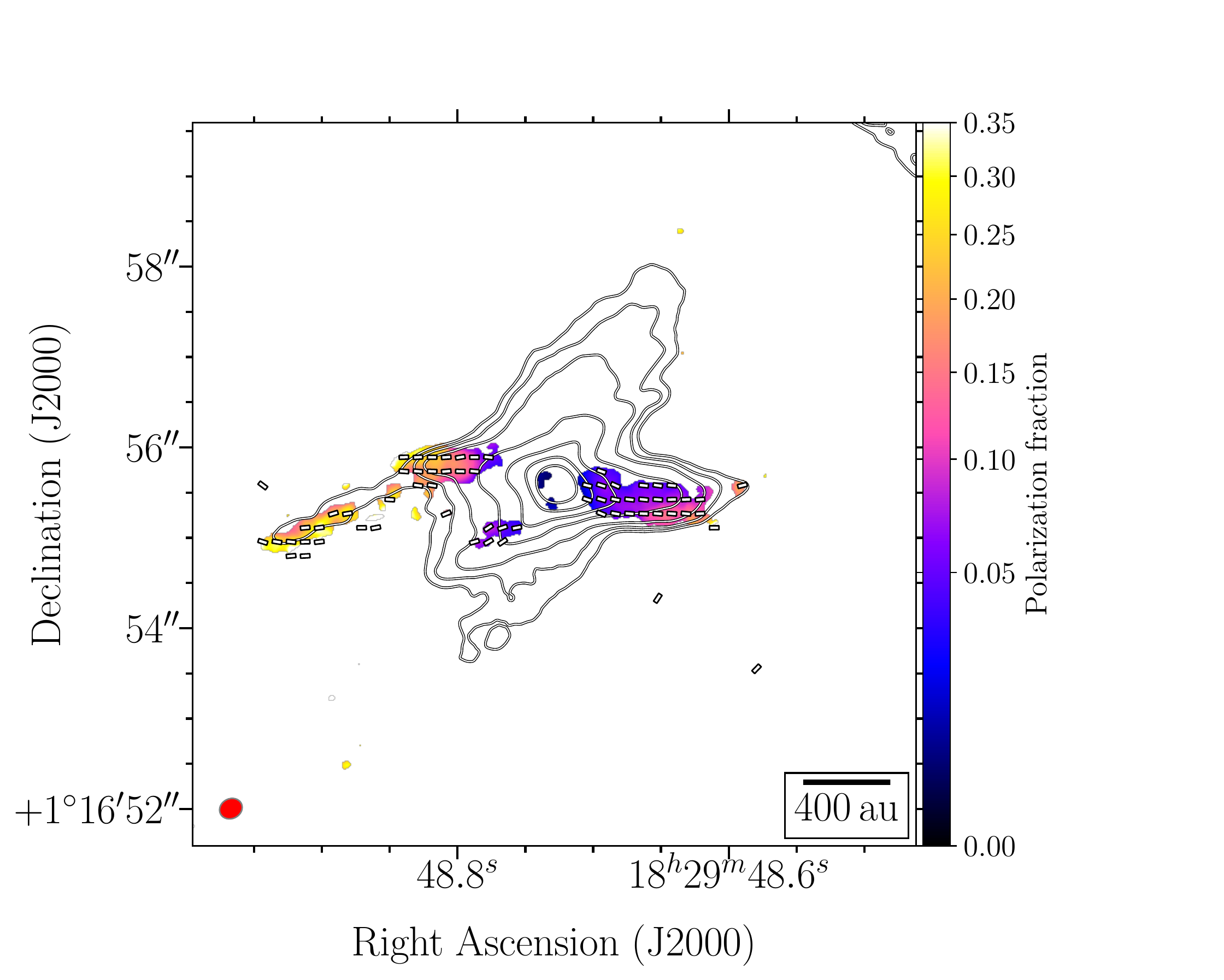}}
\caption[Dust Polarization Intensity and polarization fraction in Serpens Emb 8(N)]{\footnotesize Dust polarization intensity (top) and polarization fraction (bottom) in Serpens Emb 8(N) from Case-2. Same as Figure \ref{fig:emb8N_pola} (right) for the line segments. The color scale in the top panel is the polarized intensity $P$, shown where $P\,>\,3\sigma_P$. The color scale in the bottom panel is the polarized fraction $P_\textrm{frac}$, shown where $P\,>\,3\sigma_P$ and $I\,>\,5\sigma_I$. The peak polarized intensity is \peakPEmbEightNcaseIIUnits{}. The black contours represent the total intensity (Stokes $I$) at the following levels: 11, 16, 24, 44, 74, 128, 256 $\times$ $\sigma_{I}$, where $\sigma_I$ = \rmsIEmbEightNcaseIIUnits{}.}
\label{fig:emb8N_pol_pfrac}
\vspace{0.3cm}
\end{figure}

\begin{figure*}[!tbph]
\centering
\subfigure{\includegraphics[scale=0.4,clip,trim=0.3cm 1.2cm 6.3cm 2.4cm]{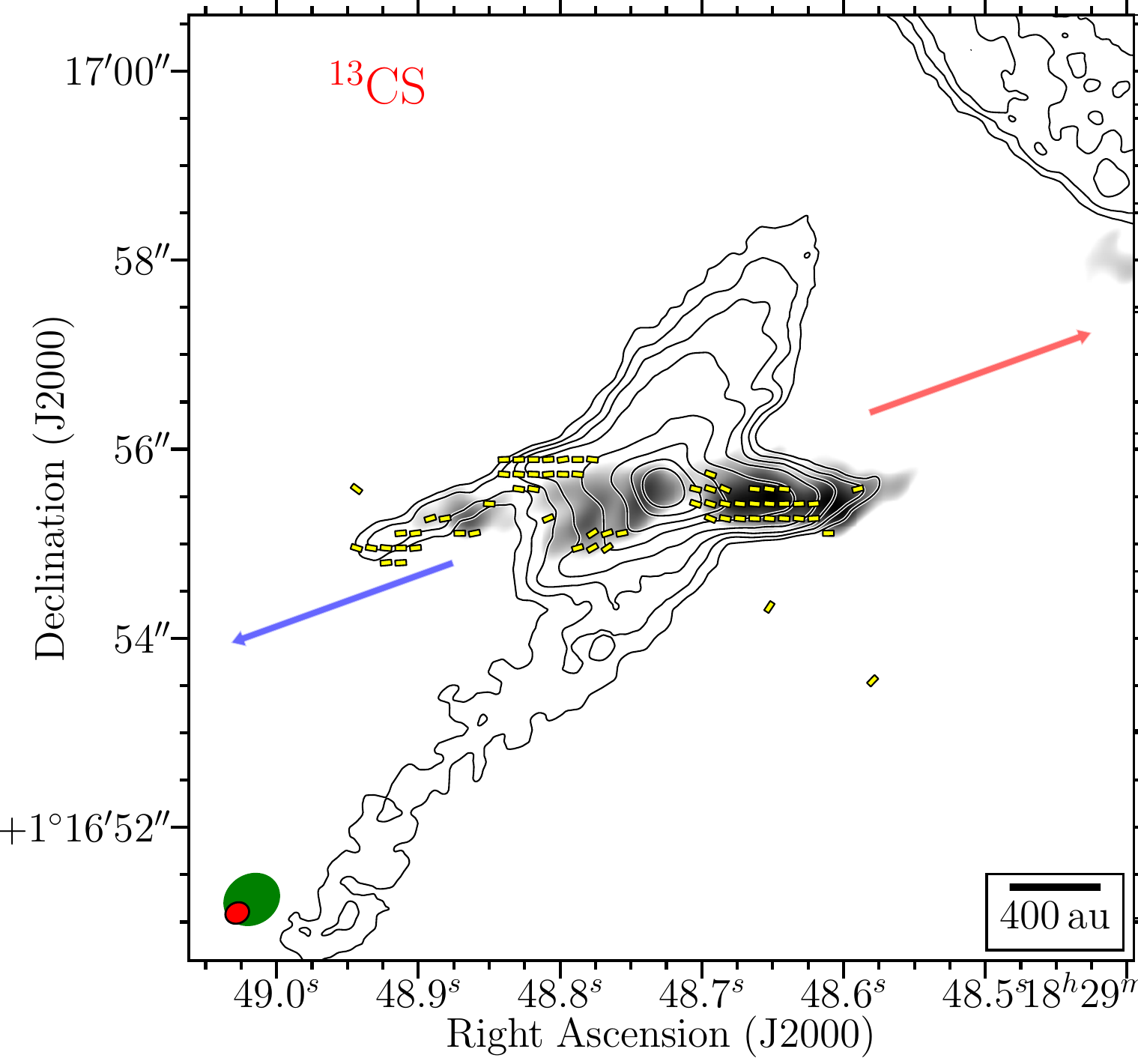}}
\subfigure{\includegraphics[scale=0.4,clip,trim=2.8cm 1.2cm 6.3cm 2.4cm]{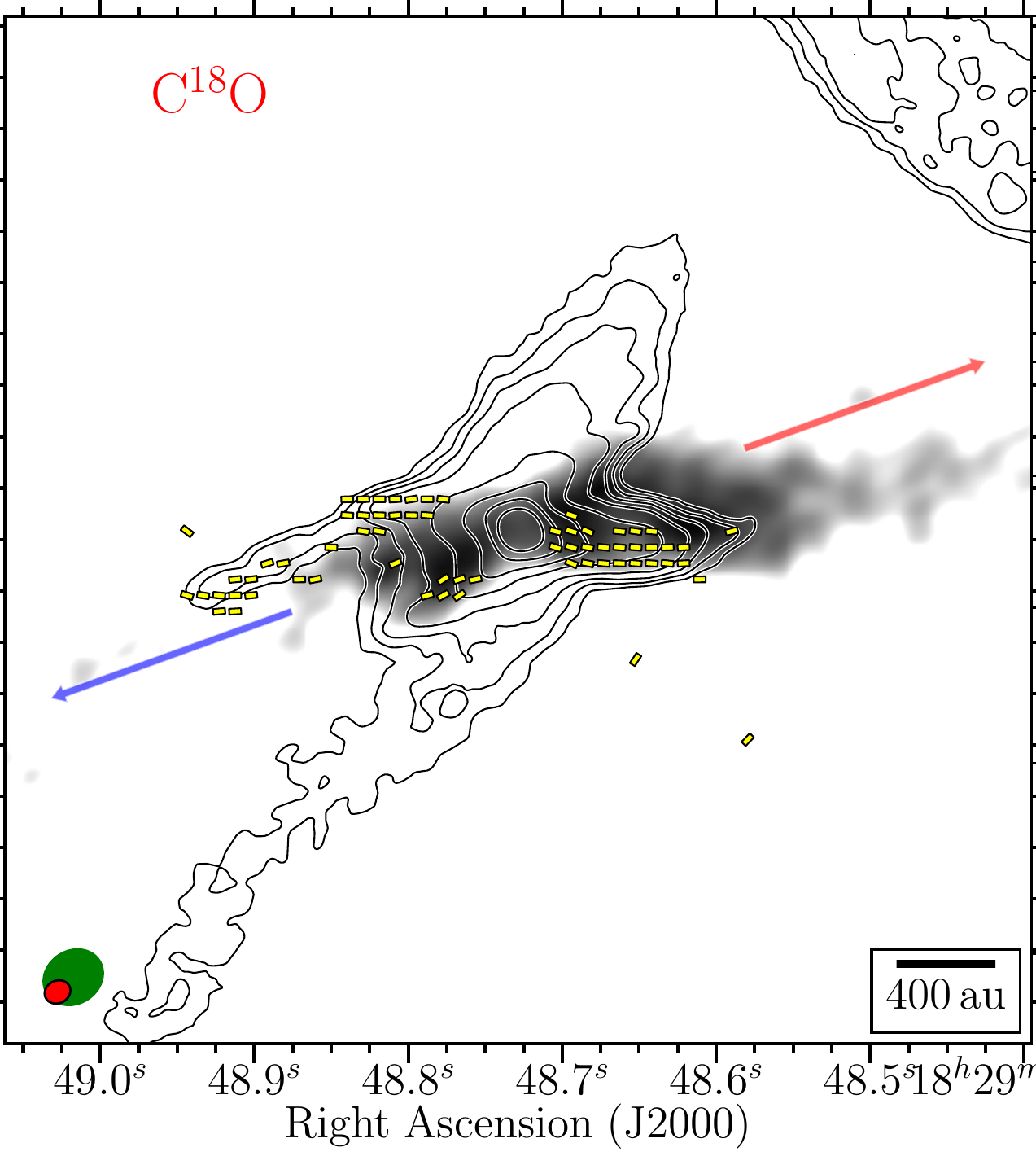}}
\subfigure{\includegraphics[scale=0.4,clip,trim=2.8cm 1.2cm 3.9cm 2.4cm]{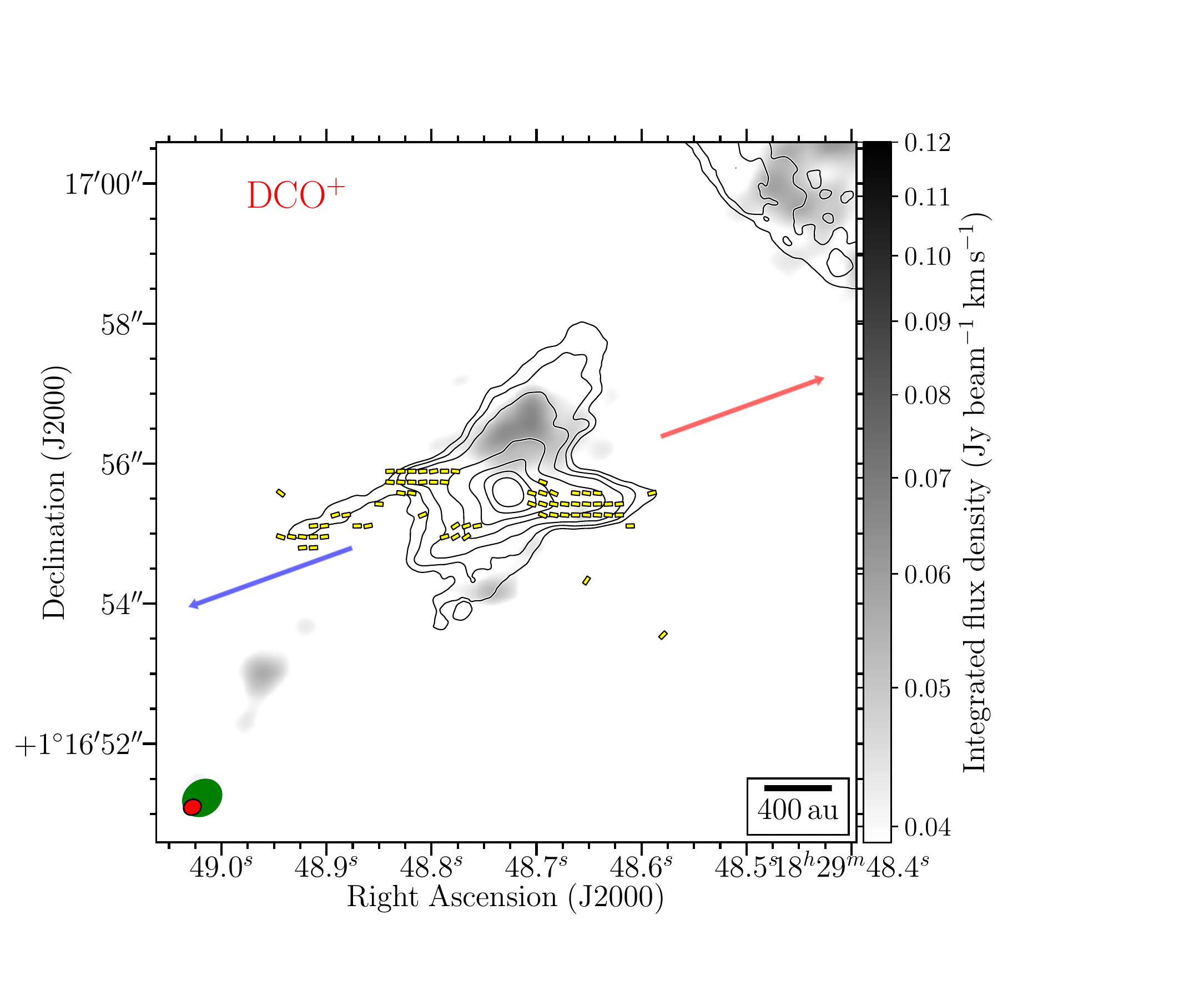}}
\caption{\footnotesize Moment 0 maps of \thirteencs{}, \ceighteeno{}, and \dco{} around Serpens Emb 8(N). The black contours represent the total intensity (Stokes $I$) at the following levels: 7, 11, 16, 24, 44, 74, 128, 256 $\times$ $\sigma_{I}$, where $\sigma_I$ = 55 \mujybm, from Case-2. The $v_\textrm{LSR}$ is about $\sim\,$8.5 \kms. \textit{Left:} Moment 0 map of \thirteencs{} in grayscale constructed by integrated emission from 5 to 12 \kms{}. The rms noise level of the moment 0 map is 16 \mjybmkms{}. The peak of the moment 0 map is 0.12\,\jybmkms{}. \textit{Middle:} Moment 0 map of \ceighteeno{} in grayscale, constructed by integrated emission from 6 to 11.25 \kms{}. The rms noise level of the moment 0 map is 12 \mjybmkms{}. The peak of the moment 0 map is 0.11\,\jybmkms{}. \textit{Right:} Moment 0 map of \dco{} in grayscale, constructed by integrated emission from 8 to 10 \kms{}. The rms noise level of the moment 0 map is 13 \mjybmkms{}. The peak of the moment 0 map is 0.067\,\jybmkms{}.
Same as Figure \ref{fig:emb8N_pola} (right) for the line segments. The red and blue arrows represent the bipolar outflow directions. The beam size of the continuum emission (red ellipse) is 0$\farcs$26 $\times$ 0$\farcs$22, with a position angle of --64$^\circ$. The green ellipse represent the resolution from the molecular line maps. Its size is 0$\farcs$53 $\times$ 0$\farcs$45.
}
\label{fig:emb8N_pol_chemistry}
\vspace{0.3cm}
\end{figure*}

Figure \ref{fig:emb8N_CO_pol} presents the integrated blue- and redshifted \co{} emission around the protostar. CO faithfully traces the outflowing gas in Class 0 protostars as it is still mostly molecular at this stage of protostellar evolution \citep{Arce2007,Panoglou2012}. The results exhibit a pristine, very high velocity,\footnote{Note that we integrated the emission in such a way that we probe both the high velocity and EHV components of the outflow: see \citet{Tychoniec2019} for details.} highly collimated molecular jet, strengthening the above assumption that the structures seen in the dust continuum trace the cavity walls, as they perfectly embrace the CO outflow. It is, however, worth noting that some vectors show up within the CO emission, particularly on the blueshifted side of the outflow. This polarized emission might be still linked to the outflow cavity wall, and simply overlaps with the CO emission because of projection effects. We shall also keep in mind the spatial resolution of the CO emission is twice as coarse as the resolution of the dust continuum. A kinematic study by \citet{Tychoniec2019} found that this collimated jet is consistent with the young age of the source, considering the narrow opening angle and the small dynamical age of the jet (the relative age of Emb 8(N) and Emb 8 is discussed in Section \ref{subsec:field_shaped_outflow}).

The polarized dust intensity and polarization fraction in Serpens Emb 8(N) are shown in Figure \ref{fig:emb8N_pol_pfrac}. As the polarization fraction comes from the ratio of polarized intensity $P$ divided by total intensity $I$, it is important to consider the S/N of both $P$ (color scale) and $I$ (contours) in order to determine whether the corresponding polarization fraction is reliable. Therefore, in order to derive the polarization fraction, we considered only emission within the zones of 3$\sigma_P$ and 5$\sigma_I$, where $\sigma_P$ and $\sigma_I$ are the rms noise level in polarized and total intensity, respectively. Figure \ref{fig:emb8N_pol_pfrac} shows a significant amount of dust polarization along the outflow cavity walls, whereas the central region is unpolarized, at this resolution, where the Stokes $I$ emission peaks. The lower resolution map (Figure \ref{fig:emb8N_pola}, top left panel), however, shows a detection of polarized emission associated with the Stokes $I$ peak, corresponding to a polarization fraction of $\sim$\,0.4$\%$.

It is common to see a ``polarization fraction hole'' where the dust continuum emission peaks. This can be due to collisional dust de-alignment in high density regimes \citep{Lazarian2005,Bethell2007,Pelkonen2009}, or higher magnetic field dispersion at high column density zones, as we know the degree of organisation of the magnetic field is a key point to allow the detection of polarized dust emission \citep{Maury2018}. In our case, we might observe this phenomenon in the inner core of Emb 8 (N).  However, line of sight effects can result in a depolarized signal in the center of the protostar as we see through zones where the redshifted and blueshifted counterparts have both affected the polarized dust emission. This effect can decrease the polarized intensity in the equatorial plane, as has been seen in synthetic observations by \citet{Frau2011}, under the threshold of the $3\,\sigma_P$ level. In addition, at our resolution, a sharp change in magnetic field orientation can produce cancellation of the polarized signal within the beam lying in the equatorial plane, leading to a beam-sized depolarization zone \citep{Kataoka2012,JLee2017,Kwon2019}. However, almost no polarized emission is detected at this location in the highest angular resolution map (Figure \ref{fig:emb8N_pola} bottom left panel), and thus the lack of polarization detection is likely due to low sensitivity at higher resolution (see Table \ref{t.imaging}). Thus, there is probably polarization toward the peak of the dust continuum emission, and more sensitive, high angular resolution observations should be able to recover it. In contrast, the polarization fraction along the outflow cavity walls, within the 5$\sigma_I$ and 3$\sigma_P$ zones, reaches 30$\%$ at 690 au and 36$\%$ at 1150 au from the center of the protostar, along the northern edge of the blueshifted outflow. The southern redshifted outflow cavity wall is less polarized with a maximum in polarization fraction of 25$\%$ at 790 au from the protostar.

Finally, Figure \ref{fig:emb8N_pol_chemistry} presents the integrated intensity maps for the emission of the three dense-gas tracers \thirteencs{}, \ceighteeno{}, and \dco{}, which we compare with the polarized intensity. It is striking to notice how both $^{13}$CS and C$^{18}$O show up roughly where we see polarized continuum emission in an E--W orientation, aligned with the outflow. C$^{18}$O is typically optically thin in protostellar cores, and thus traces high density material that is warm enough to trigger the sublimation of C$^{18}$O that was frozen onto dust grains.  The spatial extent of this molecule has been used as a tracer of protostellar accretion \citep{Visser2015,Jorgensen2015}. $^{13}$CS peaks at the same place as C$^{18}$O, but is less spatially extended, 
and seems to be very well coupled with the dust emission in the SW outflow cavity wall. The kinematics of these two lines did not reveal any evidence of rotation in the inner core, which has been seen previously in molecular line observations of Class 0 disk-envelope systems \citep{Ohashi2014, Yen2015a, LeeCF2016,Jacobsen2018,Hsieh2019}.  Rather, the kinematic information suggests that the gas is linked with the outflow motion. Finally, we present the integrated moment 0 map of the DCO$^+$ emission. This molecule is formed from a reaction between H$_2$D$^+$ and the remnant CO in the gas phase. As low temperatures are essential for deuterium fractionation, DCO$^+$ is known to be a good tracer of the cold, dense material located at the disk-envelope interface \citep{Jorgensen2004b,Jorgensen2011, Murillo2015, Murillo2018}. Its emission appears anticorrelated with the polarization, and rather seems to trace the filamentary structure mentioned above that is crossing over the protostar.  This anticorrelation with dust polarization has also been seen in emission of N$_2$D$^+$ in BHR 71 \citep{Hull2019}, suggesting that tracers of cold, dense material like DCO$^+$, N$_2$D$^+$, and N$_2$H$^+$ are good proxies for probing the conditions necessary for dust-grain alignment in protostellar cores.
Finally, we do not detect an organized velocity gradient in the DCO$^+$ at the 0.5\,\kms{} spectral resolution of our data, and thus the role of the aforementioned, large-scale filamentary structure in the formation of Serpens Emb 8(N) remains to be determined.

\subsection{Serpens SMM1}
\label{subsec:smm1}

\begin{figure*}[!tbph]
\centering
\subfigure{\includegraphics[scale=0.5,clip,trim=0.2cm 0.6cm 2.8cm 2cm]{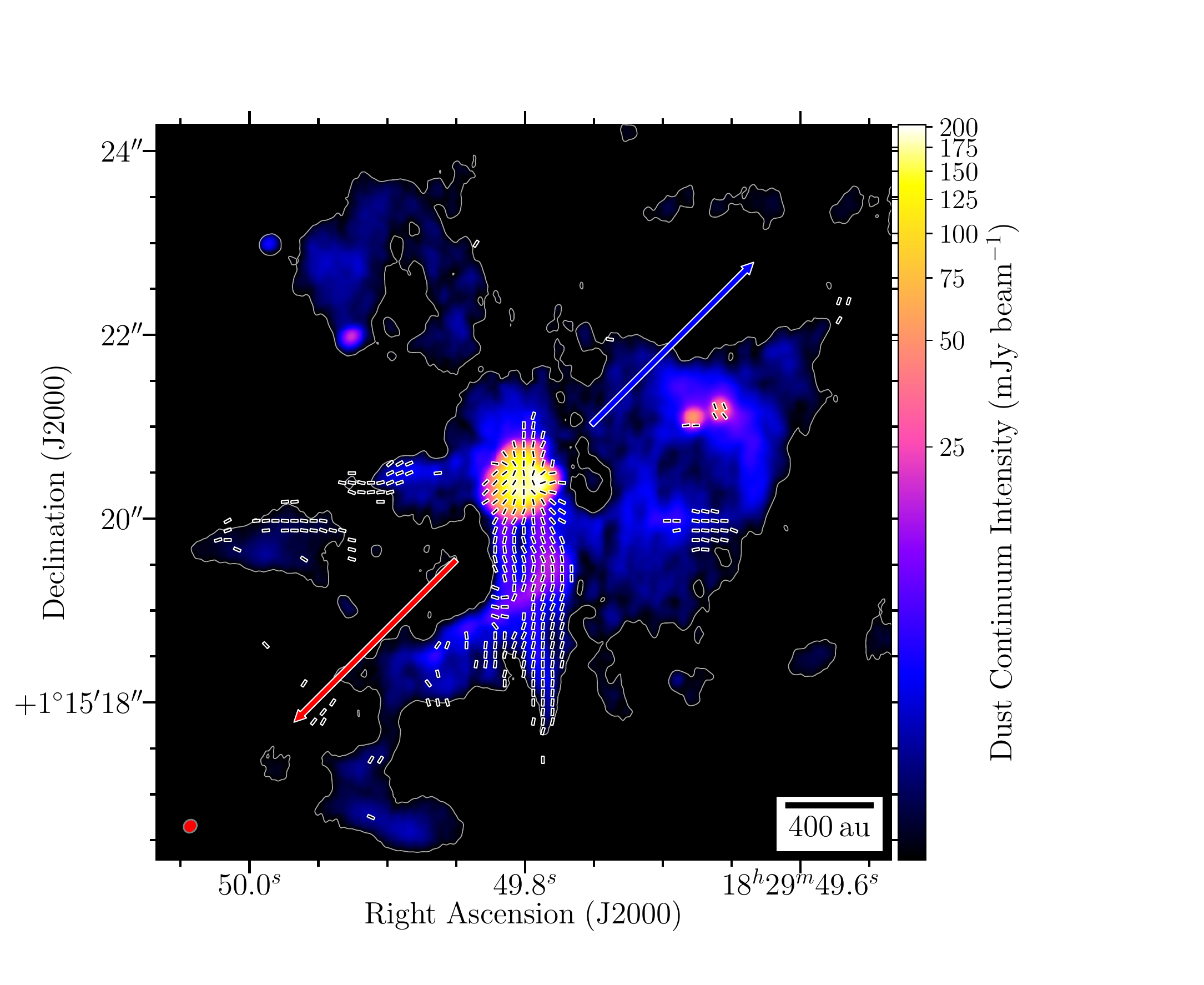}}
\subfigure{\includegraphics[scale=0.5,clip,trim=2.6cm 0.6cm 2.8cm 2cm]{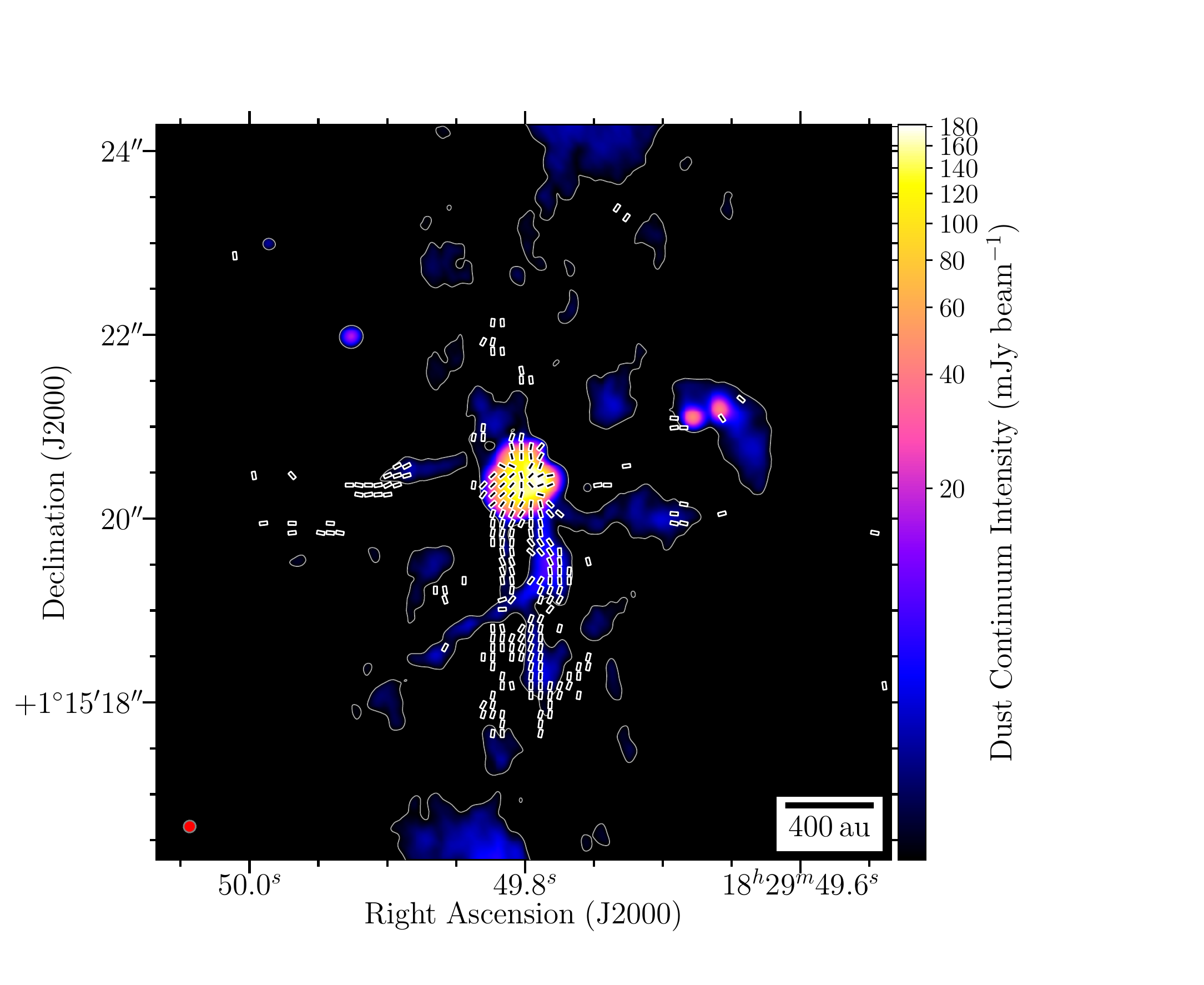}}
\caption{\footnotesize Magnetic field around Serpens SMM1. Line segments represent the magnetic field orientation, rotated by 90$^\circ$ from the dust polarization angle $\chi$ (the length of the segments does not represent any quantity). They are plotted where the polarized intensity $P\,>\,3\sigma_P$. The color scale is the total intensity (Stokes $I$) thermal dust emission, which is shown when $I\,>\,3\sigma_I$. \textit{Left}: Combination of datasets A ,B, and C, see Table \ref{t.imaging} Case-1. $\sigma_P$ = \rmsPSMMcaseIUnits{} and $\sigma_I$ = \rmsISMMcaseIUnits{}. The peak polarized and total intensities are \peakPSMMcaseIUnits{} and \peakISMMcaseIUnits{}, respectively. The red ellipse in the lower-left corner represents the beam size, \ie\xspace 0$\farcs$15 $\times$ 0$\farcs$14, with a position angle of --48.5$^\circ$. \textit{Right}: Combination of the datasets B and C, see Table \ref{t.imaging} Case-3. $\sigma_P$ = \rmsPSMMcaseIIIUnits{} and $\sigma_I$ = \rmsISMMcaseIIIUnits{}. The peak polarized and total intensities are \peakPSMMcaseIIIUnits{} and \peakISMMcaseIIIUnits{}, respectively. The red ellipse in the lower-left corner represents the beam size, \ie 0$\farcs$13 $\times$ 0$\farcs$13, with a position angle of --58.8$^\circ$. \textit{The ALMA data used to make the figures are available in the online version of this publication.}}
\label{fig:smm1_pol}
\vspace{0.3cm}
\end{figure*}

\begin{figure}[!tbph]
\centering
\includegraphics[scale=0.428,clip,trim=0.2cm -0.7cm 1cm 0.2cm]{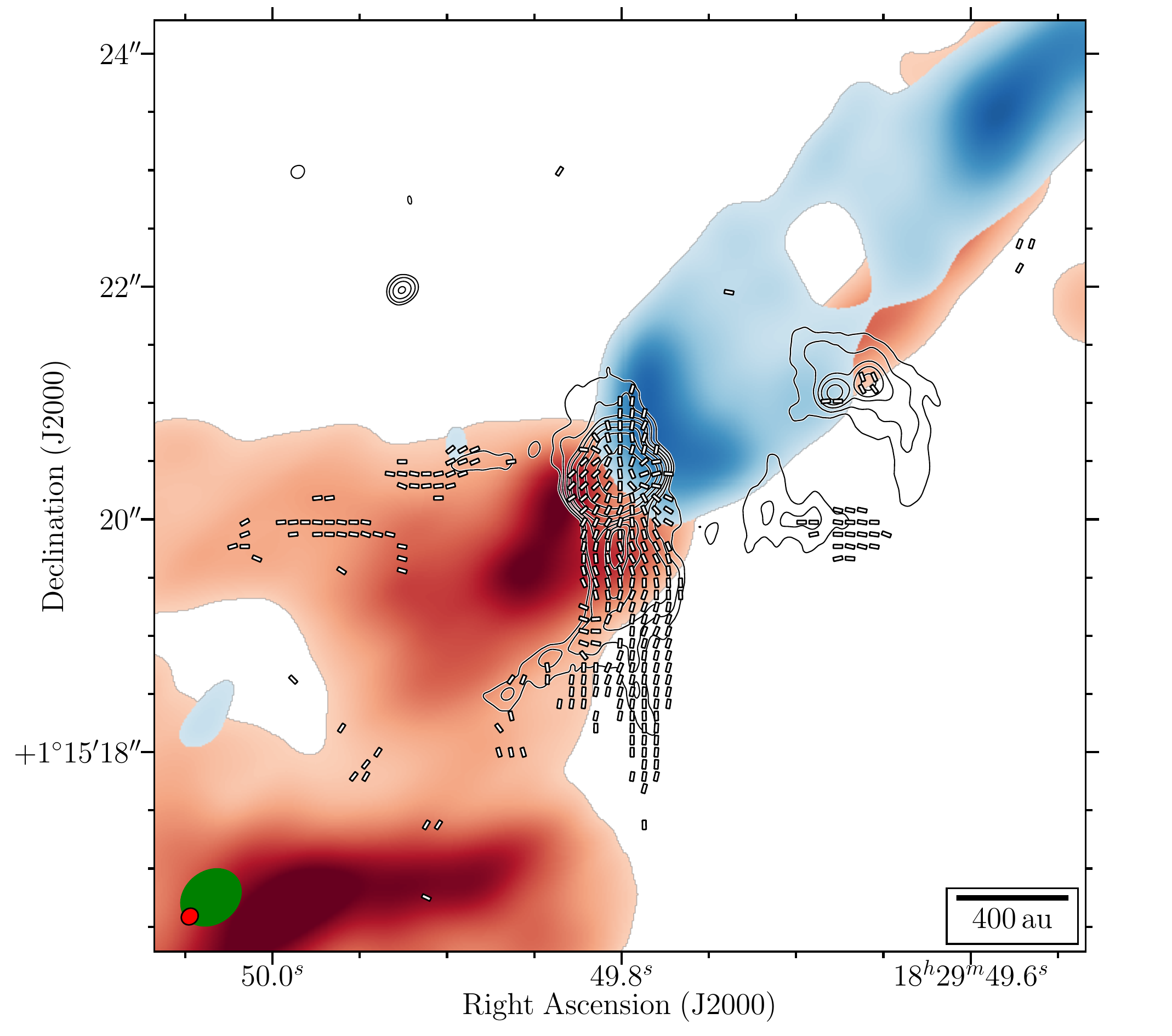}
\caption{\footnotesize Moment 0 map of \co{} in color scale overlaid with the total intensity contours and magnetic field orientations around Serpens SMM1. Same as Figure \ref{fig:smm1_pol} (left) for the line segments. The moment 0 in color scale is constructed by integrating emission from --13 to 4  \kms (blue) and from 10.5 to 30 \kms (red). The $v_\textrm{LSR}$ is $\sim$\,8.5\kms. The peaks of the red- and blueshifted moment 0 maps are 5.40\,\jybmkms{} and 3.90\,\jybmkms{}, respectively. The black contours tracing the dust continuum are 8, 12, 20, 32, 64 $\times$ the rms noise $\sigma_I$ in the Stokes $I$ map, where $\sigma_I$ = \rmsISMMcaseIUnits{}.  The red ellipse in the lower-left corner represents the synthesized beam after combining datasets A, B, and C. The beam size is 0$\farcs$15 $\times$ 0$\farcs$14, with a position angle of --48.5$^\circ$. The green ellipse represents the resolution from the molecular line map, and measures 0$\farcs$53 $\times$ 0$\farcs$45.}
\label{fig:smm1_pol_CO}
\vspace{0.3cm}
\end{figure}

\begin{figure}[!tbph]
\centering
\subfigure{\includegraphics[scale=0.46,clip,trim=0.2cm 2.2cm 3.1cm 2cm]{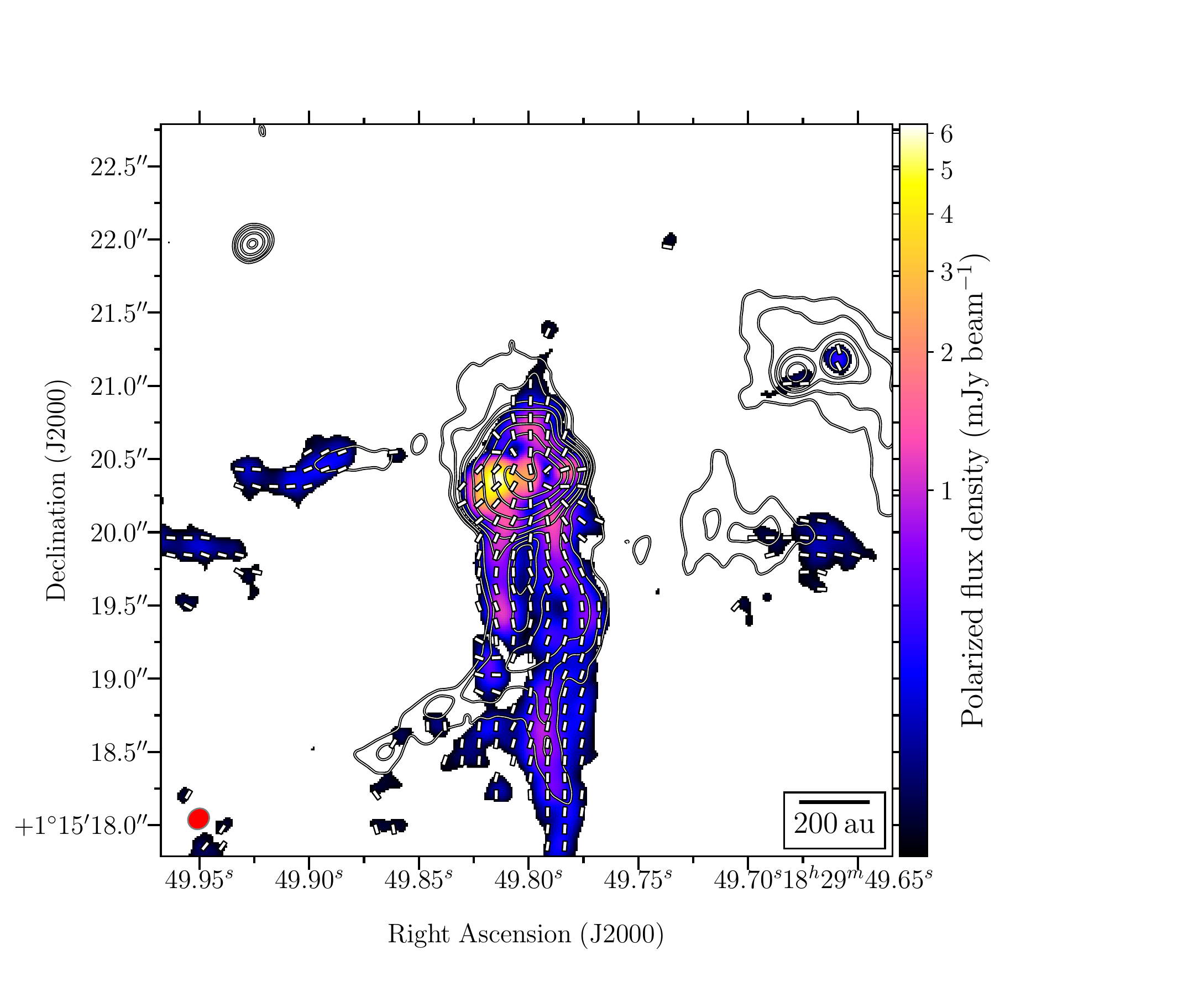}}
\subfigure{\includegraphics[scale=0.46,clip,trim=0.2cm 0.3cm 3.1cm 2cm]{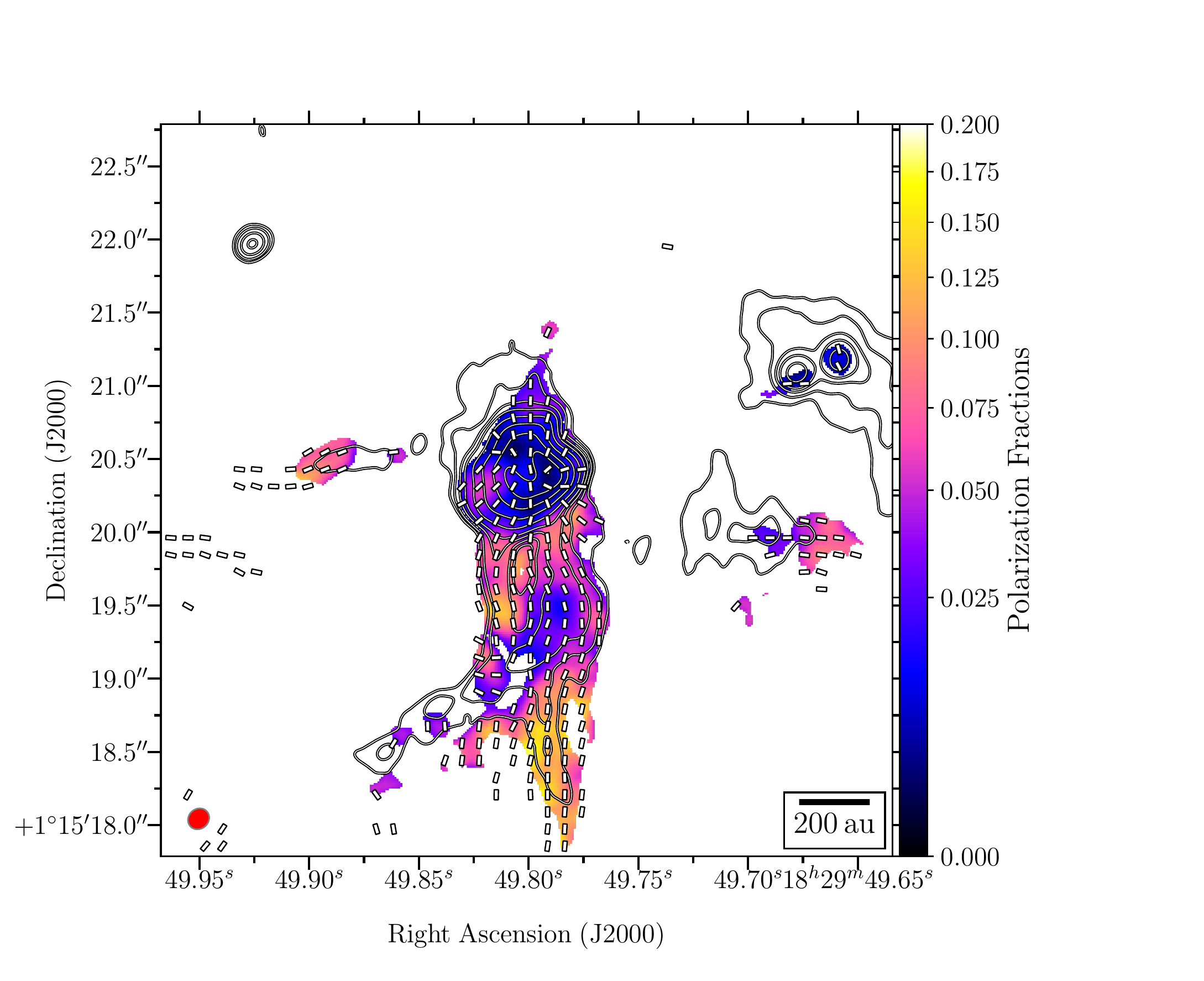}}
\caption{\footnotesize Dust polarization intensity (\textit{top}) and polarization fraction (\textit{bottom}) in SMM1a, from Case-1. Same as Figure \ref{fig:smm1_pol} (left) for the line segments. The black contours tracing the dust continuum are 8, 12, 20, 32, 64, 128, 220, 300 $\times$ the rms noise level  $\sigma_I$ in the Stokes $I$ map, where $\sigma_I$ = \rmsISMMcaseIUnits{}. The color scale in the top panel is the polarized intensity $P$, which is shown where $P\,>\,3\sigma_P$. The color scale in the bottom panel is the polarization fraction $P_\textrm{frac}$, which is shown where $P\,>\,3\sigma_P$ and $I\,>\,5\sigma_I$. The peak polarized intensity is \peakISMMcaseIUnits{}. The red ellipse in the lower-left corner represents the synthesized beam after combining datasets A, B, and C and measures 0$\farcs$15 $\times$ 0$\farcs$14, with a position angle of --48.5$^\circ$.}
\label{fig:smm1_pol_pfrac}
\vspace{0.3cm}
\end{figure}

We now present the results of our second source Serpens SMM1, an intermediate-mass Class-0 protostellar core. In Figure \ref{fig:smm1_pol} we present the magnetic field orientations and thermal dust continuum emission. The \co{} integrated emission tracing the low-velocity bipolar outflow around the protostar is shown in Figure \ref{fig:smm1_pol_CO}. Lower resolution polarization observations of this source (as well as the CO map presented here) were first published in \citet{Hull2017b}. They found that the dust along the edges of the cavity of the  wide-angle, low-velocity redshifted outflow was highly polarized. However, they did not detect any polarization that was clearly relate to the redshifted EHV jet of SMM1, reported in \citet{Hull2016a}. The results from our higher angular resolution observations in Figures \ref{fig:smm1_pol} and \ref{fig:smm1_pol_pfrac} present a more complex picture of the dust emission and magnetic field morphology, with an angular resolution reaching 0$\farcs$15$\times$0$\farcs$14 ($\sim$\,57 au).

SMM1-c, SMM1-d, and the two binary components of SMM1-b are now totally resolved (see Appendix \ref{sec:smm1_scheme} for a scheme of the SMM1 core). The polarization to the west of SMM1-a and to the south of SMM1-b has been mostly resolved out compared with the lower resolution observations. Just a few remaining polarization detections are seen toward the binary and just to the south of them, consistent with the results from \citet{Hull2017b}. To the east of SMM1-a we see weak polarization toward the E--W cavity edge, which was already relatively faint in the lower angular resolution data. Our results, however, show a clear filamentary structure to the South of SMM1-a, visible in polarization and total intensity. This structure consists of two highly polarized filaments (from now on designated as the Eastern and Western filaments, see Appendix \ref{sec:smm1_scheme} for a schematic presentation of the different features in SMM1), which have magnetic fields that clearly lie along their major axes. These two filaments observed to the South of SMM1-a appear to be connected to the central core, i.e., the resolved hot corino of SMM1-a, which exhibits a complex polarization pattern. We discuss the possible physical origin of these filaments in Section \ref{subsec:field_shaped_outflow}.

In Figure \ref{fig:smm1_pol_pfrac} we present the dust polarization intensity and polarization fraction around SMM1-a. It is immediately apparent that the two filaments to the south of SMM1-a exhibit high polarization fractions, reaching values of 10\% or higher (the Eastern filament reaches a maximum of 20\%). 
In the zone where the two filaments appear to cross, the polarization intensity and orientation suggest that the superposition in the plane of the sky of the emission emanating from the two filaments has caused the polarization to cancel. Indeed, where the two filaments cross there is a clearly depolarized zone the size of the beam (Figure \ref{fig:smm1_pol_pfrac}). This strengthens the idea that these filaments are two separate structures. Moreover, to the east of the depolarized zone the magnetic field orientation is a bit offset from the major axis of the Eastern filament. This suggests the polarization in this location is coming from both filaments, resulting in an average magnetic field orientation that is not perfectly aligned with either of the two filaments. Finally, to the South of the curved Western filament, the polarization is orientated perfectly N--S along the straight Eastern filament. The potential causes of these highly polarized filaments are discussed in Section \ref{subsec:field_shaped_outflow}.

As introduced in Section \ref{subsec:8N}, we observe the ``polarization hole'' phenomenon in the inner core of SMM1-a, where we see a clear difference in polarization fraction between the hot corino and the two southern filaments. However, inside this central region (within the $\sim$\,32$\sigma_I$ level, \ie above the 9\% level of the peak total intensity), the polarization appears quite inhomogeneous.  Both the polarized intensity and the polarization fraction (achieving a maximum of 6\%) exhibit strong peaks to the SE of the Stokes $I$ peak. This highly polarized spot clearly contrasts with the remaining area within this central zone, which on average has a polarization fraction of $\sim$\,1$\%$. At first glance, the inferred magnetic field orientation in this central region appears quite radial. We discuss in Sections \ref{subsec:pola_pattern} and \ref{subsec:poloi_field} the potential causes of this polarization pattern in SMM1-a, which we attribute primarily to a poloidal magnetic field morphology.

\begin{figure*}[!tbph]
\centering
\subfigure{\includegraphics[scale=0.5,clip,trim=0.2cm 0.6cm 3cm 2cm]{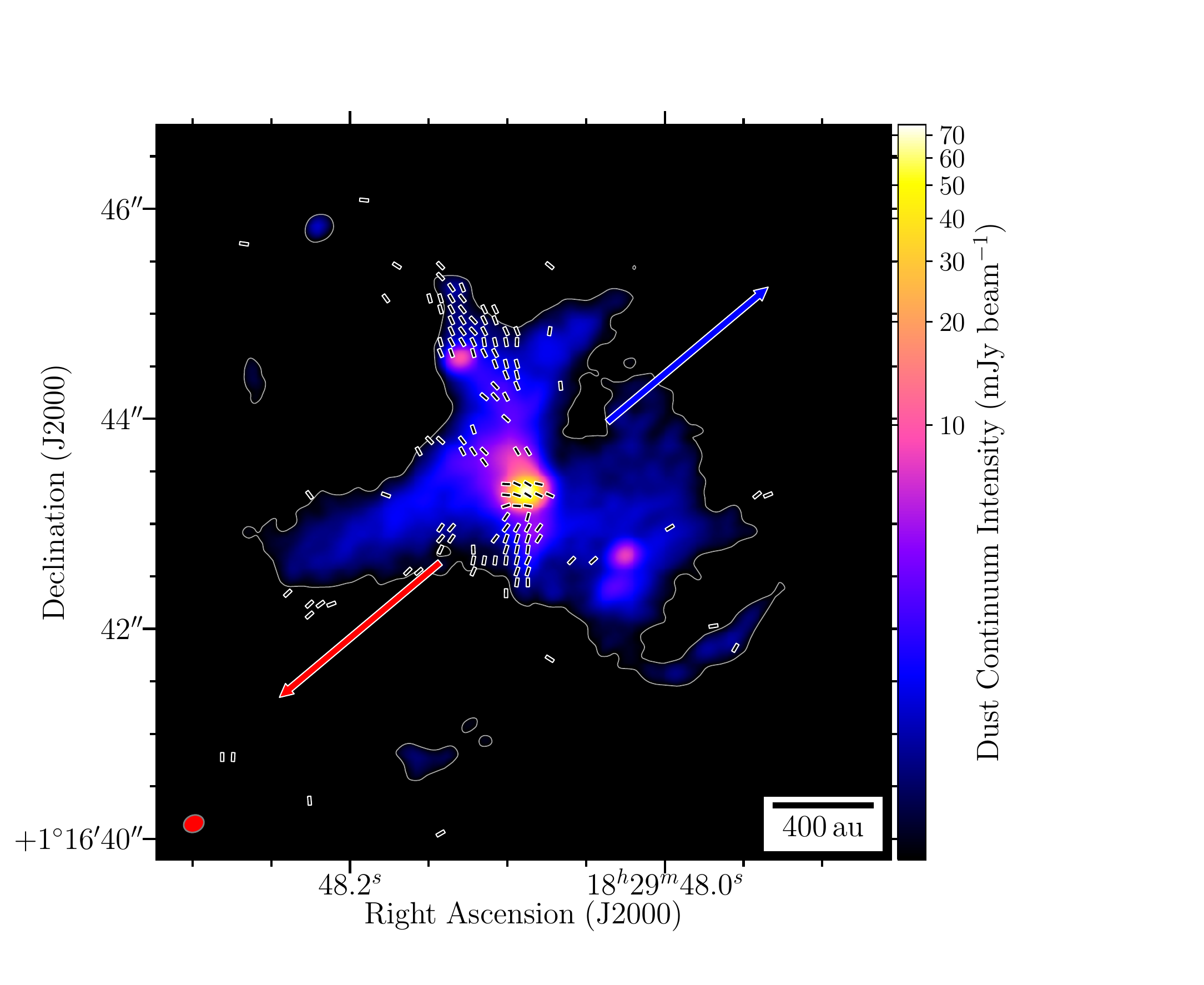}}
\subfigure{\includegraphics[scale=0.5,clip,trim=2.6cm 0.6cm 3cm 2cm]{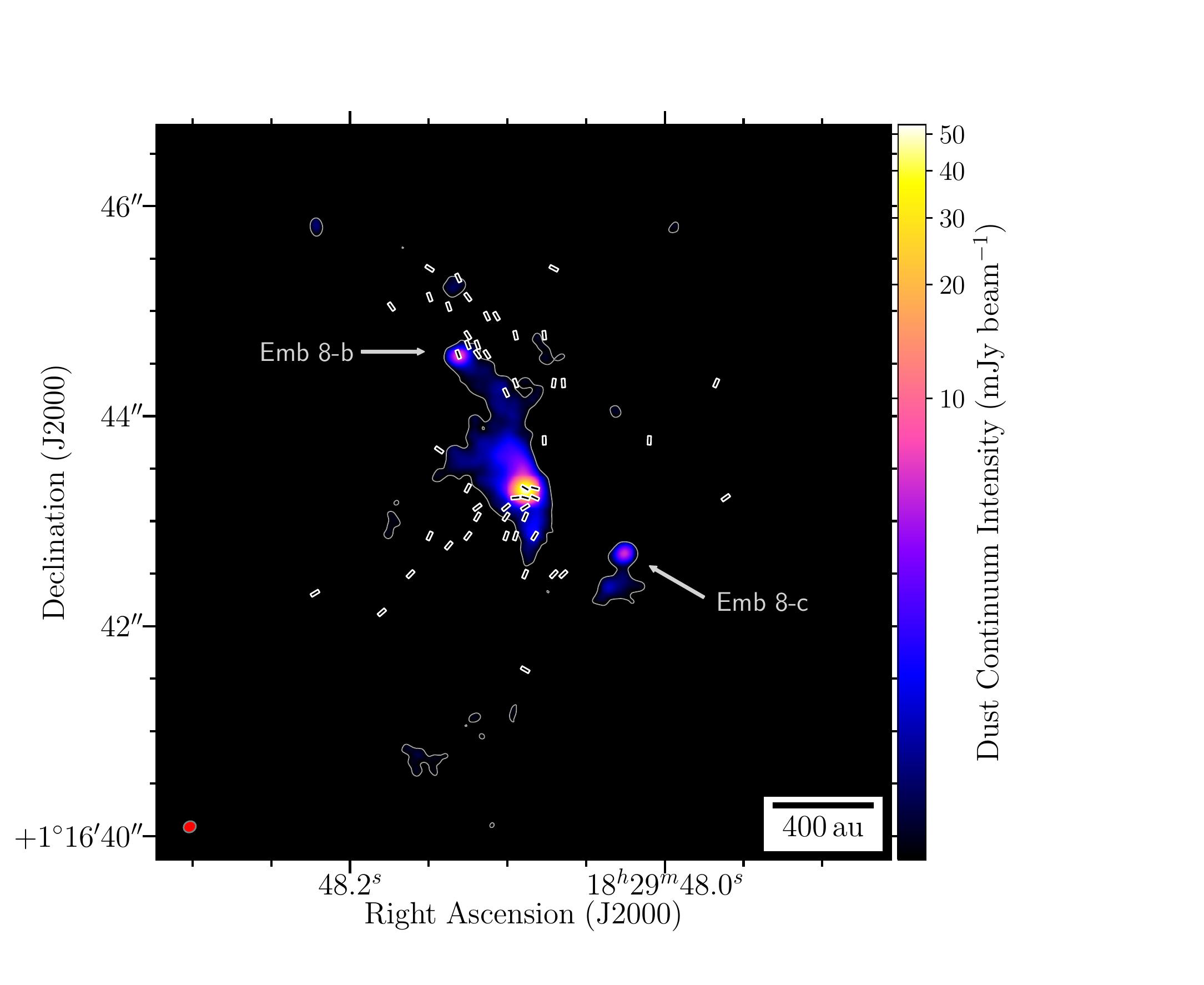}}
\caption{\footnotesize Magnetic field around Serpens Emb 8 from Case-1. Line segments represent the magnetic field orientation, rotated by 90$^\circ$ from the dust polarization angle $\chi$ (the length of the segments does not represent any quantity). They are plotted where the polarized intensity $P\,>\,3\sigma_P$. The color scale is the total intensity (Stokes $I$) thermal dust emission, which is shown where $I\,>\,3\sigma_I$. \textit{Left}: Combination of datasets A and C, see Table \ref{t.imaging} Case-2. $\sigma_P$ = \rmsPEmbEightcaseIIUnits{} and $\sigma_I$ = \rmsIEmbEightcaseIIUnits{}. The peak polarized and total intensities are \peakPEmbEightcaseIIUnits{} and \peakIEmbEightcaseIIUnits{}, respectively. The red ellipse in the lower-left corner represents the beam size, \ie 0$\farcs$20 $\times$ 0$\farcs$16, with a position angle of --67$^\circ$. \textit{Right}: Combination of datasets B and C, see Table \ref{t.imaging} Case-3. $\sigma_P$ = \rmsPEmbEightcaseIIIUnits{} and $\sigma_I$ = \rmsIEmbEightcaseIIIUnits{}. The peak polarized and total intensities are \peakPEmbEightcaseIIIUnits{} and \peakIEmbEightcaseIIIUnits{}, respectively. The red ellipse in the lower-left corner represents the beam size, \ie 0$\farcs$12 $\times$ 0$\farcs$11, with a position angle of --62.7$^\circ$. \textit{The ALMA data used to make the figures are available in the online version of this publication.}}
\label{fig:emb8_pol}
\vspace{0.3cm}
\end{figure*}

\begin{figure}[!tbh]
\centering
\includegraphics[scale=0.43,clip,trim=0.2cm 0.2cm 1cm 0.2cm]{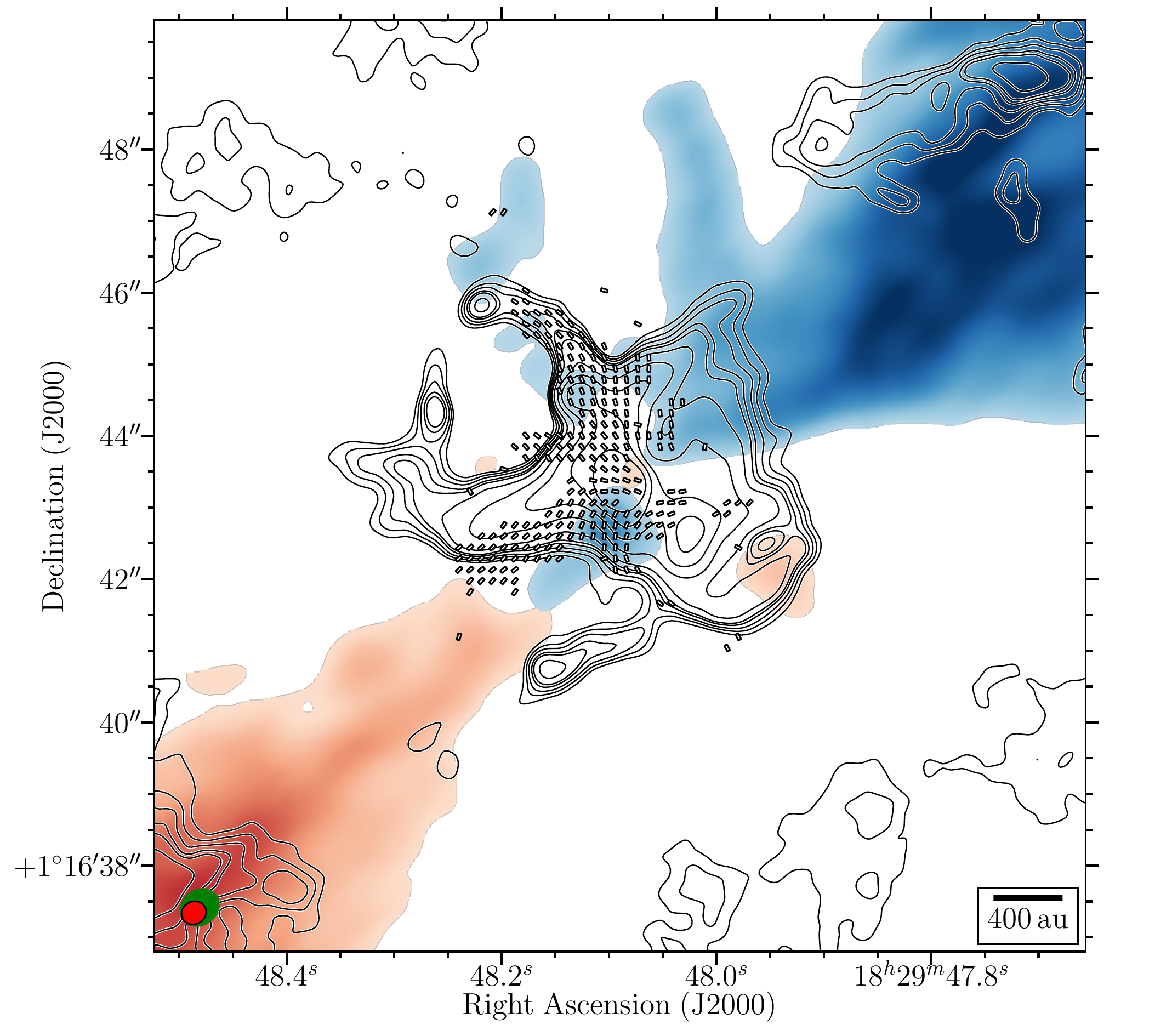}
\caption{\footnotesize Moment 0 map of \co{} in color scale overlaid with the total intensity contours and magnetic field orientations around Serpens Emb 8. Line segments represent the magnetic field orientation, rotated by 90$^\circ$ from the dust polarization angle $\chi$ (the length of the segments does not represent any quantity). They are plotted where the polarized intensity $P\,>\,3\sigma_P$, where $\sigma_P$ = \rmsPEmbEightdatAUnits{}. The moment 0 in color scale is constructed by integrating emission from --10 to 6.5 \kms (blue) and from 11 to 21 \kms (red). The $v_\textrm{LSR}$ is $\sim$\,8.5\,\kms. The peaks of the red- and blueshifted moment 0 maps are 1.65\,\jybmkms{} and 1.68\,\jybmkms{}, respectively. The black contours, which are tracing the dust continuum from the dataset A (see Table \ref{t.imaging}), are 3, 6, 9, 13, 16, 24, 44, 74, 128 $\times$ the rms noise $\sigma_I$ in the Stokes $I$ map, where $\sigma_I$ = \rmsIEmbEightdatAUnits{}.  The red ellipse in the lower-left corner represents the synthesized beam of ALMA continuum observations. The beam size is 0$\farcs$35 $\times$ 0$\farcs$32, with a position angle of --63$^\circ$. The green ellipse represents the resolution from the molecular line maps, measuring 0$\farcs$53 $\times$ 0$\farcs$45.}
\label{fig:emb8_pol_CO}
\vspace{0.3cm}
\end{figure}

\begin{figure}[!tbh]
\centering
\subfigure{\includegraphics[scale=0.42,clip,trim=0.2cm 2.1cm 2.3cm 2cm]{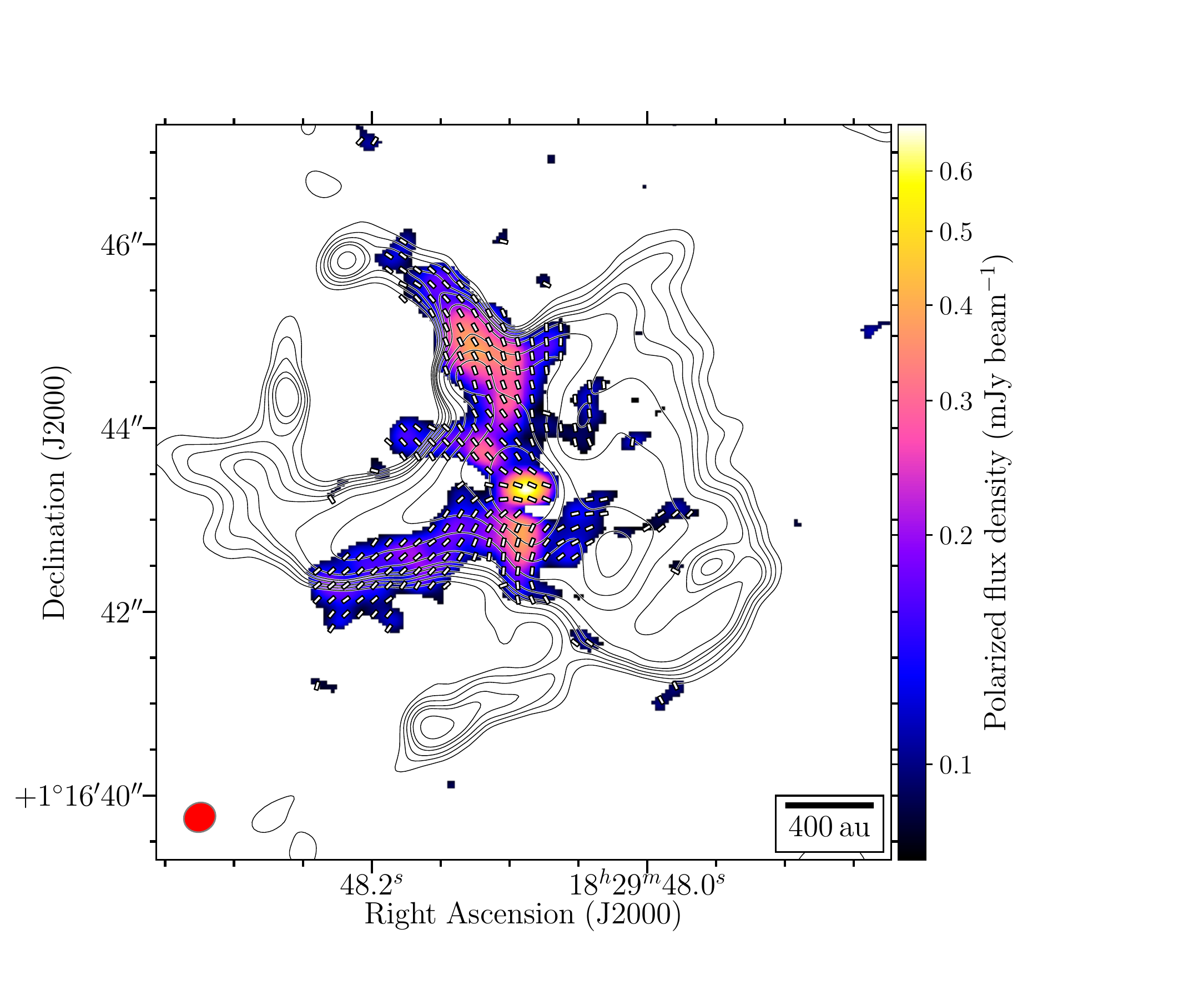}}
\subfigure{\includegraphics[scale=0.42,clip,trim=0.2cm 0.3cm 2.5cm 2cm]{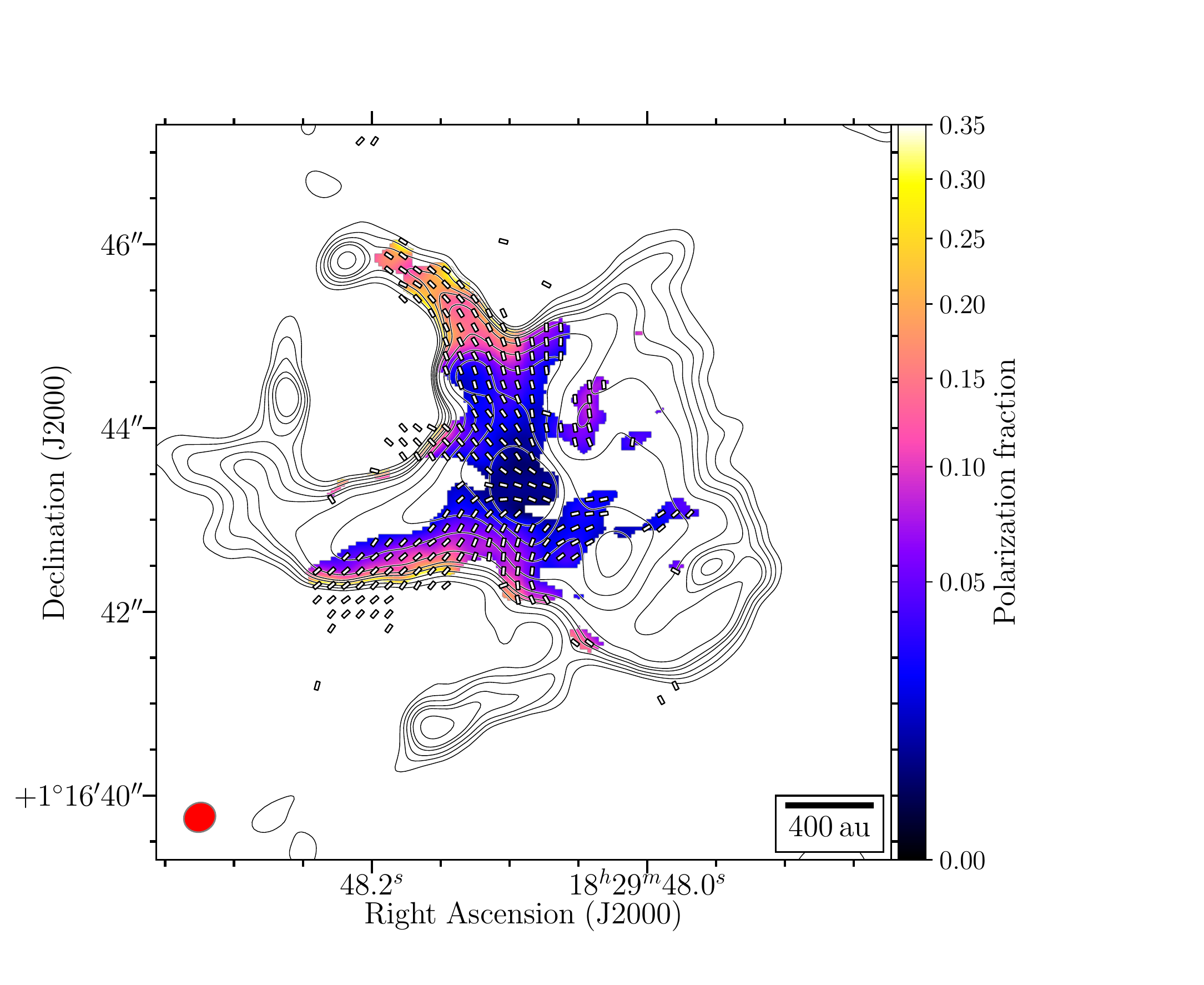}}
\caption{\footnotesize Dust polarization intensity (\textit{top}) and polarization fraction (\textit{bottom}) in Serpens Emb 8 from Dataset A. Same as Figure \ref{fig:emb8_pol_CO} for the line segments and the Stokes $I$ contours. The color scale in the top panel is the polarized intensity $P$, which is shown where $P\,>\,3\sigma_P$. The color scale in the bottom panel is the polarization fraction $P_\textrm{frac}$, shown where $P\,>\,3\sigma_P$ and $I\,>\,5\sigma_I$. The peak of the polarized intensity is \peakPEmbEightdatAUnits{}. The red ellipse in the lower-left corner represents the synthesized beam of ALMA continuum observations. The beam size is 0$\farcs$35 $\times$ 0$\farcs$32, with a position angle of --63$^\circ$.}
\label{fig:emb8_pol_pfrac}
\vspace{0.3cm}
\end{figure}

\subsection{Serpens Emb 8}
\label{subsec:emb8}

Finally, we present the ALMA dust polarization data of the protostar Serpens Emb 8, located at $\sim$\,7000 au to the SW of Serpens Emb 8(N). In Figure \ref{fig:emb8_pol} we present two high angular resolution maps of this source, with spatial resolutions of 80 and 50 au. Apart from the central source Emb 8 we now clearly resolve the two companions, which we deem Emb 8-b and Emb 8-c (see Table \ref{t.source} and the right-hand panel of Figure \ref{fig:emb8_pol}). As was the case for Serpens Emb 8(N), we attribute the loss of signal at high resolution to a lack of sensitivity.

This source was the focus of \citet{Hull2017a}, where they compare the observed magnetic field morphology of dataset A with turbulent MHD simulations. They found that Serpens Emb 8 may have formed in a weakly magnetized environment, as no obvious hourglass morphology was detected. Additionally, they found no correlation between the magnetic morphology and the gradient of the dust emission, suggesting that the field is not strong enough to shape the structure of the dust. The magnetic field morphology observed does not present major changes as we increase in resolution from dataset A \citep{Hull2017a} to Case-3 (Figure \ref{fig:emb8_pol} right-hand panel). However, we now begin to resolve the magnetic field orientation in the central core Emb 8, which exhibits an homogeneous E--W pattern. As for the dust polarization in the envelope, it mainly shows up around Emb 8-b and to the south of Emb 8, in a large arc-shaped structure that is not strongly correlated with the outflow (like Emb 8(N), for example). As for Emb 8-c, this source appears unpolarized.

In Figure \ref{fig:emb8_pol_CO} we present the magnetic field orientations from dataset A overlaid with Stokes $I$ contours and the integrated redshifted and blueshifted emission from \co{} in color scale. Contrary to the two others protostars presented above, the dust polarization is not clearly correlated with the molecular outflow. It is worth noting some hints of outflow cavity wall around the base of the blueshifted outflow, where the dust emission seems to follow the wide-angle outflow; however, there is almost no polarized dust emission in this area. On the redshifted side, there are no obvious correlations, as the CO emission is far from the detected dust emission. Nevertheless, all the magnetic field orientations to the south of Emb 8 are quite aligned with the outflow axis, suggesting that the magnetic field may not be totally uncorrelated with the bipolar outflow. Indeed, to the south of the central core, about 70\% of the magnetic field line segments are aligned with the outflow axis within an offset of $\pm\,20^\circ$.

Finally in Figure \ref{fig:emb8_pol_pfrac} we present the polarized intensity and polarization fraction maps of Emb 8 from dataset A. The polarized intensity peaks in the central core, which has a polarization fraction of 0.7$\%$. The polarization fraction within the core increases progressively to the north and south of the central core, achieving values of up to 30$\%$ inside the regions of 5$\sigma_I$ and 3$\sigma_P$.

\section{Discussion}
\label{sec:dis}

\begin{figure}[!tbph]
\centering
\includegraphics[scale=0.46,clip,trim=0.2cm 0.5cm 2.8cm 2cm]
{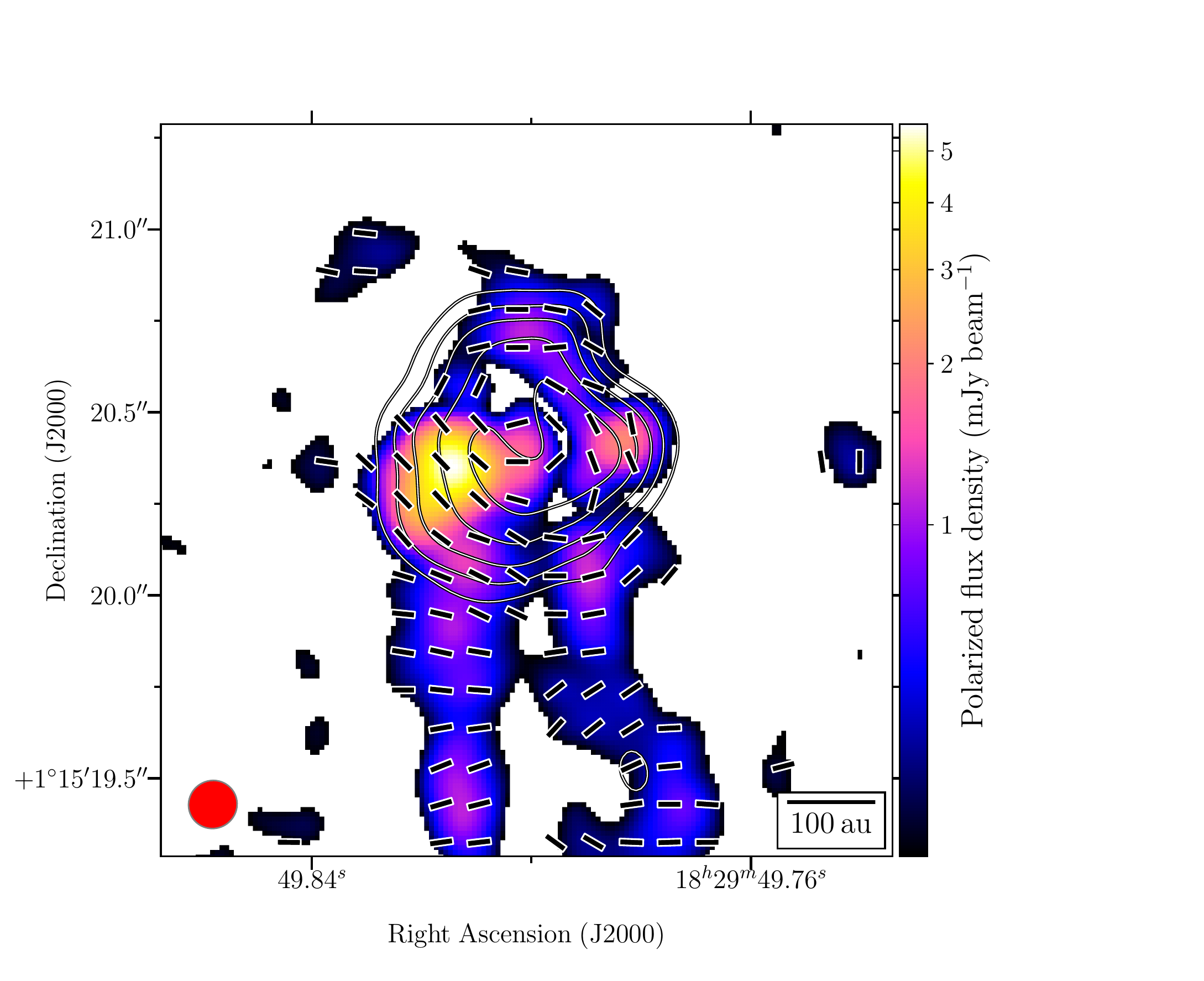}
\caption{\footnotesize Polarization orientations in SMM1-a from Case-3. Same as Figure \ref{fig:smm1_pol} (\textit{right}) for the line segments, except that in this case, they are \textit{not} rotated by 90$^\circ$, and instead represent the actual polarization orientations. The color scale is the dust polarization intensity $P$. Both the color scale and the line segments are shown where $P\,>\,3\sigma_P$, where $\sigma_P$= \rmsPSMMcaseIIIUnits{}. The black contours represent the total intensity at 16, 38, 72, 130, 200 $\times$ $\sigma_{I}$, where $\sigma_{I}$ = \rmsISMMcaseIIIUnits{}. The beam size of the continuum emission (red ellipse) is 0$\farcs$13 $\times$ 0$\farcs$13.}
\label{fig:smm1a_pola_E_vec}
\vspace{0.3cm}
\end{figure}

\begin{figure*}[!tbph]
\centering
\subfigure{\includegraphics[scale=0.48,clip,trim=4cm 0.2cm 4cm 0.8cm]
{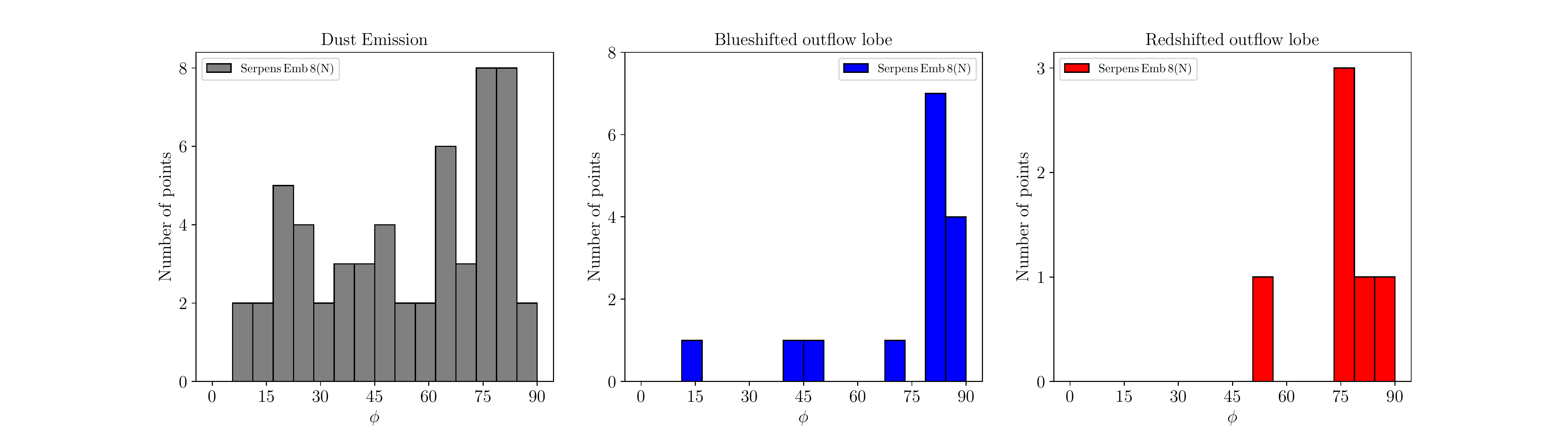}}
\subfigure{\includegraphics[scale=0.48,clip,trim=4.2cm 0.2cm 3.8cm 0.6cm]
{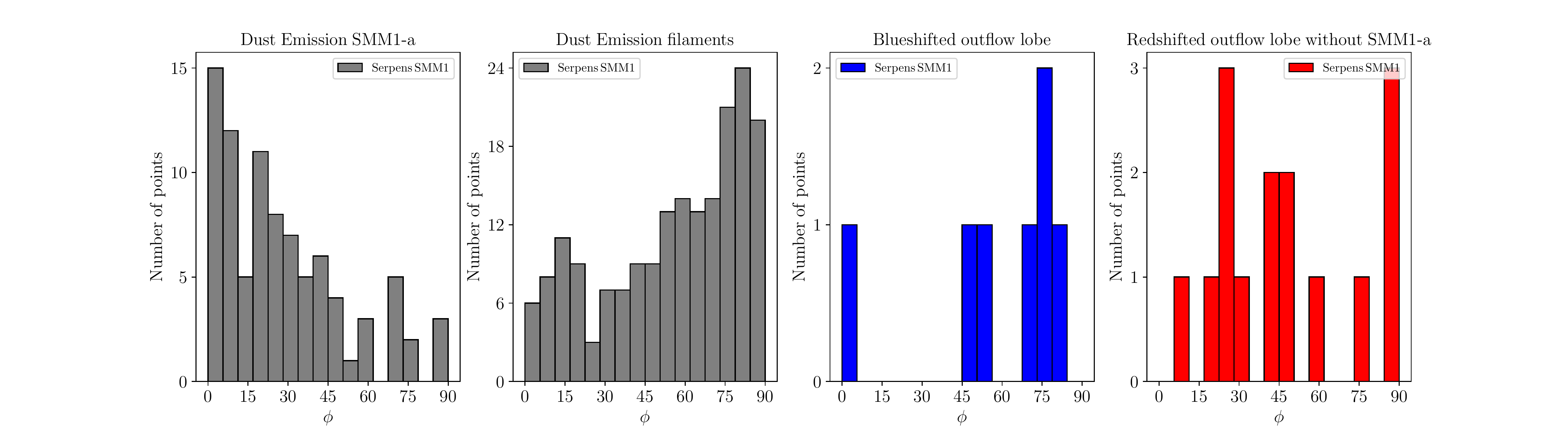}}
\caption{\footnotesize Histograms of relative orientation (HRO). Calculated HROs between the inferred magnetic field orientation and the density gradients in the total intensity maps (in grey), or in the blueshifted (in blue) and redshifted (in red) moment 0 maps of the CO\,($J$\,=\,2\,$\rightarrow\,$1) low velocity outflow in Emb 8(N) (\textit{top}) and Serpens SMM1 (\textit{bottom}). In SMM1-a, the dust emission is separated into the central hot corino and the two southern filaments. Furthermore, concerning the redshifted outflow lobe of SMM1, we did not calculate the gradient in the central zone of SMM1-a, in order to focus more clearly on the correlation between the outflow cavity walls and the magnetic field orientation. See Appendix \ref{sec:grad} for the gradient maps and a detailed explanation of how the HROs were produced.}
\label{fig:HRO}
\vspace{0.3cm}
\end{figure*}

\begin{figure}[!tbh]
\centering
\includegraphics[scale=0.63,clip,trim=0cm 0cm 1.5cm 1cm]
{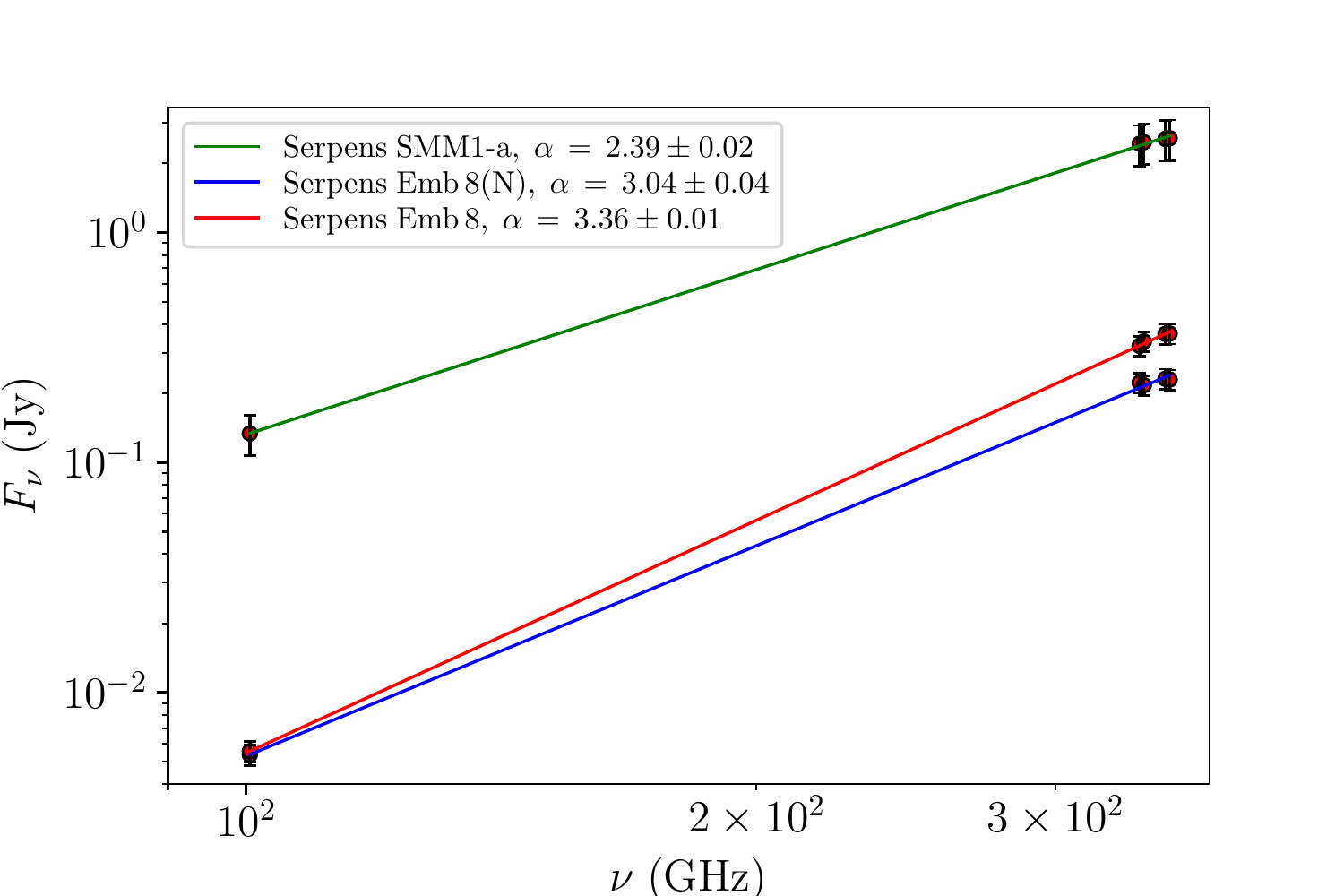}
\caption{\footnotesize Flux evolution over frequency from the best-fit visibility models. The $y$-axis is the integrated flux in Jy obtained fitting a multiple 2D-Gaussian components to the source visibilities. The $x$-axis is the frequency in GHz. Two datasets were used here, containing observations at 870\,$\mu$m (Band 7) and 3\,mm (Band 3). Each point is a fit to the visibilities from one spectral window, displayed with an error bar of $\pm \sigma$, where $\sigma$ is the error that takes into account both the fitting algorithm error on each point, as well as the 10$\%$ uncertainty in the ALMA flux calibration system.
}
\label{fig:spix}
\vspace{0.3cm}
\end{figure}

\begin{figure}[!tbh]
\centering
\includegraphics[scale=0.465,clip,trim=0.1cm 0.3cm 3cm 1.8cm]{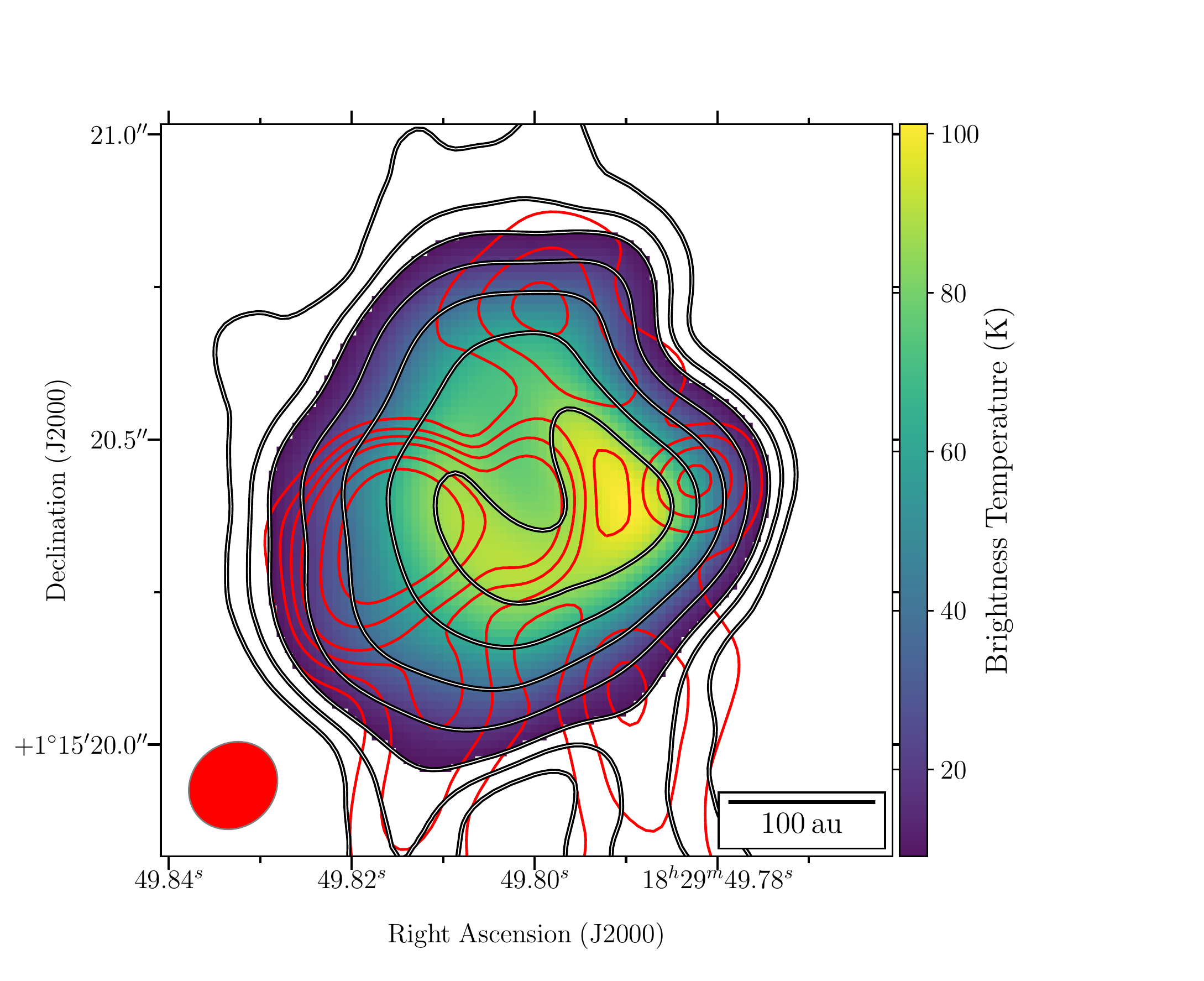}
\caption{\footnotesize Brightness temperature map of the inner core of SMM1-a from Case-1. The color scale represents the brightness temperature calculated in the Rayleigh-Jeans approximation, at a wavelength of 870\,$\mu$m. The brightness temperature peaks at 101\,K. The black contours represent the total intensity at 12, 20, 32, 64, 128, 220, 300 $\times$ $\sigma_{I}$, where $\sigma_{I}$ = \rmsISMMcaseIUnits{}. The red contours represent the polarized intensity at 3, 10, 17, 24, 32, 50, 68 $\times$ $\sigma_{P}$, where $\sigma_{P}$ = \rmsPSMMcaseIUnits{}. The red ellipse in the lower-left corner represents the synthesized beam after combining  datasets A, B, and C. The beam size is 0$\farcs$15 $\times$ 0$\farcs$14, with a position angle of --48.5$^\circ$.}
\label{fig:Tb}
\vspace{0.3cm}
\end{figure}

\begin{figure*}[!tbph]
\subfigure{\includegraphics[scale=0.653,clip,trim=0.1cm 1.5cm 8.2cm 3.3cm]
{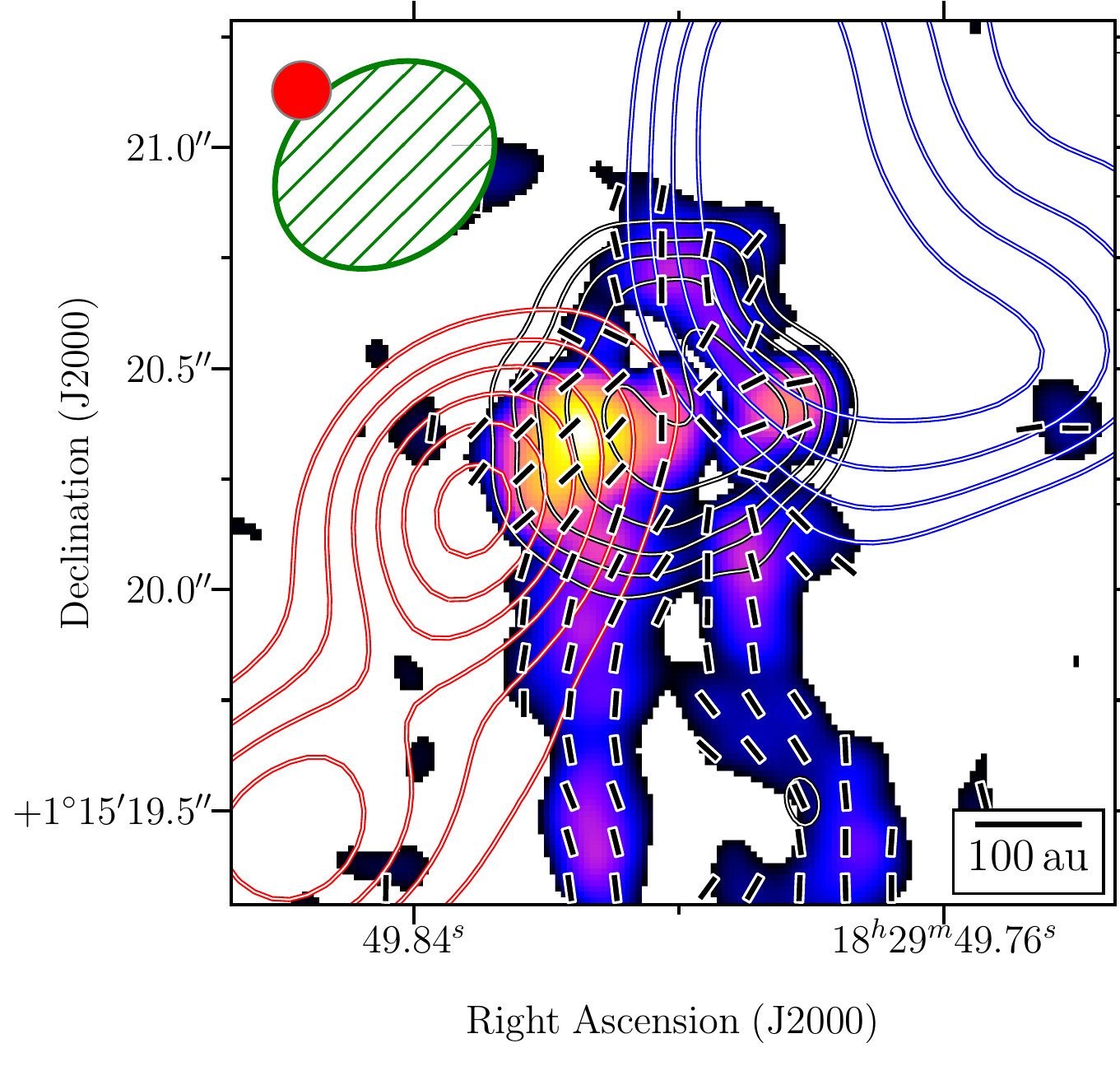}}
\subfigure{\includegraphics[scale=0.59,clip,trim=2.8cm 0.7cm 4.4cm 2.7cm]
{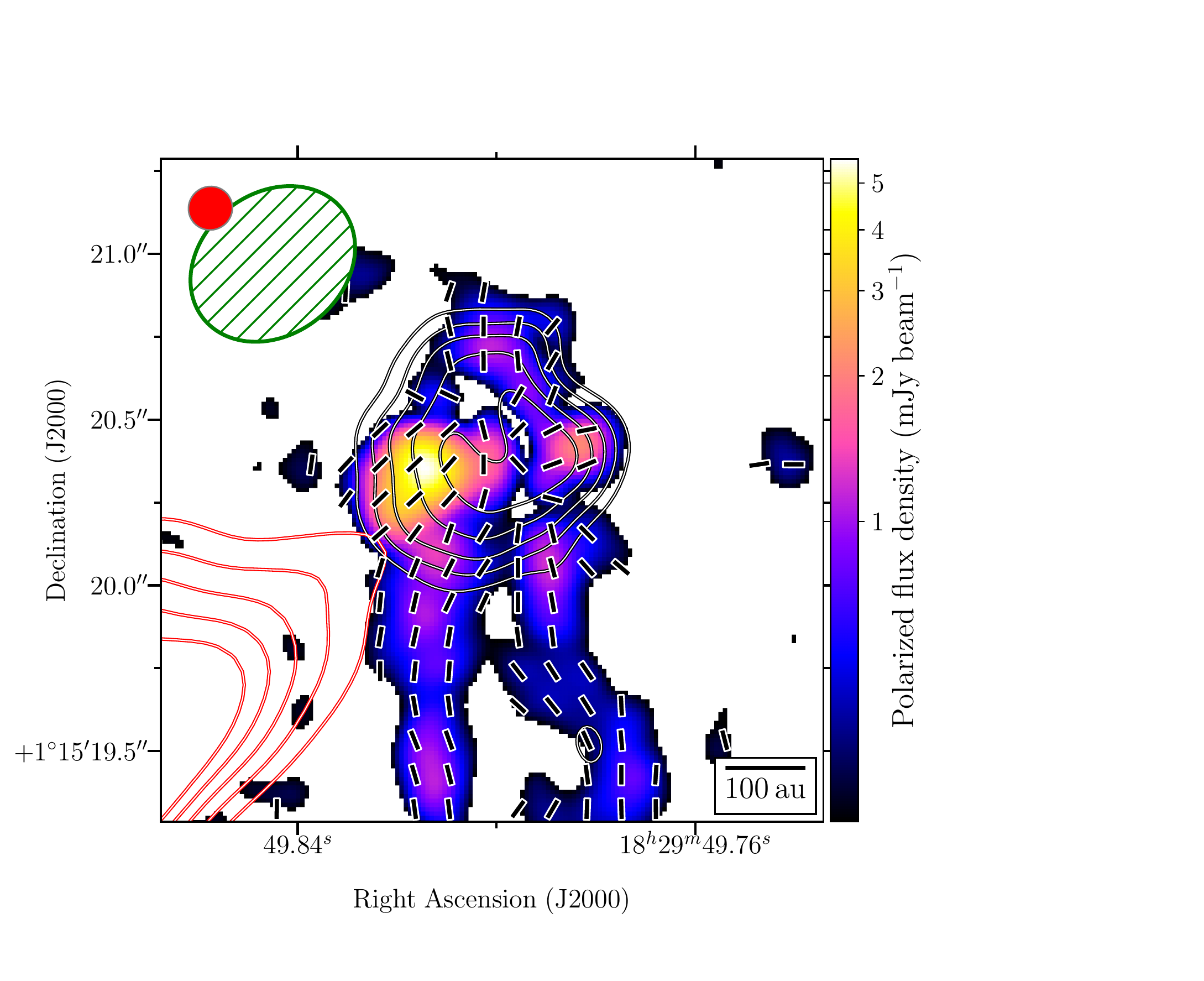}}
\caption{\footnotesize Magnetic field and outflow around SMM1-a from Case-3. Same as Figure \ref{fig:smm1_pol} (right-hand panel) for the line segments. Same as Figure \ref{fig:smm1a_pola_E_vec} for the polarized intensity in color scale and Stokes $I$ contours. The blue contours represent the moment 0 map of the low-velocity blueshifted outflow, constructed by integrating the emission of the \co{} from --13 to 4 \kms{}. The levels are 5, 6, 7, 8, 10, 12 $\times$ 0.4 \jybmkms{}, the rms noise level of the moment 0 map. The red contours in the left-hand panel represent the moment 0 map of the low-velocity redshifted outflow constructed by integrating the emission of the \co{} from 16 to 21 \kms{}. The level are 6, 7, 8, 9, 10, 11, 12 $\times$ 0.14 \jybmkms{}, the rms noise level of the moment 0 map. Finally, the red contours in the right-hand panel represent the moment 0 map of the extremely high velocity (EHV) redshifted jet constructed by integrating the emission of the \co{} from 40 to 80 \kms{}. The levels are 8, 10, 12, 14, 16 $\times$ 0.03 \jybmkms{}, the rms noise level of the moment 0 map. No blueshifted emission in shown in the right-hand panel, as we don't see any trace of EHV jet on the blueshifted side. The beam size of the continuum emission (red ellipse) is 0$\farcs$13 $\times$ 0$\farcs$13. The green ellipse represents the resolution from the molecular line maps, and measures 0$\farcs$53 $\times$ 0$\farcs$45.}
\label{fig:SMM1a_jet_pol}
\vspace{0.3cm}
\end{figure*}

Our high resolution polarimetric results from the three protostars Serpens Emb 8(N), SMM1, and Emb 8 lead us to discuss the causes of the polarization patterns in each of these sources. We investigate the different environmental conditions and try to cautiously infer the possible physical processes that would lead to the different behaviors (\eg polarization fraction and spatial distribution) of the polarized thermal dust emission. We start by focusing on the polarization pattern seen in the inner cores of our sources, investigating first the optical thickness and the correlation between structure of the continuum emission and the polarization (Section \ref{subsec:pola_pattern}). Second, under the hypothesis that the polarization reflects the magnetic field morphology, we discuss the poloidal pattern of the magnetic field visible in the inner core of two of our sources (Section \ref{subsec:poloi_field}). Third, we study the correlation between the outflow morphology and the magnetic field, especially around the outflow cavity walls (Section \ref{subsec:field_shaped_outflow}). Finally, we investigate the cause of the enhancement of the polarization along the cavity walls, focusing particularly on the role played by the radiation field (Section \ref{subsec:UV_Cav}).

\subsection{Investigating possible grain-alignment mechanisms causing polarization in inner envelopes at r\,<\,200 au scales}
\label{subsec:pola_pattern}

In protostellar envelopes, the long axes of dust grains are expected to be oriented orthogonal to the surrounding magnetic field \citep{Andersson2015}.  This effect has been the target of many observations in both low- and high-mass star-forming regions (see \citealt{HullZhang2019} for a review of interferometric polarization observations). $B$-RATs (designating dust grains magnetically aligned via Radiative Alignment Torques) is the favored mechanism to explain the  polarization at core/ISM scales, although an improved version of the theory including paramagnetic inclusions in dust grains had to be developed in order to reproduce the high ($\sim$\,20\%) polarization fractions observed observed by \textit{Planck} in the diffuse ISM; see \citet{Hoang2016}. \citet{Guillet2018} also proposed an explanation for the high polarization fraction values encountered at ISM scales. Their models revealed that, when there is a high enough mass fraction ($\sim$\,0.8--1) of aligned grains, a combined population of silicate and amorphous carbon grains can reproduce these high polarization fractions. While this grain alignment mechanism most likely continues to operate at the high column densities typical of inner envelopes, the different local conditions of radiation, temperature, opacity, and density in these regions might cause other mechanisms to contribute to the millimeter and submillimeter (hereafter, ``(sub)millimeter'') polarization signal from dust.  These effects include dust self-scattering and alignment of dust grains with respect to the radiation direction, both of which we explore below.

\subsubsection{Grain alignment via radiative torques}

The polarization orientations in SMM1-a (where no additional rotation had been performed on the polarization angle $\chi$) from the highest resolution observation (Case-3 of Table \ref{t.imaging}) are plotted in Figure \ref{fig:smm1a_pola_E_vec}.
The central region of SMM1-a, inside the 32 $\sigma_I$ contour, exhibits a generally azimuthal polarization pattern, which could be characteristic of dust grains that are aligned with respect to the local radiation field, rather than with the local magnetic field. In the theoretical study led by \citet{LazarianHoang2007} and \citet{Tazaki2017}, they found that in environments such as protoplanetary disks, the Larmor precession time scale of large dust grains ($\,\geq100\,\mu$m) can be larger than the gaseous damping time scale, which causes the grains to be aligned via RATs with respect to the gradient in the radiation field instead with respect to the magnetic field. In this case, sometimes known as ``$k$-RAT'' alignment, the long axes of the dust grains may be aligned orthogonal to the gradient in the radiation emanating from the central protostar.

In our high-angular resolution observations of SMM1-a, we compare the relative orientation between the polarization and the radiation field in the center of the protostar. To do so, we use the Stokes $I$ gradient map as a proxy for the radiation field.  In this way, we can test if the polarization orientation is perpendicular to the Stokes $I$ gradient, which would be an argument in favor of the $k$-RAT solution. A caveat of this comparison is that inhomogeneous (i.e., aspherical) conditions of temperature, density, and optical thickness may alter this correlation.  In addition, the photons that are primarily responsible for the radiative torque acting on grains at a given location are those with the largest energy density, i.e., those near the peak of spectral energy distribution (SED) at that location. These may or may not be the (sub)millimeter photons that we detect with ALMA. Nevertheless, we think this is a reasonable assumption, which has been discussed before in the interpretation of high-resolution ALMA results \citep{Sadavoy2018a}.  

The bottom-left panel of Figure \ref{fig:HRO} shows the distribution of the differences between the inferred magnetic field position angles\footnote{While here our aim is to test the $k$-RAT mechanisms by comparing the relationship between the radiation field and the polarization (which is unrelated to the magnetic field in $k$-RAT models), Figure \ref{fig:HRO} shows HROs using the inferred magnetic field, as we ultimately conclude that the polarization in all of our sources is most likely caused by magnetically aligned grains (see Section \ref{subsec:poloi_field}).} and the Stokes $I$ gradient in the inner core of SMM1-a (details of the calculations can be found in Appendix \ref{sec:grad}). The resulting distribution is single-peaked, with a maximum at 0$^\circ$, which suggests a polarization orientation perpendicular to the inferred radiation field.  This implies that it is possible that dust grains have been aligned with their minor axes along the radiation gradient. However, several caveats remain in this hypothesis: the distribution is quite broad, suggesting some imperfections in this alignment solution. In addition, the peak in polarized intensity yields a strong deviation from the perfect azimuthal pattern.

This type of azimuthal polarization pattern caused by large dust grains has been seen in the HL Tau disk at 3 mm wavelengths \citep{Kataoka2017,Stephens2017b}. However, some caveats were raised by \citet{Yang2019}, who explained that $k$-RATs should enhance the polarized emission along the major axis of an inclined disk instead of exhibiting the azimuthally symmetric polarized intensity seen in HL Tau. In the case of Class 0 protostars, the age of the system is a determining factor for the amount of grain growth that has occurred. However, the typical duration on the Class 0 stage of $\sim$0.2 Myr \citep{Dunham2015,KristensenDunham2018} seems to be large enough, as grain growth has been inferred in a few young stellar objects \citep{Chiang2012,Sadavoy2016,ChaconTanarro2017,AgurtoGangas2019} and protoplanetary disks \citep[\eg][]{Perez2012,Trotta2013,Testi2014,Perez2015,Tazzari2016,YLiu2017,Harsono2018,Huang2018}. However, based on the aforementioned results, the $\geq$\,100\,$\mu$m dust grain size invoked in \citet{Tazaki2017} is at the upper limit of the grain sizes inferred to-date in Class 0 protostars.

In order to constrain the physical conditions (and possibly dust-grain growth) within the central cores and to discuss the possible grain-alignment mechanisms, we investigated the optical thickness of our sources by measuring the spectral index $\alpha$ of the observed flux densities, given by
\begin{equation}
    F_{\nu} \approx F_{\nu_0}\left(\frac{\nu}{\nu_0}\right)^{\alpha}\,\,.
 \end{equation}
We use the continuum observations from our ALMA Band 3 (3 mm) dataset as wel as our Band 7 (870 $\mu$m) dataset B (Table \ref{t.obs}), which are separated by a period of time of three weeks. To measure the flux of SMM1-a, Emb 8(N), and Emb 8, we fit a single 2D-Gaussian component model to the visibilities of our datasets using the UVMULTIFIT tool \citep{MartiVidal2014}. Figure \ref{fig:spix} shows the fit results and the spectral index $\alpha$ for the inner core ($\sim$\,200 au) of our sources. An optically thick source would have a spectral index of $\alpha\,\approx\,2$ (black body case, in the Rayleigh-Jeans regime), whereas a source considered to be optically thin would have $\alpha\geq3$. We were not able to spatially resolve the spectral index in the inner cores of our sources as the beam of the 3\,mm dataset is significantly larger than the beam of our Band 7 observations. We are thus averaging the optical depth over the hot corino of SMM1-a with this Gaussian fitting. 

In Figure \ref{fig:Tb} we present the brightness temperature over the center of SMM1-a, overlaid with the Stokes $I$ and polarized intensity contours. The brightness temperature peaks at 101\,K and is as low as $\sim$\,20\,K at the outer edges of the hot corino. Moreover, we notice that inside the hot corino, the peaks in the polarized emission identified above are located on either side of the Stokes $I$ peak (\ie the horseshoe-shaped zone, which is likely to be an optical depth effect), which suggests that the polarized emission is not originating in the regions of highest optical depth in the central core of SMM1-a. 

The joint consideration of the brightness temperature map and the derived spectral index value suggest that the very inner $\sim$\,100 au zone is either optically thick and/or has dust emissivity properties that are significantly different from the rest of the hot corino. Dust grain growth may have begun in the very center of the hot corino; however, to confirm this we will need to further investigate the spatial distribution of the dust emissivity index and the dust temperature \citep{Bracco2017}. As \citet{Tazaki2017} predict that the $k$-RAT mechanism requires large dust grains in order to operate, this grain-alignment hypothesis may be indeed be relevant in the inner core of in SMM1-a. However, two strong contradictions lead us to discard $k$-RATs as the dominant polarization mechanism occurring here: the first one is the broadness of the distribution in the HRO histogram presented above. The second is that we don't observe the polarized intensity predicted in \citet{Yang2019}, where their model predicts that the $k$-RAT mechanism should enhance the polarized intensity along the major axis of the protoplanetary disk structure, as mentioned above. This last point is not straightforward for SMM1-a, as we do not detect any hints of a flattened, rotationally supported structure in any of our molecular line observations.

\subsubsection{Dust self-scattering}

Polarized dust emission can also arise from self-scattering of dust grains. If the dust grain sizes have reached $\sim$\,100\,$\mu$m, the self-scattering effect is expecting to be the dominant polarization pattern in dense environments such as those found in protoplanetary disks \citep{Kataoka2015}.  Self-scattering is a highly frequency dependent phenomenon \citep{Stephens2017b}, reaching a maximum efficiency for dust grains with sizes of $\sim\xspace\lambda/2\pi$ \citep{Kataoka2015}. Observations of dust scattering in protoplanetary disks have found several polarization patterns, which are highly dependent on both the optical thickness and inclination of the observed disks \citep{Kataoka2016b,Hull2018a,Bacciotti2018,Girart2018,Ohashi2018,Dent2019}. 

SMM1-a is most likely inclined (\citealt{Yildiz2015} estimate an inclination of 50$^\circ$ with respect to the line of sight), and its hot corino, as seen above, might be optically thick in the center and thus, the polarization might arise from the outer layer of this central core. \citet{Sahu2019} proposed two scenarios to describe the geometry of a different hot corino, namely that of NGC1333 IRAS 4A1: either an optically thick circumstellar disk or a temperature-stratified dense envelope. If the structure we see in SMM1-a is flattened by rotation, it is conceivable that the asymmetry in the polarized intensity (which is brighter on the near side, which we know thanks to the location of the redshifted jet) is to dust self scattering. However, several points fail to corroborate this hypothesis of self-scattering in SMM1-a. First, the typical level of polarization fraction expected from dust self-scattering is $\sim$\,1\%, whereas we observe a highly polarized spot in SMM1-a,  which exhibits a polarization fraction of 6$\%$. Second, the polarization orientations (see Figure \ref{fig:smm1a_pola_E_vec}) do not fit the prediction of self-scattering theory, especially around the highly polarized zone on the redshifted side. Finally, as mentioned above, the current lack of a detected disk structure toward SMM1-a makes it difficult to interpret our results in the context of the self-scattering theory, which requires a disk-like structure.

In their models, \citet{Yang2017} characterized the polarization emanating from inclined disks where the scattering grains had not yet settled to the disk mid-plane. In such disks, the polarized intensity becomes more asymmetric (\ie the polarized emission gets brighter on the near side of the disk) as the optical depth increases. We do see an asymmetry in the polarized intensity in SMM1-a. However, unlike the models of \citeauthor{Yang2017}, which show polarization primarily along the minor axes of the disks, we see primarily azimuthal polarization, inconsistent with dust self-scattering in an inclined disk.

In Serpens Emb 8, we do see polarization in the inner core ($\sim$ 200 au) at a level of 0.7\%. The self-scattering phenomena could indeed create this level of polarization fraction; however, as mentioned above, we would expect the polarization orientation to be along the minor axis of the source, which can be inferred from the bipolar outflow axis \citep{Cox2018}. In the case of Emb 8, the orientation of the polarization is not aligned with its inferred minor axis. Moreover, given the spectral index of its inner core ($\alpha \approx 3.36$), it is most likely optically thin. Consequently, we discard this hypothesis for Emb 8 as well.

In these sources, it is not always straightforward to determine which polarization mechanism causes the polarization from the innermost regions of our sources.  Overall, however, we conclude that neither self-scattering nor grain alignment via $k$-RATs is occurring in our sources. Moreover, a final caveat regarding the potential occurrence of these two polarization mechanisms is how the environmental conditions of the hot corino would change the dust grain size distribution. For example, the temperature and radiation from this high-column-density zone may be adequate to trigger RAdiative Torque Disruption (RATD), recently introduced in \citet{Hoang2019NatAs} and \citet{Hoang2019}. Via the RATD phenomenon, large aggregates can be spun-up to suprathermal rotation speeds and disrupted into individual icy grains, which would lower the maximum dust grain size encountered in these kinds of environments. If this is indeed the case, the resulting (smaller) dust-grain size distribution would make it likely that $B$-RATs are the dominant grain-alignment mechanism even in the bright, dense hot corino regions that we observe.

We did not discuss the case of Emb 8(N), as it exhibits almost no detection in the center. In summary, we assume that $B$-RATs are the dominant grain-alignment mechanism responsible for the polarization detected in our three sources. Under this assumption, we continue below by discussing the inferred magnetic field maps we obtain in our three sources.

\subsection{Poloidal magnetic field at outflow launching points}
\label{subsec:poloi_field}

Since our analysis presented in Section \ref{subsec:pola_pattern} suggests that most of the polarization detected toward the continuum peaks of our sources cannot be entirely due to self-scattering or alignment with radiation field, here we explore the properties of the magnetic field as inferred from the detected polarization patterns. We present maps where the polarization orientations have been rotated by 90$^\circ$ and to show the orientation of the magnetic field projected on the plane of the sky.

Both Serpens SMM1-a and Emb 8(N) seem to exhibit poloidal magnetic field morphologies in the central 200\,au of their cores. In Figure \ref{fig:SMM1a_jet_pol} we present the magnetic field orientation in SMM1-a overlaid with the low-velocity blue- and redshifted outflow and the EHV redshifted molecular jet, revealing significant correlations between the outflowing material and the magnetic field. Indeed, the magnetic field seems to present a bipolar structure that follows the molecular emission on both sides of the low-velocity CO outflow. Moreover, the redshifted EHV jet axis appears perfectly aligned with the magnetic field orientation within the highly polarized zone, \ie the SE part of the central core of SMM1-a.  This suggests a connection between the magnetic field in this region and the base of the EHV jet. Another point that corroborates the poloidal field hypothesis is the structure of the polarized intensity map. Between the two highly polarized zones in SMM1-a, respectively linked with the red- and blueshifted sides of the outflow, is a depolarized delimitation line that clearly divides the two sides of the poloidal magnetic field in SMM1-a (see the thin white region in the center of the color scale in Figure \ref{fig:SMM1a_jet_pol}). This effect is also seen in Emb 8(N) (see Section \ref{subsec:8N}), and can be explained by the sharp change of polarization orientation, which causes a line of depolarization with a width equal to the beam size \citep{Kwon2019}. Consequently, having such a depolarized line supports our hypothesis that a poloidal magnetic field configuration is the cause of the polarization in the central core of SMM1-a. The case of the central 200\,au in Emb 8(N) is less obvious, as we do not detect a large amount of polarized intensity. However, the lowest angular resolution observations (Figure \ref{fig:emb8N_pola} top-left panel) show a few detections in the center, suggesting the presence of a poloidal magnetic field in the central core of Emb 8(N).

We plotted the distribution of the angle difference between the magnetic field orientation and the gradient from the moment 0 map of the low-velocity red- and blueshifted outflow of SMM1-a in Figure \ref{fig:HRO} (see the two bottom-right panels). We compare CO and polarization data with different angular resolutions, which explains the small number of points displayed in the histograms. To derive the gradient in the HRO histograms, we select regions with strong gradients in the integrated blue- and redshifted CO maps in order to pick up only the polarization orientations associated with the edge of the outflow cavities (see Appendix \ref{sec:grad} for the gradient maps and details of the calculations). In the case of the blueshifted outflow of SMM1-a, the few beams contributing to the histogram are located to the NW of the inner core of SMM1-a (see Figure \ref{fig:SMM1a_jet_pol}). These few points, while not statistically significant, still suggest that the magnetic field is tracing the blueshifted outflow cavity. The few points in the histogram on the redshifted side seem randomly distributed, as the low-velocity redshifted outflow emission is spatially extended and does not exclusively overlap with the polarization in the central zone of SMM1-a.

The question remains, Why is the polarized intensity so different between the red- and blueshifted sides of the outflow in SMM1-a. An asymmetry is clearly visible in the polarized intensity and polarization fraction maps (Figure \ref{fig:smm1_pol_pfrac}). We see an intense peak of 6$\%$ at the base of the one-sided, redshifted molecular jet, which may be the cause of the enhanced polarization efficiency. However, \citet[][and references therein]{RodriguezKamenetzky2016} and \citet{Dionatos2014} detected ionized and atomic jets on the blueshifted side of SMM1-a, where we see no signs of enhanced polarization.

\citet{Hoang2018} developed a new theory of grain alignment by mechanical alignment torques (MATs). Unlike the mechanical alignment theorized by \citet{Gold1952}, who proposed that the major axes of dust grains could become aligned parallel to the gas-dust flow, \citeauthor{Hoang2018} found that dust grains can become aligned with respect to the magnetic field through mechanical torques induced by supersonic gas-dust drift (in an outflow or a wind from an AGB star, for example), yielding a polarization orientation that is consistent with a poloidal magnetic field aligned with the bipolar outflow axis. Consequently, the EHV redshifted jet may trigger a significant amount of grain alignment efficiency via the MAT mechanism that can contribute, along with RAT alignment, to the observed intense polarization at the base of the redshifted molecular jet. According to the RAT theory, magnetic grain alignment can easily produce this 6$\%$ level of polarization; however, such a high polarization fraction peak within the otherwise weakly polarized ``polarization hole'' is quite unexpected, and has never been seen before so clearly, to our knowledge.  In summary, MATs may be occurring at the same time as RATs, but it is not trivial to disentangle the relative roles played by these two alignment mechanisms. 

In their $\sim$\,60\,au  resolution observations of the intermediate-mass Class 0 source OMC-3 MMS 6, \citet{Takahashi2019} reported a similar asymmetry in the polarized emission at the base of the redshifted outflow from that source, where they argued that grains are most likely magnetically aligned. It is therefore possible that this type of asymmetry in the polarized intensity could be caused by inclination effects; SMM1-a and OMC-3 MMS 6 have similar estimates for their inclination angles.  While \citet{Frau2011} did see symmetric, double-peaked polarization profiles along the minor axes of their models of collapsing, magnetized protostellar cores (which did not take into account optical depth effects), they did not see the same types of asymmetries seen here and in \citet{Takahashi2019}.  

Outflow launching theories have predicted that magnetic field lines can be wrapped by the rotation of the inner envelope and the disk \citep{Hennebelle2009,Kataoka2012}, resulting in a toroidal magnetic field morphology that has tentatively been seen in a few observations of protostellar cores and disks \citep[\eg][]{Rao2014,Alves2018,Ohashi2018}.  We do not detect this morphology here, suggesting that the poloidal component of the magnetic field is dominant over any rotation in SMM1-a.  At the scales we resolve, we probe the material at the disk-envelope interface, where we do not see any evidence that of a rotationally supported disk. In such a case, a poloidal field aligned with the bipolar outflow axis is an expected result from MHD simulations of jets and outflows \citep{Fendt2006,Pudritz2007,Ramsey2019}.
A possible proof of this scenario is the work by \citet{LeeCF2018Nat}, who report linear spectral-line polarization in the SiO molecular jet emanating from the Class 0 protostar HH 212.  The shear between the wind from the disk and the ambient medium may produce a poloidal field at or near the interface, which would be consistent with what is observed here, provided that the interface region dominates the polarized emission.

Given the $\sim50^\circ$ inclination of the system, the magnetic field component orthogonal to the mid-plane $B_z$ is likely to be dominant over the radial component $B_r$ of the field, and will be reflected in the projected field morphology in the plane of the sky \citep{Harris2018}. This scenario is consistent with the magnetic field morphology we see in the center of the SMM1-a core, which is predominantly poloidal, but does have a significant radial component, leading us to consider whether SMM1 is actually more pole-on that previously thought. The estimation of the source inclination made in \citet{Yildiz2015} may not be particularly robust, as they simply examined the shape of the outflow (\ie if the outflow lobes were overlapping, and how much is the spatial extent of the low-velocity line wings) in the red- and blueshifted lobes, in order to infer inclination values within bins of 10$^\circ$, 30$^\circ$, and 70$^\circ$, with an uncertainty of 30$^\circ$. In addition, if SMM1 were close to edge-on, the fragments of the SMM1 core, SMM1-b, -c, and -d, should lie close to the equatorial plane of the core, as inferred from the SMM1-a bipolar outflow.  However, SMM1-b lies along the outflow cavity, which favors a source configuration closer to pole-on. The two others fragments SMM1-c and SMM1-d are, however, aligned with the equatorial plane projected in the plane of sky. The fact remains that, assuming SMM1 has an intrinsically poloidal magnetic field, a configuration closer to pole-on could yield the radial component of the magnetic field that we see projected in the plane of the sky toward the SMM1-a hot corino.

Finally, the presumed poloidal magnetic field morphology can also result from the magnetic field lines being pinched by the gravitational collapse of the core, typically known as the ``hourglass'' magnetic field morphology \citep{Girart2006}. Magnetic field orientations exhibiting a radial pattern similar to what we see in the central core of SMM1-a have been seen toward the high-mass star forming region W51 (albeit at larger scales and lower, $\sim$\,1000\,au resolution) in the observations of \citet{Koch2018}, who attribute the pattern to gravitational infall. While the first observations of hourglass-shaped fields were seen at the $\sim\,$1000 au scales of protostellar envelopes, it is not impossible that the detected poloidal magnetic field in the inner $\sim$\,200\,au core of SMM1-a is affected by this gravitational pinching effect. This effect has been seen, for example, in the magnetic field morphology probed in the inner few hundred au of the B335 protostellar core \citep{Maury2018}. 

Note that the collapse of magnetized material can also form a flattened structure known as a ``pseudodisk'' \citep{Galli1993a}. While the curved Western filament in SMM1-a is most likely tracing one of the redshifted outflow cavity walls, it is possible that the N--S oriented Eastern filament is linked to this pseudodisk and traces infalling material.  MHD simulations showed enhanced infall of material at 45$^\circ$ between the equatorial plane and the outflow axis of a protostar \citep{Kolligan2018}; this is similar to the orientation of the Eastern filament. Projection effects could explain why these two filaments might be of different natures---such as accretion streamers or structures associated with outflow cavity walls (see Section \ref{subsec:field_shaped_outflow})---despite being so close to one another.

\subsection{Magnetic field lines along outflow cavity walls}
\label{subsec:field_shaped_outflow}

\begin{figure*}[!tbph]
\centering
\subfigure{\includegraphics[scale=0.32,clip,trim=0cm 0.9cm 2cm 1.4cm]{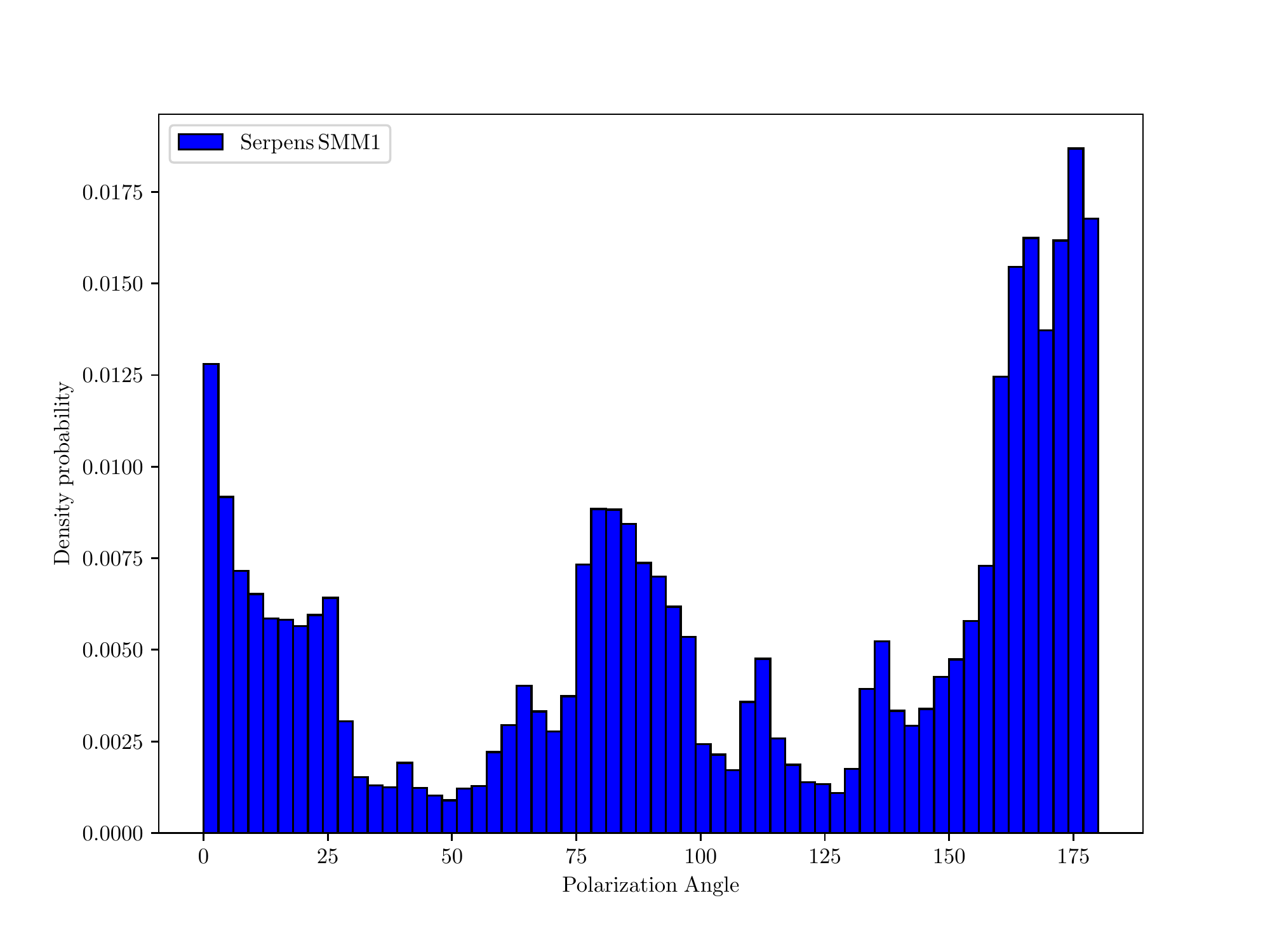}}
\subfigure{\includegraphics[scale=0.32,clip,trim=0cm 0.9cm 2cm 1.4cm]{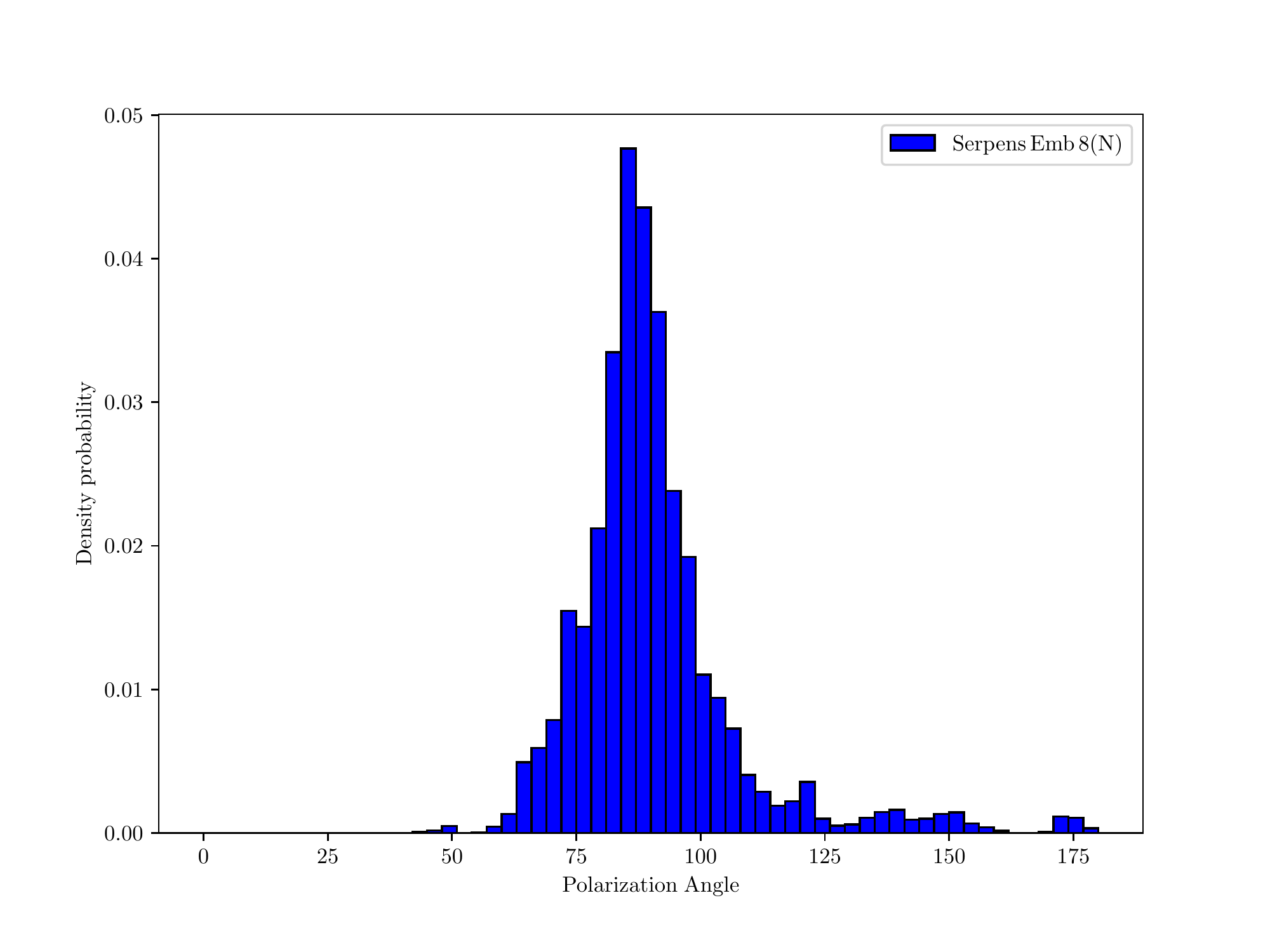}}
\subfigure{\includegraphics[scale=0.32,clip,trim=0cm 0.9cm 2cm 1.4cm]{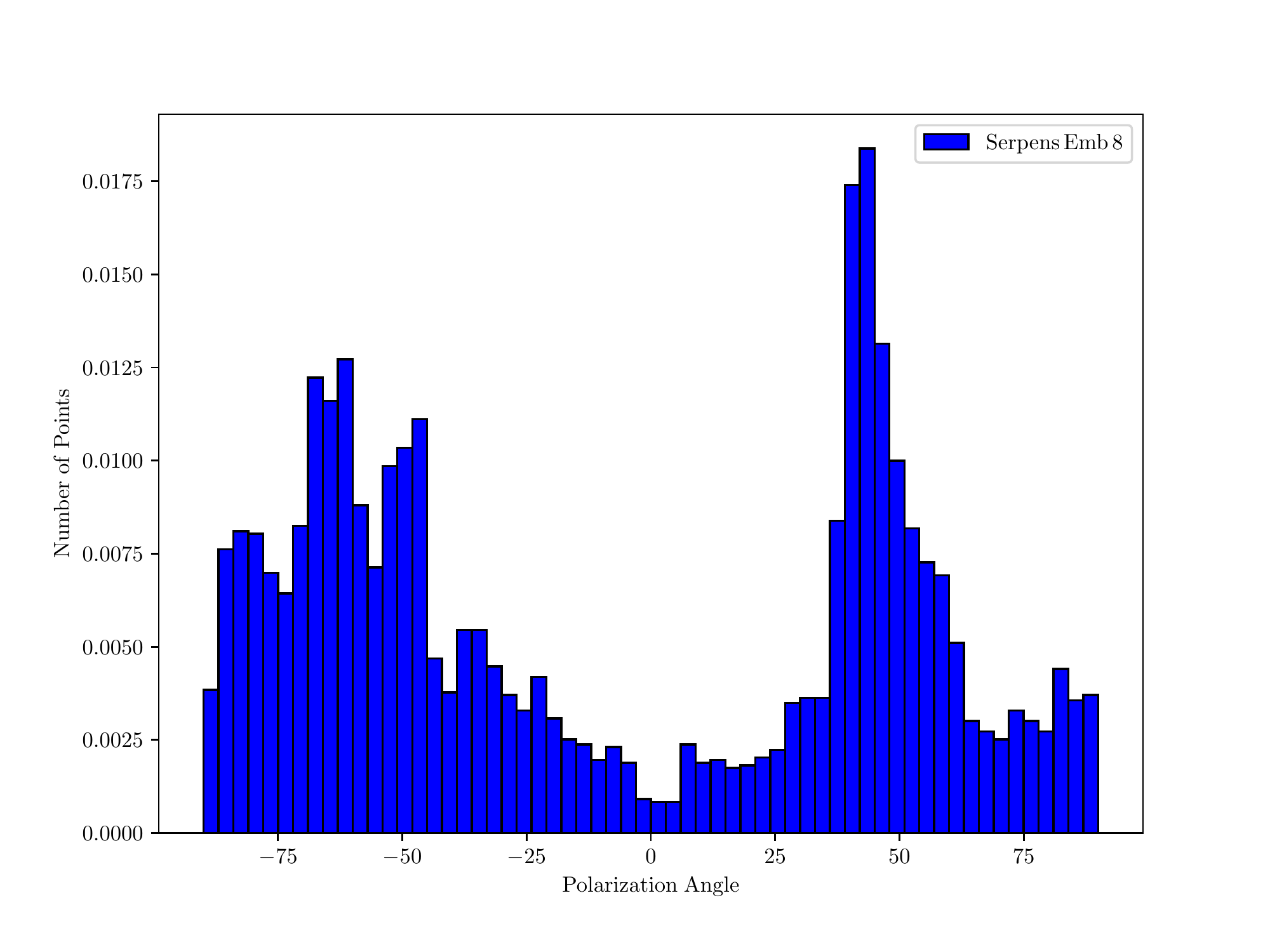}}
\caption{Distribution of the magnetic field position angles in our three protostars. We selected the position angles in the central $\sim\,4\arcsec$ zone around each protostar, and where the polarized intensity was above the 3$\sigma_P$ level, where $\sigma_P$ is the rms noise level of the polarized intensity map.
}
\label{fig:hist_B-field}
\vspace{0.3cm}
\end{figure*}

At scales larger than the central 200 au, it is clear that outflow activity is strongly linked with the magnetic field in the two protostellar sources SMM1-a and Emb 8(N), as the magnetic field morphology seems to follow the outflow cavity walls from the central 200 au up to $\sim$\,1000 au.

In SMM1, our highest angular resolution results resolve out most of the polarized emission except the two very bright filaments to the south of SMM1-a. The distribution in the HROs comparing the dust continuum emission with the magnetic field from both filaments (calculated after removing the central core emission, see Appendix \ref{sec:grad} for details; see Figure \ref{fig:HRO}, bottom row, second from the left) exhibits a distribution roughly peaking at 90$^\circ$, which confirms that the magnetic field lies along the two filamentary structures.  As mentioned in Section \ref{subsec:poloi_field}, the highly polarized Western filament follows the edge of redshifted integrated CO emission, indicating that it most likely corresponds to the outflow cavity wall. The Eastern filament is clearly shifted from the outflow axis, and does not overlap with any structure seen in the moment 0 map of CO tracing the outflow. We propose that the Eastern filament may correspond to a polarized accretion streamer that is infalling onto the central protostar. Filamentary structures seem to be common at $\gtrsim$\,1000\,au scales in protostellar cores \citep{Tobin2010,LeeK2012,Cox2018,Sadavoy2018b,Takahashi2019,Hull2019}, and many cases exhibit magnetic fields that lie along the structures' major axes.  Accretion-related filamentary structures have been seen in MHD simulations of magnetized, collapsing cores \citep[\eg][]{Seifried2015,Vaisala2019}; detailed comparisons of these types of simulations with high-resolution observations such as those we present here will paint a clearer picture of the nature of these magnetized structures and their possible link to protostellar accretion processes. 

Serpens Emb 8(N) presents a very pristine case of a magnetic field aligned along the outflow cavity walls: indeed, the HRO distributions in the top three panels of Figure \ref{fig:HRO}, comparing the magnetic field and both dust and CO moment 0 emission gradients, all peak at $\sim\xspace90^\circ$. Note that the distribution from the gradient of the continuum dust emission map (top-left panel of Figure \ref{fig:HRO}) has some points between $15^\circ$ and $45^\circ$ that correspond to the magnetic field orientations located where the dust-emission gradient is no longer dominated by the outflow cavity walls, but rather by the central core emission (on the redshifted side). The polarization patterns observed recently in B335 \citep{Maury2018} and in BHR 71 IRS2 \citep{Hull2019} appear quite similar to the case of Emb 8(N), and have magnetic field lines that are pinched in the equatorial plane and that clearly follow the outflow cavity walls. In B335, comparisons with MHD simulations  \citep{Maury2018,Yen2019} have led to the conclusion that this source most likely formed in a magnetically dominated environment. Future higher angular resolution observations of the kinematic properties and polarized emission in the central zone of Emb 8(N) will allow us to better understand the physical conditions responsible for this magnetic field structure.

The large scale polarization data of this region, observed with JCMT SCUBA polarimeter \citep{Davis2000,Matthews2009} indicate an overall E--W orientation of the magnetic field among the Serpens Main molecular cloud. It is interesting to examine the potential influence of this large scale magnetic field on our high spatial resolution results. The bipolar outflow of Emb 8(N) has an oriented that is $20^\circ$ different from purely E--W.  Even in the CARMA polarization data \citep{Hull2014}, the recovered magnetic field is purely E--W. The results we present here show an overall E--W magnetic field lying along the outflow cavity walls (see Figure \ref{fig:emb8N_pola} and Figure \ref{fig:hist_B-field} for the HRO); however, in some places the magnetic field deviates from the overall E--W orientation and clearly follows the cavity edges, which indicates that the magnetic field at these scale has been affected by the outflow activity. As for the redshifted outflow of SMM1-a, it is relevant to discuss the potential impact of the outflow on the magnetic field. We identified before the Eastern filament as a potential accretion streamer, whereas the Western filament is very likely the redshifted cavity wall, which therefore exhibits an associated magnetic field completely different from the large scale field. \citet{Hull2017b} did not find in their energetic comparisons an obvious difference between the magnetic and the outflow energies; however, we think this hypothesis of having an outflow-shaped magnetic field is reasonable, considering how the magnetic field orientations at small scales have deviated from the large-scale field morphology.

Serpens Emb 8 seems dominated by polarized emission mainly emanating from the envelope of the protostar, and thus there is not as clear a correlation with the outflow. Still, the thermal dust emission on the blueshifted side shows some hints of outflow cavity walls, but with polarization that is a bit offset to the East of this zone. To the south-east of the Stokes $I$ peak, there is a large area of polarization corresponding to a magnetic field that is perfectly aligned along the outflow axis, indicating that the outflow might have influenced the magnetic field in this zone. Emb 8 is similar in some ways to the BHR 71 IRS1 protostar \citep{Hull2019}, in the sense that the magnetic field seems to be affected by both the outflow activity and the envelope-based polarized emission, as can be seen in the northern outflow cavity of BHR 71 IRS1. 

As mentioned in Section \ref{subsec:poloi_field}, the high polarization fraction values we encounter near fast-moving material could be explained by a contribution from the MAT mechanism to the observed polarization, especially on the redshifted side of the central core of SMM1-a and in Emb 8(N), which both exhibit very high-velocity, highly collimated jets. Indeed, if we look at the polarization fraction values in Emb 8(N), which reach 36$\%$ on the blueshifted side, we see that these detections are located on the edge of the CO emission (Figure \ref{fig:emb8N_CO_pol}), indicating a possible interaction between the molecular jet and the aligned dust grains emitting this polarized emission. With RAT theory it is difficult to reproduce these high polarization fractions, even with very elongated aligned dust grains; consequently, it is conceivable that both MATs and RATs are occurring in the type of environment that we are seeing in Emb 8(N).  This hypothesis is reinforced by the fact that the polarization in the southern cavity wall of the redshifted jet of Emb 8(N) is fainter and has a lower polarization fraction precisely where the redshifted CO emission do not overlay completely with the polarized dust emission. 
This suggests that we may need to consider outflow evolution (for example, precession with respect to the outflow cavity walls) in order to better understand our observations. Finally, we note that we detect high polarization fraction values toward Serpens Emb 8 where there is no intense gas-dust flow, \ie far away from the bipolar outflow (see Figure \ref{fig:emb8_pol_pfrac}). However, Emb 8 has two little companions, Emb 8-b and Emb 8-c, that may play a role in enhancing the polarization observed in the core.  In summary, the local environmental conditions in protostars---particularly around their molecular outflows and jets---may be conducive to alignment from both MATs and RATs, thus yielding enhanced levels of dust polarization.  However, our observations are not yet able to disentangle the roles played by each of these two grain alignment mechanisms.

\subsection{Addressing the cause of enhanced polarization along cavity walls}
\label{subsec:UV_Cav}

Our results lead us to the question, What are the physical conditions necessary to achieve enhanced polarization in the walls of an outflow cavity? 

One local condition along outflow cavities that clearly distinguishes them from the rest of the envelope is that they are exposed to a strong, high-energy radiation field emanating from the central protostar. Indeed, energetic photons are expected to be generated by the accretion processes in the central protostar, leading to a high flux of UV and X-ray photons that heat the envelope via the photoelectric effect \citep{Spaans1995,Stauber2004,Stauber2005}. As outflow cavities are cleared of envelope material by the outflow itself, these energetic photons can escape from the central protostar and travel throughout the cavity. The shape of the cavity edges (\eg parabolic) allows this high-energy radiation to impinge upon the cavity walls \citep{Visser2012}.  Indeed, ionized cavity walls have been seen toward SMM1-a in \citet{Hull2016a}, who analyzed free-free emission seen in the VLA observations of \citet{RodriguezKamenetzky2016}.

The inner surfaces of the cavity walls also see their chemical composition evolve, thanks to molecular photodissociation by UV photons \citep{Drozdovskaya2015}. Observations of light hydrides, water, and high-$J$ transitions of CO have led to a better understanding of the effect of the UV field on the chemistry in highly irradiated protostellar outflows \citep{vanKempen2009,Yildiz2012,LeeS2015,Benz2016}.  The physics of UV-irradiated shocks in these same regions has also been addressed in a number of studies focusing on \textit{Herschel} observations    \citep{Goicoechea2012,Kristensen2013a,Kristensen2017, Mottram2014, Mottram2017,Karska2018}.  Note that \citet{Tychoniec2019} report that the redshifted molecular jet of Emb 8(N) has a bow shock visible in SiO\,($J = 5 \rightarrow 4$) emission $\sim$\,3000\,au from the central source; this type of shock could contribute to the irradiation of the outflow cavity walls and thus to the enhanced polarization we see. 

\begin{figure}
\centering
\includegraphics[scale=0.51,clip,trim=2.8cm 1.2cm 3.9cm 2.4cm]{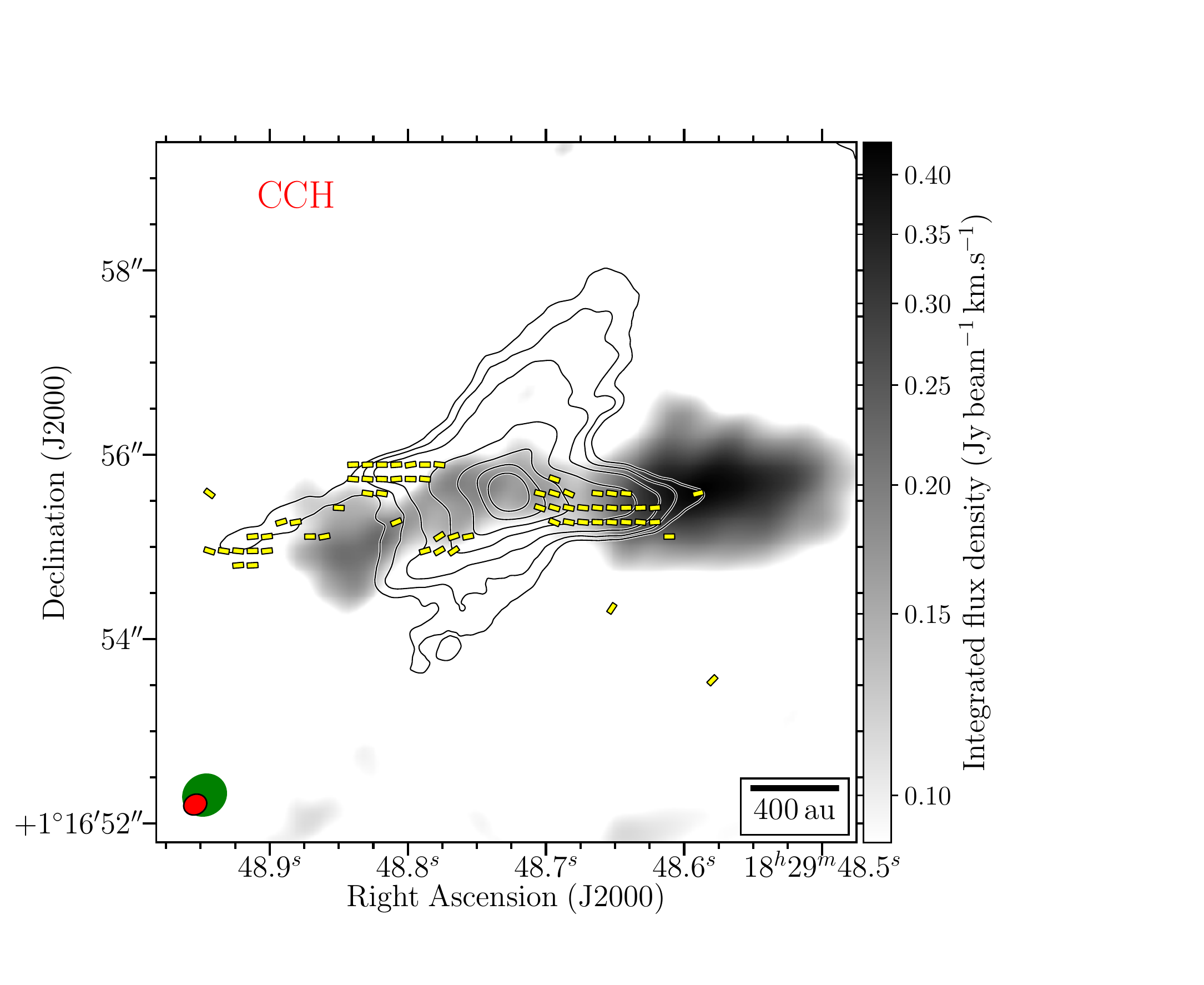}
\caption{\footnotesize Moment 0 map of CCH ($N=3\rightarrow2,\,J=7/2\rightarrow5/2,\,F=4\rightarrow3$) around Serpens Emb 8(N). The black contours represent the total intensity (Stokes $I$) at the following levels: 11, 16, 24, 44, 74, 128, 256 $\times$ the rms level in the Stokes $I$ dust emission map, where $\sigma_{I}$, where $\sigma_I$ = 55 \mujybm, from Case-2. The $v_\textrm{LSR}$ is $\sim\,$8.5 \kms.  The grayscale is the moment 0 map of CCH constructed by integrated emission from 5.4 to 12.8 \kms{}. The rms noise level of the moment 0 map is 0.32 \mjybmkms{}. Same as Figure \ref{fig:emb8N_pola} (right) for the line segments, which represent the magnetic field. The red and blue arrows represent the bipolar outflow directions. The beam size of the continuum emission (red ellipse) is 0$\farcs$26 $\times$ 0$\farcs$22, with a position angle of --64$^\circ$. The green ellipse represents the resolution of the molecular line maps, and measures 0$\farcs$46 $\times$ 0$\farcs$41.}
\label{fig:emb8N_CCH}
\vspace{0.3cm}
\end{figure}

UV-irradiated cavity walls toward Class 0 protostars have been inferred from observations of UV-tracing molecules such as \textit{c}-C$_3$H$_2$ and C$_2$H, which are expected to be seen in photon-dominated regions \citep[PDRs;][]{Murillo2018}. These two molecules have been seen toward others highly irradiated regions such as the Orion bar \citep{Cuadrado2015}, the Horse-Head nebula \citep{Guzman2015}, and protoplanetary disks \citep{Kastner2015,Bergin2016}. Of the Class 0 sources whose cavity walls show both UV-tracing molecules and dust polarization, there are two so far that exhibit an excellent correlation between the polarization and the molecular emission. The first is B335; \citet{Imai2016} found CCH and \textit{c}-C$_3$H$_2$ emission along the cavity walls, which also show enhanced polarization \citep{Maury2018}.  The second is Ser-emb 8(N), discussed above and shown in Figure \ref{fig:emb8N_CCH}.\footnote{Note that Serpens Emb 8 also shows CCH (van Gelder et al., in prep.; ALMA project 2017.1.01174.S, PI E. van Dishoeck) toward much of the polarized emission published in \citet{Hull2017a}, and shown at higher resolution in Figure \ref{fig:emb8_pol_pfrac}.  However, in this source, which has a chaotic magnetic field morphology, the outflow cavities are not clear.  Further work is necessary to tell the full story of the relationship between UV-sensitive chemistry and the enhanced polarization in Emb 8.}

The UV radiation field emanating from the material accreting onto the central protostar is a good candidate to explain enhanced polarization in the cavity walls via the RAT grain-alignment mechanism. We see what appear to be thick layers of polarization from magnetically aligned dust grains in the outflow cavity walls of Emb 8(N) ($\sim$\,200\,au thick) and BHR 71 IRS2 ($\sim$\,300\,au thick). In an attempt to reveal the nature of the radiation responsible for aligning the grains in outflow cavity walls, \citet{Hull2019} calculated the depth to which UV and longer-wavelength photons could penetrate in order to align grains to the depth that they see in their observations, and also to align a sufficient quantity of grains to produce the typical levels of polarization fractions that they (and we) detect. 

Following the procedure of \citet{Hull2019}, we estimate the depth to which UV and longer-wavelength photons can penetrate into the walls of the outflow cavities of Serpens Emb 8(N). From their PDR model results, \citet{Girart2005} found that UV radiation is fully extincted at a visual magnitude of $\sim$\,1, which corresponds a column density between $10^{21}$ and $10^{22}$ cm$^{-2}$ \citep{Bohlin1978}. We calculate the gas mass inside a circular area of 300\,au located at a distance of 600\,au from the protostar along the SW cavity wall of Emb 8(N), using the following relation between the flux density and the gas mass:
\begin{equation}
    M_{\textrm{gas}}=\frac{S_{\nu}d^2}{\kappa_\nu B_\nu(T_\textrm{d})}\,\,,
\end{equation}
where $B_\nu(T_\textrm{d})$ is the Planck function at the 343.5\,GHz frequency of our observations, the distance $d= 440$\,pc, the estimated dust temperature $T_\textrm{d}\,\approx\,20$K \citep{Hull2017b}, and the opacity $\kappa_\nu$ at a wavelength of 1\,mm is 2.74\,$\textrm{cm}^2$/g \citep{Ossenkopf1994}.  We assume a gas-to-dust ratio of 100. The flux density along the cavity walls of Emb 8(N) is not constant and slowly decreases the further we look from the central protostar. We selected the SW outflow cavity wall as it not only the brightest cavity wall, but it also presents resolved polarized dust emission that matches very well the CCH emission line integrated intensity, as shown in Figure \ref{fig:emb8N_CCH}. Note, however, that the CCH emission is even brighter at >\,600\,au distances from the central source, suggesting that UV radiation efficiently drives UV-sensitive chemistry even where the column density of the cavity wall has decreased.

We derive the column density in the cavity wall of Emb 8(N) as follows:
\begin{equation}
    N_{\textrm{H}_2}=\frac{M_{\textrm{gas}}} {\mu_{\textrm{H}_2} m_H A}\,\,,
\end{equation}
where $m_H$ is the mass of an hydrogen atom, $\mu_{\textrm{H}_2}$ is the mean molecular weight per hydrogen molecule ($\mu_{\textrm{H}_2}=2.8$ for gas composed of 71\% of hydrogen, 27\% of helium, and 2\% of metal mass; \citealt{Kauffmann2008}), and $A$ is the area over which we derived the flux density. Given a flux density of $\sim$\,5.5\,mJy measured in a 200\,au-diameter circle located at 600 au from the dust continuum peak of Emb 8(N), we obtain column density values between $\sim\,3.4\,\times\,10^{22}\,\textrm{cm}^{-2}$ and $\sim\,2.5\,\times\,10^{23}\,\textrm{cm}^{-2}$, using 100\,K and 20\,K dust temperatures, respectively. Note that the gas temperature is usually much higher than the dust temperature, as has been seen in models of UV-irradiated outflow cavity walls \citep{Visser2012,Drozdovskaya2015}; however, the two temperatures are largely decoupled.  Note that these values of the envelope column density at small scales are at best lower limits, as a result of the interferometric filtering of the optically thin dust continuum emission.  As a result, these values strongly suggest that UV radiation cannot penetrate deep inside the cavity walls.

Finally, we estimate the penetration depth of longer wavelength photons impinging on the cavity walls. \citet{Ossenkopf1994} calculated opacity values $\kappa_\nu$ at different wavelengths for gas number densities of $10^6$ and $10^8$\,$\textrm{cm}^{-3}$ \footnote{Our two derived gas densities for the SW cavity wall of Emb 8(N) are $1.24\times10^8$\,cm$^{-3}$ at 20\,K and $1.74\times10^7$\,cm$^{-3}$ at 100\,K}, and thus we use the $\kappa_\nu$ values from \citet{Ossenkopf1994} corresponding to gas densities of $10^6$\,cm$^{-3}$.  The optical depth is given by the following equation:
\begin{equation}
    \tau_\nu = \int_{0}^{s}\kappa_\nu \rho_\textrm{dust} d s\,\,.
\end{equation}

If we assume constant values of dust opacity $\kappa_\nu$ and dust mass density $\rho_\textrm{dust}$ throughout the thickness of the cavity wall, we can derive the path length $s$ at an optical depth of $\tau=1$, following that $s\,=\,1/\kappa_\nu \rho_\textrm{dust}$. Note the values of $\kappa_\nu$ in \citet{Ossenkopf1994} are normalized by the dust mass density, and thus we multiply these values by $\rho_\textrm{dust}$ in our calculations\footnote{The chosen $\kappa$ values from \citet{Ossenkopf1994} are the following: $\kappa_{1\textrm{\micron}}\,=\,1.18\times10^4\,\textrm{cm}^2\,\textrm{g}^{-1}$, $\kappa_{10\textrm{\micron}}\,=\,2.35\times10^3\,\textrm{cm}^2\,\textrm{g}^{-1}$, $\kappa_{100\textrm{\micron}}\,=\,5.92\times10^1\,\textrm{cm}^2\,\textrm{g}^{-1}$, and $\kappa_{1\textrm{mm}}\,=\,2.74\,\textrm{cm}^2\,\textrm{g}^{-1}$.}. 
For 1\,$\mu$m photons, the penetration depths assuming
20\,K and 100\,K temperatures are $\sim$\,1\,au and $\sim$\,7\,au, respectively. 
For 10\,$\mu$m photons, the depths are $\sim$\,5\,au and $\sim$\,35\,au. 
For 100\,$\mu$m photons, the depths are $\sim$\,200\,au and $\sim$\,1400\,au. 
Finally, for 1\,mm photons, the depths are $\sim$\,4200\,au and $\sim$\,30,000\,au.

RAT theory suggests that dust grains can only be spun-up efficiently by photons whose wavelengths are comparable to the dust grain size \citep{LazarianHoang2007}. Assuming this is true, and given that most short wavelength ($\lesssim$\,10\,$\micron$) photons do not penetrate to depths larger than a few tens of au, our observations suggest that a population of large, >\,10\,$\micron$-sized dust grains should be present in the cavity walls of Serpens Emb 8(N) to explain the $\sim$\,300\,au thickness of the polarized emission. Note that, in their study of the polarized outflow cavities of BHR 71 IRS2, \citet{Hull2019} also find that only long-wavelength (\ie mid- to far-infrared) photons can penetrate to depths similar to those where we see polarization in the cavity walls of Serpens Emb 8(N).

\citeauthor{Hull2019} conclude that this scenario of a ``thick'' layer of aligned dust grains in BHR 71 is more likely than the scenario where the polarization is produced by an extremely thin layer of grains aligned only by UV/optical photons.  However, even given the likelihood of the thick-wall scenario, the numbers we calculate above are still an upper limit to the penetration depth.  This is because the ``thick'' wall we see will have its thickness increased, at least to a small degree, by the projection of the curved cavity wall onto the plane of the sky.

Beyond outflow cavity walls, the high polarization fractions observed at small radii, where dense protostellar envelopes are optically thick to short wavelength photons, require that grains be large (>\,10\,$\micron$) in order to be to aligned via $B$-RATs with the local (sub-)mm radiation field, as shown for the first time in \citet{Valdivia2019}. While dust is known to grow to such sizes in circumstellar disks \citep[\eg][]{Perez2012, Testi2014, Stephens2017b, Hull2018a}, this degree of grain growth at large ($\sim$\,500--1000\,au) distances from the central source is unexpected. However, as was proposed in \citet{Wong2016}, outflow activity may be sufficient to transport large dust grains from the innermost part of the core (where the large grains are most likely formed) to the envelope, where we can detect their presence via the dust polarization we observe in outflow cavity walls (in the case of our calculations, at 600\,au from the center of the protostar).

As the average size of the dust grains probed changes with observing wavelength, it would be interesting to perform longer wavelength measurements of the polarized features we detect that are presumably not being irradiated by a strong UV field. Such observations would allow us to characterize the environmental conditions (in terms of dust grain population and radiation field) necessary for RATs to occur. Indeed, the Eastern filament in SMM1-a, the polarization to the north of the central core of Emb 8, the equatorial plane polarization detection in B335 \citep{Maury2018}), the ``arm-like'' in OMC-3 MMS 6 \citep{Takahashi2019}, the magnetized bridge in IRAS 16293 \citep{Sadavoy2018b}, and the sharp filamentary structure to the NE of BHR 71 IRS1 \citep{Hull2019} are unlikely to be strongly illuminated by UV radiation from the central protostar, as these structures are not associated with the outflow cavity walls.  This leaves us with the question, Where does the irradiation and/or anisotropic radiation field necessary to align these grains via RATs come from? While the answer is not yet clear, comparing dust polarization observations with synthetic observations of MHD models (see \citealt{Valdivia2019}) will enable a deeper understanding of the dust grain populations in the innermost regions of Class 0 envelopes.

\section{Conclusions}
\label{sec:con}

We have presented ALMA dust polarization observations of the three Class-0 protostars Serpens SMM1, Emb 8(N), and Emb 8, at spatial scales from $\sim$150 au down to $\sim$40 au. The conclusions we draw are as follows:

\begin{enumerate}
    \item Serpens Emb 8(N) exhibits strong polarization along its outflow cavity walls, with magnetic field orientations aligned with the major axis of the dust emission.
    
    \item In Emb 8(N), there is an obvious anti-correlation between areas where we see polarized dust emission and emission from warm dense-gas tracers such as C$^{18}$O and $^{13}$CS, and regions of cold, dense gas traced by DCO$^+$.
    
    \item Serpens SMM1-a presents several interesting polarized emission zones. To the south of the central $\sim$\,200\,au core are two polarized filaments with magnetic field orientations along the filamentary structure. We identified one filament as being the cavity wall of the redshifted outflow; we speculate that the other one may be a potential infalling accretion streamer.
    
    \item The inner cores of SMM1-a and Emb 8(N) exhibit poloidal magnetic fields that are perfectly aligned with the bipolar outflows; no hints of toroidal magnetic fields wrapped by core rotation have been found.
    
    \item The polarized intensity map of the inner core of SMM1-a presents a clear asymmetry, exhibiting an intense polarized spot that peaks at 6$\%$ in polarization fraction. This asymmetry in the polarized emission may be linked to the fact we only see an EHV molecular jet on the redshifted side of the bipolar outflow, which could contribute to the high efficiency of the grain alignment in this zone via a combination of the MAT and RAT alignment mechanisms.
    
    \item We propose that the enhanced polarization seen along outflow cavity walls in Class 0 protostars is caused by irradiation of the cavities, which have been cleared of material by the outflow activity. Several studies (including the spectral-line observations of Emb 8(N) that we show here) have reported UV-tracing molecules along outflow cavity walls, pointing chemistry driven by UV irradiation in these regions. However, to align dust grains deep within the walls of the cavity and to cause the high polarization fraction that we see, longer-wavelength (\ie mid- to far-infrared) photons may be necessary, which would most effectively align large (>\,10\,$\micron$-sized) grains.  This challenges our current understanding of grain growth in Class 0 sources.
    
    \item High polarization fractions are seen in highly embedded areas of Class 0 sources that are unlikely to be irradiated by the central protostar.  These regions include the possible accretion streamer at the South of SMM1-a, the polarized emission we see to the north of the central core of Emb 8, and magnetized filamentary structures detected toward several other sources.  The question of what would trigger the enhanced polarization in these regions---which are highly embedded, and far away from any obvious source of strong irradiation---remains open. 
    
\end{enumerate}

More work is needed to better understand the relationships among the radiation field, chemistry, and dust-grain alignment in young protostellar sources.  Understanding better the environmental conditions that trigger enhanced polarization in embedded protostars will require working with next-generation full-polarization radiative transfer such as POLARIS \citep{Brauer2016,Reissl2017}; this future work will ultimately improve our understanding of the role played by the magnetic field in protostellar evolution.

\acknowledgments

The authors acknowledge the support of the North American ALMA Science Center. 
V.J.M.L.G. and C.L.H.H. acknowledge the ESO Studentship Program, and the guidance and support of Eric Villard.
C.L.H.H. acknowledges the support of both the NAOJ Fellowship as well as JSPS KAKENHI grant 18K13586. 
This work has received support from the European Research Council (ERC Starting Grant MagneticYSOs with grant agreement no. 679937).
J.M.G. acknowledges the support of the Spanish MINECO AYA2017-84390-C2-R grant, and the Joint ALMA Observatory Visitor Program.
Astrochemistry in Leiden is supported by the Netherlands Research School for Astronomy (NOVA), by a Royal Netherlands Academy of Arts and Sciences (KNAW) professor prize, and by the European Union A-ERC
grant 291141 CHEMPLAN.  
L.E.K. acknowledges a research grant (19127) from VILLUM FONDEN.
Z.-Y.L. is supported in part by NASA 80NSSC18K1095 and NSF AST-1716259 and AST-1815784.
This paper makes use of the following ALMA data: ADS/JAO.ALMA\#2016.1.00710.S, ADS/JAO.ALMA\#2013.1.00726.S, ADS/JAO.ALMA\#2015.1.00768.S, and ADS/JAO.ALMA\#2017.1.01174.S. ALMA is a partnership of ESO (representing its member states), NSF (USA) and NINS (Japan), together with NRC (Canada), MOST and ASIAA (Taiwan), and KASI (Republic of Korea), in cooperation with the Republic of Chile. The Joint ALMA Observatory is operated by ESO, AUI/NRAO and NAOJ.
The National Radio Astronomy Observatory is a facility of the National Science Foundation operated under cooperative agreement by Associated Universities, Inc.

\textit{Facilities:} ALMA.

\textit{Software:} APLpy, an open-source plotting package for Python hosted at \url{http://aplpy.github.com} \citep{Robitaille2012}.  CASA \citep{McMullin2007}.  Astropy \citep{Astropy2018}.

\bibliography{ms}

\begin{thebibliography}{}
\expandafter\ifx\csname natexlab\endcsname\relax\def\natexlab#1{#1}\fi

\bibitem[{{Agurto-Gangas} {et~al.}(2019){Agurto-Gangas}, {Pineda},
  {Sz{\H{u}}cs}, {Testi}, {Tazzari}, {Miotello}, {Caselli}, {Dunham},
  {Stephens}, \& {Bourke}}]{AgurtoGangas2019}
{Agurto-Gangas}, C., {Pineda}, J.~E., {Sz{\H{u}}cs}, L., {et~al.} 2019, \aap,
  623, A147

\bibitem[{{Alves} {et~al.}(2008){Alves}, {Franco}, \& {Girart}}]{Alves2008}
{Alves}, F.~O., {Franco}, G.~A.~P., \& {Girart}, J.~M. 2008, \aap, 486, L13

\bibitem[{{Alves} {et~al.}(2018){Alves}, {Girart}, {Padovani}, {Galli},
  {Franco}, {Caselli}, {Vlemmings}, {Zhang}, \& {Wiesemeyer}}]{Alves2018}
{Alves}, F.~O., {Girart}, J.~M., {Padovani}, M., {et~al.} 2018, \aap, 616, A56

\bibitem[{{Andersson} {et~al.}(2015){Andersson}, {Lazarian}, \&
  {Vaillancourt}}]{Andersson2015}
{Andersson}, B.-G., {Lazarian}, A., \& {Vaillancourt}, J.~E. 2015, \araa, 53,
  501

\bibitem[{{Andr{\'e}} {et~al.}(2014){Andr{\'e}}, {Di Francesco},
  {Ward-Thompson}, {Inutsuka}, {Pudritz}, \& {Pineda}}]{Andre2014}
{Andr{\'e}}, P., {Di Francesco}, J., {Ward-Thompson}, D., {et~al.} 2014, in
  Protostars and Planets VI, ed. H.~{Beuther}, R.~S. {Klessen}, C.~P.
  {Dullemond}, \& T.~{Henning}, 27

\bibitem[{{Andr{\'e}} {et~al.}(2000){Andr{\'e}}, {Ward-Thompson}, \&
  {Barsony}}]{Andre2000}
{Andr{\'e}}, P., {Ward-Thompson}, D., \& {Barsony}, M. 2000, Protostars and
  Planets IV, 59

\bibitem[{{Arce} \& {Sargent}(2006)}]{ArceSargent2006}
{Arce}, H.~G., \& {Sargent}, A.~I. 2006, \apj, 646, 1070

\bibitem[{{Arce} {et~al.}(2007){Arce}, {Shepherd}, {Gueth}, {Lee}, {Bachiller},
  {Rosen}, \& {Beuther}}]{Arce2007}
{Arce}, H.~G., {Shepherd}, D., {Gueth}, F., {et~al.} 2007, in Protostars and
  Planets V, ed. B.~{Reipurth}, D.~{Jewitt}, \& K.~{Keil}, 245

\bibitem[{{Astropy Collaboration} {et~al.}(2018){Astropy Collaboration},
  {Price-Whelan}, {Sip{\H o}cz}, {G{\"u}nther}, {Lim}, {Crawford}, {Conseil},
  {Shupe}, {Craig}, {Dencheva}, {Ginsburg}, {VanderPlas}, {Bradley},
  {P{\'e}rez-Su{\'a}rez}, {de Val-Borro}, {Aldcroft}, {Cruz}, {Robitaille},
  {Tollerud}, {Ardelean}, {Babej}, {Bach}, {Bachetti}, {Bakanov}, {Bamford},
  {Barentsen}, {Barmby}, {Baumbach}, {Berry}, {Biscani}, {Boquien}, {Bostroem},
  {Bouma}, {Brammer}, {Bray}, {Breytenbach}, {Buddelmeijer}, {Burke},
  {Calderone}, {Cano Rodr{\'{\i}}guez}, {Cara}, {Cardoso}, {Cheedella},
  {Copin}, {Corrales}, {Crichton}, {D'Avella}, {Deil}, {Depagne}, {Dietrich},
  {Donath}, {Droettboom}, {Earl}, {Erben}, {Fabbro}, {Ferreira}, {Finethy},
  {Fox}, {Garrison}, {Gibbons}, {Goldstein}, {Gommers}, {Greco}, {Greenfield},
  {Groener}, {Grollier}, {Hagen}, {Hirst}, {Homeier}, {Horton}, {Hosseinzadeh},
  {Hu}, {Hunkeler}, {Ivezi{\'c}}, {Jain}, {Jenness}, {Kanarek}, {Kendrew},
  {Kern}, {Kerzendorf}, {Khvalko}, {King}, {Kirkby}, {Kulkarni}, {Kumar},
  {Lee}, {Lenz}, {Littlefair}, {Ma}, {Macleod}, {Mastropietro}, {McCully},
  {Montagnac}, {Morris}, {Mueller}, {Mumford}, {Muna}, {Murphy}, {Nelson},
  {Nguyen}, {Ninan}, {N{\"o}the}, {Ogaz}, {Oh}, {Parejko}, {Parley}, {Pascual},
  {Patil}, {Patil}, {Plunkett}, {Prochaska}, {Rastogi}, {Reddy Janga},
  {Sabater}, {Sakurikar}, {Seifert}, {Sherbert}, {Sherwood-Taylor}, {Shih},
  {Sick}, {Silbiger}, {Singanamalla}, {Singer}, {Sladen}, {Sooley},
  {Sornarajah}, {Streicher}, {Teuben}, {Thomas}, {Tremblay}, {Turner},
  {Terr{\'o}n}, {van Kerkwijk}, {de la Vega}, {Watkins}, {Weaver}, {Whitmore},
  {Woillez}, {Zabalza}, \& {Astropy Contributors}}]{Astropy2018}
{Astropy Collaboration}, {Price-Whelan}, A.~M., {Sip{\H o}cz}, B.~M., {et~al.}
  2018, \aj, 156, 123

\bibitem[{{Bacciotti} {et~al.}(2018){Bacciotti}, {Girart}, {Padovani}, {Podio},
  {Paladino}, {Testi}, {Bianchi}, {Galli}, {Codella}, {Coffey}, {Favre}, \&
  {Fedele}}]{Bacciotti2018}
{Bacciotti}, F., {Girart}, J.~M., {Padovani}, M., {et~al.} 2018, \apjl, 865,
  L12

\bibitem[{{Bally}(2016)}]{Bally2016}
{Bally}, J. 2016, \araa, 54, 491

\bibitem[{{Benz} {et~al.}(2016){Benz}, {Bruderer}, {van Dishoeck}, {Melchior},
  {Wampfler}, {van der Tak}, {Goicoechea}, {Indriolo}, {Kristensen}, {Lis},
  {Mottram}, {Bergin}, {Caselli}, {Herpin}, {Hogerheijde}, {Johnstone},
  {Liseau}, {Nisini}, {Tafalla}, {Visser}, \& {Wyrowski}}]{Benz2016}
{Benz}, A.~O., {Bruderer}, S., {van Dishoeck}, E.~F., {et~al.} 2016, \aap, 590,
  A105

\bibitem[{{Bergin} {et~al.}(2016){Bergin}, {Du}, {Cleeves}, {Blake}, {Schwarz},
  {Visser}, \& {Zhang}}]{Bergin2016}
{Bergin}, E.~A., {Du}, F., {Cleeves}, L.~I., {et~al.} 2016, \apj, 831, 101

\bibitem[{{Bethell} {et~al.}(2007){Bethell}, {Chepurnov}, {Lazarian}, \&
  {Kim}}]{Bethell2007}
{Bethell}, T.~J., {Chepurnov}, A., {Lazarian}, A., \& {Kim}, J. 2007, \apj,
  663, 1055

\bibitem[{{Bohlin} {et~al.}(1978){Bohlin}, {Savage}, \& {Drake}}]{Bohlin1978}
{Bohlin}, R.~C., {Savage}, B.~D., \& {Drake}, J.~F. 1978, \apj, 224, 132

\bibitem[{{Bracco} {et~al.}(2017){Bracco}, {Palmeirim}, {Andr{\'e}}, {Adam},
  {Ade}, {Bacmann}, {Beelen}, {Beno{\^i}t}, {Bideaud}, {Billot}, {Bourrion},
  {Calvo}, {Catalano}, {Coiffard}, {Comis}, {D'Addabbo}, {D{\'e}sert},
  {Didelon}, {Doyle}, {Goupy}, {K{\"o}nyves}, {Kramer}, {Lagache}, {Leclercq},
  {Mac{\'{\i}}as-P{\'e}rez}, {Maury}, {Mauskopf}, {Mayet}, {Monfardini},
  {Motte}, {Pajot}, {Pascale}, {Peretto}, {Perotto}, {Pisano}, {Ponthieu},
  {Rev{\'e}ret}, {Rigby}, {Ritacco}, {Rodriguez}, {Romero}, {Roy}, {Ruppin},
  {Schuster}, {Sievers}, {Triqueneaux}, {Tucker}, \& {Zylka}}]{Bracco2017}
{Bracco}, A., {Palmeirim}, P., {Andr{\'e}}, P., {et~al.} 2017, \aap, 604, A52

\bibitem[{{Brauer} {et~al.}(2016){Brauer}, {Wolf}, \& {Reissl}}]{Brauer2016}
{Brauer}, R., {Wolf}, S., \& {Reissl}, S. 2016, \aap, 588, A129

\bibitem[{{Chac{\'o}n-Tanarro} {et~al.}(2017){Chac{\'o}n-Tanarro}, {Caselli},
  {Bizzocchi}, {Pineda}, {Harju}, {Spaans}, \&
  {D{\'e}sert}}]{ChaconTanarro2017}
{Chac{\'o}n-Tanarro}, A., {Caselli}, P., {Bizzocchi}, L., {et~al.} 2017, \aap,
  606, A142

\bibitem[{{Chiang} {et~al.}(2012){Chiang}, {Looney}, \& {Tobin}}]{Chiang2012}
{Chiang}, H.-F., {Looney}, L.~W., \& {Tobin}, J.~J. 2012, \apj, 756, 168

\bibitem[{{Cox} {et~al.}(2018){Cox}, {Harris}, {Looney}, {Li}, {Yang}, {Tobin},
  \& {Stephens}}]{Cox2018}
{Cox}, E.~G., {Harris}, R.~J., {Looney}, L.~W., {et~al.} 2018, \apj, 855, 92

\bibitem[{{Cuadrado} {et~al.}(2015){Cuadrado}, {Goicoechea}, {Pilleri},
  {Cernicharo}, {Fuente}, \& {Joblin}}]{Cuadrado2015}
{Cuadrado}, S., {Goicoechea}, J.~R., {Pilleri}, P., {et~al.} 2015, \aap, 575,
  A82

\bibitem[{{Davis} {et~al.}(2000){Davis}, {Chrysostomou}, {Matthews}, {Jenness},
  \& {Ray}}]{Davis2000}
{Davis}, C.~J., {Chrysostomou}, A., {Matthews}, H.~E., {Jenness}, T., \& {Ray},
  T.~P. 2000, \apjl, 530, L115

\bibitem[{{Dent} {et~al.}(2019){Dent}, {Pinte}, {Cortes}, {M{\'e}nard},
  {Hales}, {Fomalont}, \& {de Gregorio-Monsalvo}}]{Dent2019}
{Dent}, W.~R.~F., {Pinte}, C., {Cortes}, P.~C., {et~al.} 2019, \mnras, 482, L29

\bibitem[{{Dionatos} {et~al.}(2014){Dionatos}, {J{\o}rgensen}, {Teixeira},
  {G{\"u}del}, \& {Bergin}}]{Dionatos2014}
{Dionatos}, O., {J{\o}rgensen}, J.~K., {Teixeira}, P.~S., {G{\"u}del}, M., \&
  {Bergin}, E. 2014, \aap, 563, A28

\bibitem[{{Dionatos} {et~al.}(2010){Dionatos}, {Nisini}, {Codella}, \&
  {Giannini}}]{Dionatos2010b}
{Dionatos}, O., {Nisini}, B., {Codella}, C., \& {Giannini}, T. 2010, \aap, 523,
  A29

\bibitem[{{Drozdovskaya} {et~al.}(2015){Drozdovskaya}, {Walsh}, {Visser},
  {Harsono}, \& {van Dishoeck}}]{Drozdovskaya2015}
{Drozdovskaya}, M.~N., {Walsh}, C., {Visser}, R., {Harsono}, D., \& {van
  Dishoeck}, E.~F. 2015, \mnras, 451, 3836

\bibitem[{{Duarte-Cabral} {et~al.}(2011){Duarte-Cabral}, {Dobbs}, {Peretto}, \&
  {Fuller}}]{DuarteCabral2011}
{Duarte-Cabral}, A., {Dobbs}, C.~L., {Peretto}, N., \& {Fuller}, G.~A. 2011,
  \aap, 528, A50

\bibitem[{{Duarte-Cabral} {et~al.}(2010){Duarte-Cabral}, {Fuller}, {Peretto},
  {Hatchell}, {Ladd}, {Buckle}, {Richer}, \& {Graves}}]{DuarteCabral2010}
{Duarte-Cabral}, A., {Fuller}, G.~A., {Peretto}, N., {et~al.} 2010, \aap, 519,
  A27

\bibitem[{Dunham {et~al.}(2015)Dunham, Allen, II, Broekhoven-Fiene, Cieza,
  Francesco, Gutermuth, Harvey, Hatchell, Heiderman, Huard, Johnstone, Kirk,
  Matthews, Miller, Peterson, \& Young}]{Dunham2015}
Dunham, M.~M., Allen, L.~E., II, N. J.~E., {et~al.} 2015, The Astrophysical
  Journal Supplement Series, 220, 11

\bibitem[{{Enoch} {et~al.}(2009){Enoch}, {Evans}, {Sargent}, \&
  {Glenn}}]{Enoch2009b}
{Enoch}, M.~L., {Evans}, II, N.~J., {Sargent}, A.~I., \& {Glenn}, J. 2009,
  \apj, 692, 973

\bibitem[{{Enoch} {et~al.}(2011){Enoch}, {Corder}, {Duch{\^e}ne}, {Bock},
  {Bolatto}, {Culverhouse}, {Kwon}, {Lamb}, {Leitch}, {Marrone}, {Muchovej},
  {P{\'e}rez}, {Scott}, {Teuben}, {Wright}, \& {Zauderer}}]{Enoch2011}
{Enoch}, M.~L., {Corder}, S., {Duch{\^e}ne}, G., {et~al.} 2011, \apjs, 195, 21

\bibitem[{{Fendt}(2006)}]{Fendt2006}
{Fendt}, C. 2006, \apj, 651, 272

\bibitem[{{Frank} {et~al.}(2014){Frank}, {Ray}, {Cabrit}, {Hartigan}, {Arce},
  {Bacciotti}, {Bally}, {Benisty}, {Eisl{\"o}ffel}, {G{\"u}del}, {Lebedev},
  {Nisini}, \& {Raga}}]{Frank2014}
{Frank}, A., {Ray}, T.~P., {Cabrit}, S., {et~al.} 2014, in Protostars and
  Planets VI, ed. H.~{Beuther}, R.~S. {Klessen}, C.~P. {Dullemond}, \&
  T.~{Henning} (Tucson, Arizona: University of Arizona Press), 451--474

\bibitem[{{Frau} {et~al.}(2011){Frau}, {Galli}, \& {Girart}}]{Frau2011}
{Frau}, P., {Galli}, D., \& {Girart}, J.~M. 2011, \aap, 535, A44

\bibitem[{{Galametz} {et~al.}(2018){Galametz}, {Maury}, {Girart}, {Rao},
  {Zhang}, {Gaudel}, {Valdivia}, {Keto}, \& {Lai}}]{Galametz2018}
{Galametz}, M., {Maury}, A., {Girart}, J.~M., {et~al.} 2018, \aap, 616, A139

\bibitem[{{Galli} \& {Shu}(1993)}]{Galli1993a}
{Galli}, D., \& {Shu}, F.~H. 1993, \apj, 417, 220

\bibitem[{{Girart} {et~al.}(2006){Girart}, {Rao}, \& {Marrone}}]{Girart2006}
{Girart}, J.~M., {Rao}, R., \& {Marrone}, D.~P. 2006, Science, 313, 812

\bibitem[{{Girart} {et~al.}(2005){Girart}, {Viti}, {Estalella}, \&
  {Williams}}]{Girart2005}
{Girart}, J.~M., {Viti}, S., {Estalella}, R., \& {Williams}, D.~A. 2005, \aap,
  439, 601

\bibitem[{{Girart} {et~al.}(2018){Girart}, {Fern{\'a}ndez-L{\'o}pez}, {Li},
  {Yang}, {Estalella}, {Anglada}, {{\'A}{\~n}ez-L{\'o}pez}, {Busquet},
  {Carrasco-Gonz{\'a}lez}, {Curiel}, {Galvan-Madrid}, {G{\'o}mez}, {de
  Gregorio-Monsalvo}, {Jim{\'e}nez-Serra}, {Krasnopolsky}, {Mart{\'\i}},
  {Osorio}, {Padovani}, {Rao}, {Rodr{\'\i}guez}, \& {Torrelles}}]{Girart2018}
{Girart}, J.~M., {Fern{\'a}ndez-L{\'o}pez}, M., {Li}, Z.~Y., {et~al.} 2018,
  \apj, 856, L27

\bibitem[{{Goicoechea} {et~al.}(2012){Goicoechea}, {Cernicharo}, {Karska},
  {Herczeg}, {Polehampton}, {Wampfler}, {Kristensen}, {van Dishoeck},
  {Etxaluze}, {Bern{\'e}}, \& {Visser}}]{Goicoechea2012}
{Goicoechea}, J.~R., {Cernicharo}, J., {Karska}, A., {et~al.} 2012, \aap, 548,
  A77

\bibitem[{{Gold}(1952)}]{Gold1952}
{Gold}, T. 1952, \mnras, 112, 215

\bibitem[{{Guillet} {et~al.}(2018){Guillet}, {Fanciullo}, {Verstraete},
  {Boulanger}, {Jones}, {Miville-Desch{\^e}nes}, {Ysard}, {Levrier}, \&
  {Alves}}]{Guillet2018}
{Guillet}, V., {Fanciullo}, L., {Verstraete}, L., {et~al.} 2018, \aap, 610, A16

\bibitem[{{Guzm{\'a}n} {et~al.}(2015){Guzm{\'a}n}, {Pety}, {Goicoechea},
  {Gerin}, {Roueff}, {Gratier}, \& {{\"O}berg}}]{Guzman2015}
{Guzm{\'a}n}, V.~V., {Pety}, J., {Goicoechea}, J.~R., {et~al.} 2015, \apjl,
  800, L33

\bibitem[{{Harris} {et~al.}(2018){Harris}, {Cox}, {Looney}, {Li}, {Yang},
  {Fern{\'a}ndez-L{\'o}pez}, {Kwon}, {Sadavoy}, {Segura-Cox}, {Stephens}, \&
  {Tobin}}]{Harris2018}
{Harris}, R.~J., {Cox}, E.~G., {Looney}, L.~W., {et~al.} 2018, \apj, 861, 91

\bibitem[{{Harsono} {et~al.}(2018){Harsono}, {Bjerkeli}, {van der Wiel},
  {Ramsey}, {Maud}, {Kristensen}, \& {J{\o}rgensen}}]{Harsono2018}
{Harsono}, D., {Bjerkeli}, P., {van der Wiel}, M. H.~D., {et~al.} 2018, Nature
  Astronomy, 2, 646

\bibitem[{{Hennebelle} \& {Ciardi}(2009)}]{Hennebelle2009}
{Hennebelle}, P., \& {Ciardi}, A. 2009, \aap, 506, L29

\bibitem[{{Hennebelle} {et~al.}(2016){Hennebelle}, {Commer{\c c}on},
  {Chabrier}, \& {Marchand}}]{Hennebelle2016}
{Hennebelle}, P., {Commer{\c c}on}, B., {Chabrier}, G., \& {Marchand}, P. 2016,
  \apjl, 830, L8

\bibitem[{{Hoang} {et~al.}(2018){Hoang}, {Cho}, \& {Lazarian}}]{Hoang2018}
{Hoang}, T., {Cho}, J., \& {Lazarian}, A. 2018, \apj, 852, 129

\bibitem[{{Hoang} \& {Lazarian}(2016)}]{Hoang2016}
{Hoang}, T., \& {Lazarian}, A. 2016, \apj, 831, 159

\bibitem[{{Hoang} \& {Tram}(2019)}]{Hoang2019}
{Hoang}, T., \& {Tram}, L.~N. 2019, arXiv e-prints, arXiv:1902.06438

\bibitem[{{Hoang} {et~al.}(2019){Hoang}, {Tram}, {Lee}, \&
  {Ahn}}]{Hoang2019NatAs}
{Hoang}, T., {Tram}, L.~N., {Lee}, H., \& {Ahn}, S.-H. 2019, Nature Astronomy,
  319

\bibitem[{{Hsieh} {et~al.}(2019){Hsieh}, {Hirano}, {Belloche}, {Lee}, {Aso}, \&
  {Lai}}]{Hsieh2019}
{Hsieh}, T.-H., {Hirano}, N., {Belloche}, A., {et~al.} 2019, \apj, 871, 100

\bibitem[{{Hsieh} {et~al.}(2017){Hsieh}, {Lai}, \& {Belloche}}]{Hsieh2017}
{Hsieh}, T.-H., {Lai}, S.-P., \& {Belloche}, A. 2017, \aj, 153, 173

\bibitem[{{Huang} {et~al.}(2018){Huang}, {Andrews}, {Cleeves}, {{\"O}berg},
  {Wilner}, {Bai}, {Birnstiel}, {Carpenter}, {Hughes}, {Isella}, {P{\'e}rez},
  {Ricci}, \& {Zhu}}]{Huang2018}
{Huang}, J., {Andrews}, S.~M., {Cleeves}, L.~I., {et~al.} 2018, \apj, 852, 122

\bibitem[{{Hull} {et~al.}(2019){Hull}, {Le Gouellec}, {Girart}, \&
  {Tobin}}]{Hull2019}
{Hull}, C.~L.~H., {Le Gouellec}, V.~J.~M., {Girart}, J.~M., \& {Tobin}, J.~J.
  2019, ApJ, in prep

\bibitem[{{Hull} \& {Plambeck}(2015)}]{Hull2015b}
{Hull}, C.~L.~H., \& {Plambeck}, R.~L. 2015, Journal of Astronomical
  Instrumentation, 4, 1550005

\bibitem[{{Hull} \& {Zhang}(2019)}]{HullZhang2019}
{Hull}, C. L.~H., \& {Zhang}, Q. 2019, Frontiers in Astronomy and Space
  Sciences, 6, 3

\bibitem[{{Hull} {et~al.}(2014){Hull}, {Plambeck}, {Kwon}, {Bower},
  {Carpenter}, {Crutcher}, {Fiege}, {Franzmann}, {Hakobian}, {Heiles}, {Houde},
  {Hughes}, {Lamb}, {Looney}, {Marrone}, {Matthews}, {Pillai}, {Pound},
  {Rahman}, {Sandell}, {Stephens}, {Tobin}, {Vaillancourt}, {Volgenau}, \&
  {Wright}}]{Hull2014}
{Hull}, C.~L.~H., {Plambeck}, R.~L., {Kwon}, W., {et~al.} 2014, \apjs, 213, 13

\bibitem[{{Hull} {et~al.}(2016){Hull}, {Girart}, {Kristensen}, {Dunham},
  {Rodr{\'{\i}}guez-Kamenetzky}, {Carrasco-Gonz{\'a}lez}, {Cort{\'e}s}, {Li},
  \& {Plambeck}}]{Hull2016a}
{Hull}, C.~L.~H., {Girart}, J.~M., {Kristensen}, L.~E., {et~al.} 2016, \apjl,
  823, L27

\bibitem[{{Hull} {et~al.}(2017{\natexlab{a}}){Hull}, {Girart}, {Tychoniec},
  {Rao}, {Cort{\'e}s}, {Pokhrel}, {Zhang}, {Houde}, {Dunham}, {Kristensen},
  {Lai}, {Li}, \& {Plambeck}}]{Hull2017b}
{Hull}, C.~L.~H., {Girart}, J.~M., {Tychoniec}, {\L}., {et~al.}
  2017{\natexlab{a}}, \apj, 847, 92

\bibitem[{{Hull} {et~al.}(2017{\natexlab{b}}){Hull}, {Mocz}, {Burkhart},
  {Goodman}, {Girart}, {Cort{\'e}s}, {Hernquist}, {Springel}, {Li}, \&
  {Lai}}]{Hull2017a}
{Hull}, C.~L.~H., {Mocz}, P., {Burkhart}, B., {et~al.} 2017{\natexlab{b}},
  \apjl, 842, L9

\bibitem[{{Hull} {et~al.}(2018){Hull}, {Yang}, {Li}, {Kataoka}, {Stephens},
  {Andrews}, {Bai}, {Cleeves}, {Hughes}, {Looney}, {P{\'e}rez}, \&
  {Wilner}}]{Hull2018a}
{Hull}, C.~L.~H., {Yang}, H., {Li}, Z.-Y., {et~al.} 2018, \apj, 860, 82

\bibitem[{{Imai} {et~al.}(2016){Imai}, {Sakai}, {Oya}, {L{\'o}pez-Sepulcre},
  {Watanabe}, {Ceccarelli}, {Lefloch}, {Caux}, {Vastel}, {Kahane}, {Sakai},
  {Hirota}, {Aikawa}, \& {Yamamoto}}]{Imai2016}
{Imai}, M., {Sakai}, N., {Oya}, Y., {et~al.} 2016, \apj, 830, L37

\bibitem[{{Jacobsen} {et~al.}(2018){Jacobsen}, {J{\o}rgensen}, {Di Francesco},
  {Evans}, {Choi}, \& {Lee}}]{Jacobsen2018}
{Jacobsen}, S.~K., {J{\o}rgensen}, J.~K., {Di Francesco}, J., {et~al.} 2018,
  arXiv e-prints, arXiv:1809.00390

\bibitem[{{J{\o}rgensen} {et~al.}(2011){J{\o}rgensen}, {Bourke}, {Nguyen
  Luong}, \& {Takakuwa}}]{Jorgensen2011}
{J{\o}rgensen}, J.~K., {Bourke}, T.~L., {Nguyen Luong}, Q., \& {Takakuwa}, S.
  2011, \aap, 534, A100

\bibitem[{{J{\o}rgensen} {et~al.}(2004){J{\o}rgensen}, {Sch{\"o}ier}, \& {van
  Dishoeck}}]{Jorgensen2004b}
{J{\o}rgensen}, J.~K., {Sch{\"o}ier}, F.~L., \& {van Dishoeck}, E.~F. 2004,
  \aap, 416, 603

\bibitem[{{J{\o}rgensen} {et~al.}(2015){J{\o}rgensen}, {Visser}, {Williams}, \&
  {Bergin}}]{Jorgensen2015}
{J{\o}rgensen}, J.~K., {Visser}, R., {Williams}, J.~P., \& {Bergin}, E.~A.
  2015, \aap, 579, A23

\bibitem[{{Karska} {et~al.}(2018){Karska}, {Kaufman}, {Kristensen}, {van
  Dishoeck}, {Herczeg}, {Mottram}, {Tychoniec}, {Lindberg}, {Evans}, {Green},
  {Yang}, {Gusdorf}, {Itrich}, \& {Si{\'o}dmiak}}]{Karska2018}
{Karska}, A., {Kaufman}, M.~J., {Kristensen}, L.~E., {et~al.} 2018, \apjs, 235,
  30

\bibitem[{{Kastner} {et~al.}(2015){Kastner}, {Qi}, {Gorti}, {Hily-Blant},
  {Oberg}, {Forveille}, {Andrews}, \& {Wilner}}]{Kastner2015}
{Kastner}, J.~H., {Qi}, C., {Gorti}, U., {et~al.} 2015, \apj, 806, 75

\bibitem[{{Kataoka} {et~al.}(2012){Kataoka}, {Machida}, \&
  {Tomisaka}}]{Kataoka2012}
{Kataoka}, A., {Machida}, M.~N., \& {Tomisaka}, K. 2012, \apj, 761, 40

\bibitem[{{Kataoka} {et~al.}(2017){Kataoka}, {Tsukagoshi}, {Pohl}, {Muto},
  {Nagai}, {Stephens}, {Tomisaka}, \& {Momose}}]{Kataoka2017}
{Kataoka}, A., {Tsukagoshi}, T., {Pohl}, A., {et~al.} 2017, \apjl, 844, L5

\bibitem[{{Kataoka} {et~al.}(2015){Kataoka}, {Muto}, {Momose}, {Tsukagoshi},
  {Fukagawa}, {Shibai}, {Hanawa}, {Murakawa}, \& {Dullemond}}]{Kataoka2015}
{Kataoka}, A., {Muto}, T., {Momose}, M., {et~al.} 2015, \apj, 809, 78

\bibitem[{{Kataoka} {et~al.}(2016){Kataoka}, {Tsukagoshi}, {Momose}, {Nagai},
  {Muto}, {Dullemond}, {Pohl}, {Fukagawa}, {Shibai}, {Hanawa}, \&
  {Murakawa}}]{Kataoka2016b}
{Kataoka}, A., {Tsukagoshi}, T., {Momose}, M., {et~al.} 2016, \apjl, 831, L12

\bibitem[{{Kauffmann} {et~al.}(2008){Kauffmann}, {Bertoldi}, {Bourke}, {Evans},
  \& {Lee}}]{Kauffmann2008}
{Kauffmann}, J., {Bertoldi}, F., {Bourke}, T.~L., {Evans}, II, N.~J., \& {Lee},
  C.~W. 2008, \aap, 487, 993

\bibitem[{{Koch} {et~al.}(2018){Koch}, {Tang}, {Ho}, {Yen}, {Su}, \&
  {Takakuwa}}]{Koch2018}
{Koch}, P.~M., {Tang}, Y.-W., {Ho}, P. T.~P., {et~al.} 2018, \apj, 855, 39

\bibitem[{{K{\"o}lligan} \& {Kuiper}(2018)}]{Kolligan2018}
{K{\"o}lligan}, A., \& {Kuiper}, R. 2018, \aap, 620, A182

\bibitem[{{Kristensen} \& {Dunham}(2018)}]{KristensenDunham2018}
{Kristensen}, L.~E., \& {Dunham}, M.~M. 2018, \aap, 618, A158

\bibitem[{{Kristensen} {et~al.}(2017){Kristensen}, {Gusdorf}, {Mottram},
  {Karska}, {Visser}, {Wiesemeyer}, {G{\"u}sten}, \& {Simon}}]{Kristensen2017}
{Kristensen}, L.~E., {Gusdorf}, A., {Mottram}, J.~C., {et~al.} 2017, \aap, 601,
  L4

\bibitem[{{Kristensen} {et~al.}(2013){Kristensen}, {Klaassen}, {Mottram},
  {Schmalzl}, \& {Hogerheijde}}]{Kristensen2013a}
{Kristensen}, L.~E., {Klaassen}, P.~D., {Mottram}, J.~C., {Schmalzl}, M., \&
  {Hogerheijde}, M.~R. 2013, \aap, 549, L6

\bibitem[{{Kristensen} {et~al.}(2010){Kristensen}, {van Dishoeck}, {van
  Kempen}, {Cuppen}, {Brinch}, {J{\o}rgensen}, \&
  {Hogerheijde}}]{Kristensen2010}
{Kristensen}, L.~E., {van Dishoeck}, E.~F., {van Kempen}, T.~A., {et~al.} 2010,
  \aap, 516, A57

\bibitem[{{Kristensen} {et~al.}(2012){Kristensen}, {van Dishoeck}, {Bergin},
  {Visser}, {Y{\i}ld{\i}z}, {San Jose-Garcia}, {J{\o}rgensen}, {Herczeg},
  {Johnstone}, {Wampfler}, {Benz}, {Bruderer}, {Cabrit}, {Caselli}, {Doty},
  {Harsono}, {Herpin}, {Hogerheijde}, {Karska}, {van Kempen}, {Liseau},
  {Nisini}, {Tafalla}, {van der Tak}, \& {Wyrowski}}]{Kristensen2012}
{Kristensen}, L.~E., {van Dishoeck}, E.~F., {Bergin}, E.~A., {et~al.} 2012,
  \aap, 542, A8

\bibitem[{{Kwon} {et~al.}(2019){Kwon}, {Stephens}, {Tobin}, {Looney}, {Li},
  {van der Tak}, \& {Crutcher}}]{Kwon2019}
{Kwon}, W., {Stephens}, I.~W., {Tobin}, J.~J., {et~al.} 2019, \apj, 879, 25

\bibitem[{{Lazarian}(2005)}]{Lazarian2005}
{Lazarian}, A. 2005, in American Institute of Physics Conference Series, Vol.
  784, Magnetic Fields in the Universe: From Laboratory and Stars to Primordial
  Structures., ed. E.~M. {de Gouveia dal Pino}, G.~{Lugones}, \& A.~{Lazarian},
  42--53

\bibitem[{{Lazarian}(2007)}]{Lazarian2007}
{Lazarian}, A. 2007, J.~Quant.~Spec.~Radiat.~Transf., 106, 225

\bibitem[{{Lazarian} \& {Hoang}(2007)}]{LazarianHoang2007}
{Lazarian}, A., \& {Hoang}, T. 2007, \mnras, 378, 910

\bibitem[{{Lee} {et~al.}(2018){Lee}, {Hwang}, {Ching}, {Hirano}, {Lai}, {Rao},
  \& {Ho}}]{LeeCF2018Nat}
{Lee}, C.-F., {Hwang}, H.-C., {Ching}, T.-C., {et~al.} 2018, Nature
  Communications, 9, 4636

\bibitem[{Lee {et~al.}(2016)Lee, Hwang, \& Li}]{LeeCF2016}
Lee, C.-F., Hwang, H.-C., \& Li, Z.-Y. 2016, The Astrophysical Journal, 826,
  213

\bibitem[{{Lee} {et~al.}(2017){Lee}, {Hull}, \& {Offner}}]{JLee2017}
{Lee}, J.~W.~Y., {Hull}, C.~L.~H., \& {Offner}, S.~S.~R. 2017, \apj, 834, 201

\bibitem[{{Lee} {et~al.}(2012){Lee}, {Looney}, {Johnstone}, \&
  {Tobin}}]{LeeK2012}
{Lee}, K., {Looney}, L., {Johnstone}, D., \& {Tobin}, J. 2012, \apj, 761, 171

\bibitem[{{Lee} {et~al.}(2014){Lee}, {Fern{\'a}ndez-L{\'o}pez}, {Storm},
  {Looney}, {Mundy}, {Segura-Cox}, {Teuben}, {Rosolowsky}, {Arce}, {Ostriker},
  {Shirley}, {Kwon}, {Kauffmann}, {Tobin}, {Plunkett}, {Pound}, {Salter},
  {Volgenau}, {Chen}, {Tassis}, {Isella}, {Crutcher}, {Gammie}, \&
  {Testi}}]{Lee2014}
{Lee}, K.~I., {Fern{\'a}ndez-L{\'o}pez}, M., {Storm}, S., {et~al.} 2014, \apj,
  797, 76

\bibitem[{{Lee} {et~al.}(2015){Lee}, {Lee}, \& {Bergin}}]{LeeS2015}
{Lee}, S., {Lee}, J.-E., \& {Bergin}, E.~A. 2015, The Astrophysical Journal
  Supplement Series, 217, 30

\bibitem[{{Li} {et~al.}(2014){Li}, {Goodman}, {Sridharan}, {Houde}, {Li},
  {Novak}, \& {Tang}}]{HBLi2014}
{Li}, H.-B., {Goodman}, A., {Sridharan}, T.~K., {et~al.} 2014, Protostars and
  Planets VI, 101

\bibitem[{{Liu} {et~al.}(2017){Liu}, {Henning}, {Carrasco-Gonz{\'a}lez},
  {Chandler}, {Linz}, {Birnstiel}, {van Boekel}, {P{\'e}rez}, {Flock}, {Testi},
  {Rodr{\'{\i}}guez}, \& {Galv{\'a}n-Madrid}}]{YLiu2017}
{Liu}, Y., {Henning}, T., {Carrasco-Gonz{\'a}lez}, C., {et~al.} 2017, \aap,
  607, A74

\bibitem[{{Machida} {et~al.}(2005){Machida}, {Matsumoto}, {Hanawa}, \&
  {Tomisaka}}]{Machida2005a}
{Machida}, M.~N., {Matsumoto}, T., {Hanawa}, T., \& {Tomisaka}, K. 2005,
  \mnras, 362, 382

\bibitem[{{Machida} {et~al.}(2006){Machida}, {Matsumoto}, {Hanawa}, \&
  {Tomisaka}}]{Machida2006}
---. 2006, \apj, 645, 1227

\bibitem[{{Mart{\'{\i}}-Vidal} {et~al.}(2014){Mart{\'{\i}}-Vidal}, {Vlemmings},
  {Muller}, \& {Casey}}]{MartiVidal2014}
{Mart{\'{\i}}-Vidal}, I., {Vlemmings}, W.~H.~T., {Muller}, S., \& {Casey}, S.
  2014, \aap, 563, A136

\bibitem[{{Matthews} {et~al.}(2009){Matthews}, {McPhee}, {Fissel}, \&
  {Curran}}]{Matthews2009}
{Matthews}, B.~C., {McPhee}, C.~A., {Fissel}, L.~M., \& {Curran}, R.~L. 2009,
  \apjs, 182, 143

\bibitem[{{Maury} {et~al.}(2014){Maury}, {Belloche}, {Andr{\'e}}, {Maret},
  {Gueth}, {Codella}, {Cabrit}, {Testi}, \& {Bontemps}}]{Maury2014}
{Maury}, A.~J., {Belloche}, A., {Andr{\'e}}, P., {et~al.} 2014, \aap, 563, L2

\bibitem[{{Maury} {et~al.}(2018){Maury}, {Girart}, {Zhang}, {Hennebelle},
  {Keto}, {Rao}, {Lai}, {Ohashi}, \& {Galametz}}]{Maury2018}
{Maury}, A.~J., {Girart}, J.~M., {Zhang}, Q., {et~al.} 2018, \mnras, 477, 2760

\bibitem[{{Maury} {et~al.}(2019){Maury}, {Andr{\'e}}, {Testi}, {Maret},
  {Belloche}, {Hennebelle}, {Cabrit}, {Codella}, {Gueth}, {Podio}, {Anderl},
  {Bacmann}, {Bontemps}, {Gaudel}, {Ladjelate}, {Lef{\`e}vre}, {Tabone}, \&
  {Lefloch}}]{Maury2019}
{Maury}, A.~J., {Andr{\'e}}, P., {Testi}, L., {et~al.} 2019, \aap, 621, A76

\bibitem[{{McMullin} {et~al.}(2007){McMullin}, {Waters}, {Schiebel}, {Young},
  \& {Golap}}]{McMullin2007}
{McMullin}, J.~P., {Waters}, B., {Schiebel}, D., {Young}, W., \& {Golap}, K.
  2007, in Astronomical Society of the Pacific Conference Series, Vol. 376,
  Astronomical Data Analysis Software and Systems XVI, ed. R.~A. {Shaw},
  F.~{Hill}, \& D.~J. {Bell}, 127

\bibitem[{{Mottram} {et~al.}(2014){Mottram}, {Kristensen}, {van Dishoeck},
  {Bruderer}, {San Jos{\'e}-Garc{\'\i}a}, {Karska}, {Visser}, {Santangelo},
  {Benz}, {Bergin}, {Caselli}, {Herpin}, {Hogerheijde}, {Johnstone}, {van
  Kempen}, {Liseau}, {Nisini}, {Tafalla}, {van der Tak}, \&
  {Wyrowski}}]{Mottram2014}
{Mottram}, J.~C., {Kristensen}, L.~E., {van Dishoeck}, E.~F., {et~al.} 2014,
  \aap, 572, A21

\bibitem[{{Mottram} {et~al.}(2017){Mottram}, {van Dishoeck}, {Kristensen},
  {Karska}, {San Jos{\'e}-Garc{\'\i}a}, {Khanna}, {Herczeg}, {Andr{\'e}},
  {Bontemps}, {Cabrit}, {Carney}, {Drozdovskaya}, {Dunham}, {Evans}, {Fedele},
  {Green}, {Harsono}, {Johnstone}, {J{\o}rgensen}, {K{\"o}nyves}, {Nisini},
  {Persson}, {Tafalla}, {Visser}, \& {Y{\i}ld{\i}z}}]{Mottram2017}
{Mottram}, J.~C., {van Dishoeck}, E.~F., {Kristensen}, L.~E., {et~al.} 2017,
  \aap, 600, A99

\bibitem[{{Murillo} {et~al.}(2015){Murillo}, {Bruderer}, {van Dishoeck},
  {Walsh}, {Harsono}, {Lai}, \& {Fuchs}}]{Murillo2015}
{Murillo}, N.~M., {Bruderer}, S., {van Dishoeck}, E.~F., {et~al.} 2015, \aap,
  579, A114

\bibitem[{{Murillo} {et~al.}(2018){Murillo}, {van Dishoeck}, {van der Wiel},
  {J{\o}rgensen}, {Drozdovskaya}, {Calcutt}, \& {Harsono}}]{Murillo2018}
{Murillo}, N.~M., {van Dishoeck}, E.~F., {van der Wiel}, M.~H.~D., {et~al.}
  2018, \aap, 617, A120

\bibitem[{{Nagai} {et~al.}(2016){Nagai}, {Nakanishi}, {Paladino}, {Hull},
  {Cortes}, {Moellenbrock}, {Fomalont}, {Asada}, \& {Hada}}]{Nagai2016}
{Nagai}, H., {Nakanishi}, K., {Paladino}, R., {et~al.} 2016, \apj, 824, 132

\bibitem[{{{\"O}berg} {et~al.}(2011){{\"O}berg}, {van der Marel}, {Kristensen},
  \& {van Dishoeck}}]{Oberg2011}
{{\"O}berg}, K.~I., {van der Marel}, N., {Kristensen}, L.~E., \& {van
  Dishoeck}, E.~F. 2011, \apj, 740, 14

\bibitem[{{Offner} \& {Chaban}(2017)}]{Offner2017}
{Offner}, S.~S.~R., \& {Chaban}, J. 2017, \apj, 847, 104

\bibitem[{{Offner} {et~al.}(2016){Offner}, {Dunham}, {Lee}, {Arce}, \&
  {Fielding}}]{Offner2016}
{Offner}, S.~S.~R., {Dunham}, M.~M., {Lee}, K.~I., {Arce}, H.~G., \&
  {Fielding}, D.~B. 2016, \apjl, 827, L11

\bibitem[{{Ohashi} {et~al.}(2014){Ohashi}, {Saigo}, {Aso}, {Aikawa},
  {Koyamatsu}, {Machida}, {Saito}, {Takahashi}, {Takakuwa}, {Tomida},
  {Tomisaka}, \& {Yen}}]{Ohashi2014}
{Ohashi}, N., {Saigo}, K., {Aso}, Y., {et~al.} 2014, \apj, 796, 131

\bibitem[{{Ohashi} {et~al.}(2018){Ohashi}, {Kataoka}, {Nagai}, {Momose},
  {Muto}, {Hanawa}, {Fukagawa}, {Tsukagoshi}, {Murakawa}, \&
  {Shibai}}]{Ohashi2018}
{Ohashi}, S., {Kataoka}, A., {Nagai}, H., {et~al.} 2018, \apj, 864, 81

\bibitem[{{Ortiz-Le{\'o}n} {et~al.}(2017){Ortiz-Le{\'o}n}, {Dzib}, {Kounkel},
  {Loinard}, {Mioduszewski}, {Rodr{\'{\i}}guez}, {Torres}, {Pech}, {Rivera},
  {Hartmann}, {Boden}, {Evans}, {Brice{\~n}o}, {Tobin}, \&
  {Galli}}]{OrtizLeon2017b}
{Ortiz-Le{\'o}n}, G.~N., {Dzib}, S.~A., {Kounkel}, M.~A., {et~al.} 2017, \apj,
  834, 143

\bibitem[{{Ossenkopf} \& {Henning}(1994)}]{Ossenkopf1994}
{Ossenkopf}, V., \& {Henning}, T. 1994, \aap, 291, 943

\bibitem[{{Panoglou} {et~al.}(2012){Panoglou}, {Cabrit}, {Pineau Des
  For{\^e}ts}, {Garcia}, {Ferreira}, \& {Casse}}]{Panoglou2012}
{Panoglou}, D., {Cabrit}, S., {Pineau Des For{\^e}ts}, G., {et~al.} 2012, \aap,
  538, A2

\bibitem[{{Pattle} {et~al.}(2017){Pattle}, {Ward-Thompson}, {Berry},
  {Hatchell}, {Chen}, {Pon}, {Koch}, {Kwon}, {Kim}, {Bastien}, {Cho},
  {Coud{\'e}}, {Di Francesco}, {Fuller}, {Furuya}, {Graves}, {Johnstone},
  {Kirk}, {Kwon}, {Lee}, {Matthews}, {Mottram}, {Parsons}, {Sadavoy},
  {Shinnaga}, {Soam}, {Hasegawa}, {Lai}, {Qiu}, \& {Friberg}}]{Pattle2017}
{Pattle}, K., {Ward-Thompson}, D., {Berry}, D., {et~al.} 2017, \apj, 846, 122

\bibitem[{{Pelkonen} {et~al.}(2009){Pelkonen}, {Juvela}, \&
  {Padoan}}]{Pelkonen2009}
{Pelkonen}, V.-M., {Juvela}, M., \& {Padoan}, P. 2009, \aap, 502, 833

\bibitem[{{P{\'e}rez} {et~al.}(2012){P{\'e}rez}, {Carpenter}, {Chandler},
  {Isella}, {Andrews}, {Ricci}, {Calvet}, {Corder}, {Deller}, {Dullemond},
  {Greaves}, {Harris}, {Henning}, {Kwon}, {Lazio}, {Linz}, {Mundy}, {Sargent},
  {Storm}, {Testi}, \& {Wilner}}]{Perez2012}
{P{\'e}rez}, L.~M., {Carpenter}, J.~M., {Chandler}, C.~J., {et~al.} 2012,
  \apjl, 760, L17

\bibitem[{{P{\'e}rez} {et~al.}(2015){P{\'e}rez}, {Chandler}, {Isella},
  {Carpenter}, {Andrews}, {Calvet}, {Corder}, {Deller}, {Dullemond}, {Greaves},
  {Harris}, {Henning}, {Kwon}, {Lazio}, {Linz}, {Mundy}, {Ricci}, {Sargent},
  {Storm}, {Tazzari}, {Testi}, \& {Wilner}}]{Perez2015}
{P{\'e}rez}, L.~M., {Chandler}, C.~J., {Isella}, A., {et~al.} 2015, \apj, 813,
  41

\bibitem[{{Planck Collaboration} {et~al.}(2016){Planck Collaboration}, {Ade},
  {Aghanim}, {Alves}, {Arnaud}, {Arzoumanian}, {Aumont}, {Baccigalupi},
  {Banday}, {Barreiro}, {Bartolo}, {Battaner}, {Benabed}, {Benoit-L{\'e}vy},
  {Bernard}, {Bern{\'e}}, {Bersanelli}, {Bielewicz}, {Bonaldi}, {Bonavera},
  {Bond}, {Borrill}, {Bouchet}, {Boulanger}, {Bracco}, {Burigana}, {Calabrese},
  {Cardoso}, {Catalano}, {Chamballu}, {Chiang}, {Christensen}, {Clements},
  {Colombi}, {Colombo}, {Combet}, {Couchot}, {Crill}, {Curto}, {Cuttaia},
  {Danese}, {Davies}, {Davis}, {de Bernardis}, {de Rosa}, {de Zotti},
  {Delabrouille}, {Dickinson}, {Diego}, {Donzelli}, {Dor{\'e}}, {Douspis},
  {Ducout}, {Dupac}, {Elsner}, {En{\ss}lin}, {Eriksen}, {Falgarone},
  {Ferri{\`e}re}, {Finelli}, {Forni}, {Frailis}, {Fraisse}, {Franceschi},
  {Frejsel}, {Galeotta}, {Galli}, {Ganga}, {Ghosh}, {Giard},
  {Giraud-H{\'e}raud}, {Gjerl{\o}w}, {Gonz{\'a}lez-Nuevo}, {G{\'o}rski},
  {Gregorio}, {Gruppuso}, {Guillet}, {Hansen}, {Hanson}, {Harrison},
  {Hern{\'a}ndez-Monteagudo}, {Herranz}, {Hildebrandt}, {Hivon}, {Hobson},
  {Holmes}, {Huffenberger}, {Hurier}, {Jaffe}, {Jaffe}, {Jones}, {Juvela},
  {Keskitalo}, {Kisner}, {Knoche}, {Kunz}, {Kurki-Suonio}, {Lagache},
  {Lamarre}, {Lasenby}, {Lawrence}, {Leonardi}, {Levrier}, {Liguori}, {Lilje},
  {Linden-V{\o}rnle}, {L{\'o}pez-Caniego}, {Lubin}, {Mac{\'{\i}}as-P{\'e}rez},
  {Maffei}, {Mandolesi}, {Mangilli}, {Maris}, {Martin},
  {Mart{\'{\i}}nez-Gonz{\'a}lez}, {Masi}, {Matarrese}, {Mazzotta},
  {Melchiorri}, {Mendes}, {Mennella}, {Migliaccio}, {Mitra},
  {Miville-Desch{\^e}nes}, {Moneti}, {Montier}, {Morgante}, {Mortlock},
  {Munshi}, {Murphy}, {Naselsky}, {Nati}, {Natoli}, {N{\o}rgaard-Nielsen},
  {Noviello}, {Novikov}, {Novikov}, {Oppermann}, {Pagano}, {Pajot}, {Paladini},
  {Paoletti}, {Pasian}, {Perrotta}, {Pettorino}, {Piacentini}, {Piat},
  {Pierpaoli}, {Pietrobon}, {Plaszczynski}, {Pointecouteau}, {Polenta},
  {Pratt}, {Puget}, {Rachen}, {Rebolo}, {Reinecke}, {Remazeilles}, {Renault},
  {Renzi}, {Ricciardi}, {Ristorcelli}, {Rocha}, {Rosset}, {Rossetti},
  {Roudier}, {Rubi{\~n}o-Mart{\'{\i}}n}, {Rusholme}, {Sandri}, {Savelainen},
  {Savini}, {Scott}, {Soler}, {Stolyarov}, {Sutton}, {Suur-Uski}, {Sygnet},
  {Tauber}, {Terenzi}, {Toffolatti}, {Tomasi}, {Tristram}, {Tucci},
  {Valenziano}, {Valiviita}, {Van Tent}, {Vielva}, {Villa}, {Wade}, {Wandelt},
  {Yvon}, {Zacchei}, \& {Zonca}}]{PlanckCollaborationXXXIII2016}
{Planck Collaboration}, {Ade}, P.~A.~R., {Aghanim}, N., {et~al.} 2016, \aap,
  586, A136

\bibitem[{{Pudritz} {et~al.}(2007){Pudritz}, {Ouyed}, {Fendt}, \&
  {Brandenburg}}]{Pudritz2007}
{Pudritz}, R.~E., {Ouyed}, R., {Fendt}, C., \& {Brandenburg}, A. 2007,
  Protostars and Planets V, 277

\bibitem[{{Pudritz} {et~al.}(2006){Pudritz}, {Rogers}, \&
  {Ouyed}}]{Pudritz2006}
{Pudritz}, R.~E., {Rogers}, C.~S., \& {Ouyed}, R. 2006, \mnras, 365, 1131

\bibitem[{{Ramsey} \& {Clarke}(2019)}]{Ramsey2019}
{Ramsey}, J.~P., \& {Clarke}, D.~A. 2019, \mnras, 484, 2364

\bibitem[{{Rao} {et~al.}(2014){Rao}, {Girart}, {Lai}, \& {Marrone}}]{Rao2014}
{Rao}, R., {Girart}, J.~M., {Lai}, S.-P., \& {Marrone}, D.~P. 2014, \apjl, 780,
  L6

\bibitem[{{Reissl} {et~al.}(2017){Reissl}, {Seifried}, {Wolf}, {Banerjee}, \&
  {Klessen}}]{Reissl2017}
{Reissl}, S., {Seifried}, D., {Wolf}, S., {Banerjee}, R., \& {Klessen}, R.~S.
  2017, \aap, 603, A71

\bibitem[{{Robitaille} \& {Bressert}(2012)}]{Robitaille2012}
{Robitaille}, T., \& {Bressert}, E. 2012, {APLpy: Astronomical Plotting Library
  in Python}, Astrophysics Source Code Library, ascl:1208.017

\bibitem[{{Rodr{\'{\i}}guez-Kamenetzky}
  {et~al.}(2016){Rodr{\'{\i}}guez-Kamenetzky}, {Carrasco-Gonz{\'a}lez},
  {Araudo}, {Torrelles}, {Anglada}, {Mart{\'{\i}}}, {Rodr{\'{\i}}guez}, \&
  {Valotto}}]{RodriguezKamenetzky2016}
{Rodr{\'{\i}}guez-Kamenetzky}, A., {Carrasco-Gonz{\'a}lez}, C., {Araudo}, A.,
  {et~al.} 2016, \apj, 818, 27

\bibitem[{{Sadavoy} {et~al.}(2016){Sadavoy}, {Stutz}, {Schnee}, {Mason}, {Di
  Francesco}, \& {Friesen}}]{Sadavoy2016}
{Sadavoy}, S.~I., {Stutz}, A.~M., {Schnee}, S., {et~al.} 2016, \aap, 588, A30

\bibitem[{{Sadavoy} {et~al.}(2018{\natexlab{a}}){Sadavoy}, {Myers}, {Stephens},
  {Tobin}, {Commer{\c{c}}on}, {Henning}, {Looney}, {Kwon}, {Segura-Cox}, \&
  {Harris}}]{Sadavoy2018a}
{Sadavoy}, S.~I., {Myers}, P.~C., {Stephens}, I.~W., {et~al.}
  2018{\natexlab{a}}, \apj, 859, 165

\bibitem[{{Sadavoy} {et~al.}(2018{\natexlab{b}}){Sadavoy}, {Myers}, {Stephens},
  {Tobin}, {Kwon}, {Segura-Cox}, {Henning}, {Commer{\c c}on}, \&
  {Looney}}]{Sadavoy2018b}
---. 2018{\natexlab{b}}, \apj, 869, 115

\bibitem[{{Sahu} {et~al.}(2019){Sahu}, {Liu}, {Su}, {Li}, {Lee}, {Hirano}, \&
  {Takakuwa}}]{Sahu2019}
{Sahu}, D., {Liu}, S.-Y., {Su}, Y.-N., {et~al.} 2019, \apj, 872, 196

\bibitem[{{Segura-Cox} {et~al.}(2015){Segura-Cox}, {Looney}, {Stephens},
  {Fern{\'a}ndez-L{\'o}pez}, {Kwon}, {Tobin}, {Li}, \&
  {Crutcher}}]{SeguraCox2015}
{Segura-Cox}, D.~M., {Looney}, L.~W., {Stephens}, I.~W., {et~al.} 2015, \apjl,
  798, L2

\bibitem[{{Seifried} {et~al.}(2015){Seifried}, {Banerjee}, {Pudritz}, \&
  {Klessen}}]{Seifried2015}
{Seifried}, D., {Banerjee}, R., {Pudritz}, R.~E., \& {Klessen}, R.~S. 2015,
  \mnras, 446, 2776

\bibitem[{{Spaans} {et~al.}(1995){Spaans}, {Hogerheijde}, {Mundy}, \& {van
  Dishoeck}}]{Spaans1995}
{Spaans}, M., {Hogerheijde}, M.~R., {Mundy}, L.~G., \& {van Dishoeck}, E.~F.
  1995, \apj, 455, L167

\bibitem[{{St{\"a}uber} {et~al.}(2005){St{\"a}uber}, {Doty}, {van Dishoeck}, \&
  {Benz}}]{Stauber2005}
{St{\"a}uber}, P., {Doty}, S.~D., {van Dishoeck}, E.~F., \& {Benz}, A.~O. 2005,
  \aap, 440, 949

\bibitem[{{St{\"a}uber} {et~al.}(2004){St{\"a}uber}, {Doty}, {van Dishoeck},
  {J{\o}rgensen}, \& {Benz}}]{Stauber2004}
{St{\"a}uber}, P., {Doty}, S.~D., {van Dishoeck}, E.~F., {J{\o}rgensen}, J.~K.,
  \& {Benz}, A.~O. 2004, \aap, 425, 577

\bibitem[{{Stephens} {et~al.}(2013){Stephens}, {Looney}, {Kwon}, {Hull},
  {Plambeck}, {Crutcher}, {Chapman}, {Novak}, {Davidson}, {Vaillancourt},
  {Shinnaga}, \& {Matthews}}]{Stephens2013}
{Stephens}, I.~W., {Looney}, L.~W., {Kwon}, W., {et~al.} 2013, \apjl, 769, L15

\bibitem[{{Stephens} {et~al.}(2017){Stephens}, {Yang}, {Li}, {Looney},
  {Kataoka}, {Kwon}, {Fern{\'a}ndez-L{\'o}pez}, {Hull}, {Hughes}, {Segura-Cox},
  {Mundy}, {Crutcher}, \& {Rao}}]{Stephens2017b}
{Stephens}, I.~W., {Yang}, H., {Li}, Z.-Y., {et~al.} 2017, \apj, 851, 55

\bibitem[{{Takahashi} {et~al.}(2019){Takahashi}, {Machida}, {Tomisaka}, {Ho},
  {Fomalont}, {Nakanishi}, \& {Girart}}]{Takahashi2019}
{Takahashi}, S., {Machida}, M.~N., {Tomisaka}, K., {et~al.} 2019, \apj, 872, 70

\bibitem[{{Tazaki} {et~al.}(2017){Tazaki}, {Lazarian}, \&
  {Nomura}}]{Tazaki2017}
{Tazaki}, R., {Lazarian}, A., \& {Nomura}, H. 2017, \apj, 839, 56

\bibitem[{{Tazzari} {et~al.}(2016){Tazzari}, {Testi}, {Ercolano}, {Natta},
  {Isella}, {Chandler}, {P{\'e}rez}, {Andrews}, {Wilner}, {Ricci}, {Henning},
  {Linz}, {Kwon}, {Corder}, {Dullemond}, {Carpenter}, {Sargent}, {Mundy},
  {Storm}, {Calvet}, {Greaves}, {Lazio}, \& {Deller}}]{Tazzari2016}
{Tazzari}, M., {Testi}, L., {Ercolano}, B., {et~al.} 2016, \aap, 588, A53

\bibitem[{{Testi} {et~al.}(2014){Testi}, {Birnstiel}, {Ricci}, {Andrews},
  {Blum}, {Carpenter}, {Dominik}, {Isella}, {Natta}, {Williams}, \&
  {Wilner}}]{Testi2014}
{Testi}, L., {Birnstiel}, T., {Ricci}, L., {et~al.} 2014, Protostars and
  Planets VI, 339

\bibitem[{{Tobin} {et~al.}(2010){Tobin}, {Hartmann}, {Looney}, \&
  {Chiang}}]{Tobin2010}
{Tobin}, J.~J., {Hartmann}, L., {Looney}, L.~W., \& {Chiang}, H.-F. 2010, \apj,
  712, 1010

\bibitem[{{Trotta} {et~al.}(2013){Trotta}, {Testi}, {Natta}, {Isella}, \&
  {Ricci}}]{Trotta2013}
{Trotta}, F., {Testi}, L., {Natta}, A., {Isella}, A., \& {Ricci}, L. 2013,
  \aap, 558, A64

\bibitem[{{Tychoniec} {et~al.}(2019){Tychoniec}, {Hull}, {Kristensen}, {Tobin},
  {Le Gouellec}, \& {van Dishoeck}}]{Tychoniec2019}
{Tychoniec}, {\L}., {Hull}, C.~L.~H., {Kristensen}, L.~E., {et~al.} 2019,
  Astronomy and Astrophysics, in prep.

\bibitem[{{Tychoniec} {et~al.}(2018){Tychoniec}, {Hull}, {Tobin}, \& {van
  Dishoeck}}]{Tychoniec2018}
{Tychoniec}, {\L}., {Hull}, C.~L.~H., {Tobin}, J.~J., \& {van Dishoeck}, E.~F.
  2018, in IAU Symposium, Vol. 332, IAU Symposium, ed. M.~{Cunningham},
  T.~{Millar}, \& Y.~{Aikawa}, 249--253

\bibitem[{{Vaillancourt}(2006)}]{Vaillancourt2006}
{Vaillancourt}, J.~E. 2006, \pasp, 118, 1340

\bibitem[{{V{\"a}is{\"a}l{\"a}} {et~al.}(2019){V{\"a}is{\"a}l{\"a}}, {Shang},
  {Krasnopolsky}, {Liu}, {Lam}, \& {Li}}]{Vaisala2019}
{V{\"a}is{\"a}l{\"a}}, M.~S., {Shang}, H., {Krasnopolsky}, R., {et~al.} 2019,
  \apj, 873, 114

\bibitem[{{Valdivia} {et~al.}(2019){Valdivia}, {Maury}, {Brauer}, {Hennebelle},
  {Galametz}, {Guillet}, \& {Reissl}}]{Valdivia2019}
{Valdivia}, V., {Maury}, A., {Brauer}, R., {et~al.} 2019, arXiv e-prints,
  arXiv:1907.10945

\bibitem[{{van Kempen} {et~al.}(2009){van Kempen}, {Wilner}, \&
  {Gurwell}}]{vanKempen2009}
{van Kempen}, T.~A., {Wilner}, D., \& {Gurwell}, M. 2009, \apjl, 706, L22

\bibitem[{{Velusamy} {et~al.}(2014){Velusamy}, {Langer}, \&
  {Thompson}}]{Velusamy2014}
{Velusamy}, T., {Langer}, W.~D., \& {Thompson}, T. 2014, \apj, 783, 6

\bibitem[{{Visser} {et~al.}(2015){Visser}, {Bergin}, \&
  {J{\o}rgensen}}]{Visser2015}
{Visser}, R., {Bergin}, E.~A., \& {J{\o}rgensen}, J.~K. 2015, \aap, 577, A102

\bibitem[{{Visser} {et~al.}(2012){Visser}, {Kristensen}, {Bruderer}, {van
  Dishoeck}, {Herczeg}, {Brinch}, {Doty}, {Harsono}, \& {Wolfire}}]{Visser2012}
{Visser}, R., {Kristensen}, L.~E., {Bruderer}, S., {et~al.} 2012, \aap, 537,
  A55

\bibitem[{{Walsh} {et~al.}(2014){Walsh}, {Millar}, {Nomura}, {Herbst}, {Widicus
  Weaver}, {Aikawa}, {Laas}, \& {Vasyunin}}]{Walsh2014}
{Walsh}, C., {Millar}, T.~J., {Nomura}, H., {et~al.} 2014, \aap, 563, A33

\bibitem[{{Wong} {et~al.}(2016){Wong}, {Hirashita}, \& {Li}}]{Wong2016}
{Wong}, Y. H.~V., {Hirashita}, H., \& {Li}, Z.-Y. 2016, \pasj, 68, 67

\bibitem[{{Wurster} \& {Li}(2018)}]{Wurster2018}
{Wurster}, J., \& {Li}, Z.-Y. 2018, Frontiers in Astronomy and Space Sciences,
  5, 39

\bibitem[{{Yang} {et~al.}(2017){Yang}, {Li}, {Looney}, {Girart}, \&
  {Stephens}}]{Yang2017}
{Yang}, H., {Li}, Z.-Y., {Looney}, L.~W., {Girart}, J.~M., \& {Stephens}, I.~W.
  2017, \mnras, 472, 373

\bibitem[{{Yang} {et~al.}(2019){Yang}, {Li}, {Stephens}, {Kataoka}, \&
  {Looney}}]{Yang2019}
{Yang}, H., {Li}, Z.-Y., {Stephens}, I.~W., {Kataoka}, A., \& {Looney}, L.
  2019, \mnras, 483, 2371

\bibitem[{{Yen} {et~al.}(2015){Yen}, {Koch}, {Takakuwa}, {Ho}, {Ohashi}, \&
  {Tang}}]{Yen2015a}
{Yen}, H.-W., {Koch}, P.~M., {Takakuwa}, S., {et~al.} 2015, \apj, 799, 193

\bibitem[{{Yen} {et~al.}(2019){Yen}, {Zhao}, {Hsieh}, {Koch}, {Krasnopolsky},
  {Lee}, {Li}, {Liu}, {Ohashi}, {Takakuwa}, \& {Tang}}]{Yen2019}
{Yen}, H.-W., {Zhao}, B., {Hsieh}, I.~T., {et~al.} 2019, \apj, 871, 243

\bibitem[{{Y{\i}ld{\i}z} {et~al.}(2012){Y{\i}ld{\i}z}, {Kristensen}, {van
  Dishoeck}, {Belloche}, {van Kempen}, {Hogerheijde}, {G{\"u}sten}, \& {van der
  Marel}}]{Yildiz2012}
{Y{\i}ld{\i}z}, U.~A., {Kristensen}, L.~E., {van Dishoeck}, E.~F., {et~al.}
  2012, \aap, 542, A86

\bibitem[{{Y{\i}ld{\i}z} {et~al.}(2015){Y{\i}ld{\i}z}, {Kristensen}, {van
  Dishoeck}, {Hogerheijde}, {Karska}, {Belloche}, {Endo}, {Frieswijk},
  {G{\"u}sten}, {van Kempen}, {Leurini}, {Nagy}, {P{\'e}rez-Beaupuits},
  {Risacher}, {van der Marel}, {van Weeren}, \& {Wyrowski}}]{Yildiz2015}
---. 2015, \aap, 576, A109

\end{thebibliography}
\bibliographystyle{apj}


\newpage

\appendix
\addcontentsline{toc}{section}{Appendix}
\renewcommand{\thesubsection}{\Alph{subsection}}

\subsection{\normalfont{\ref{sec:smm1_scheme}. SCHEME OF THE SMM1 CORE}}
\label{sec:smm1_scheme}

\begin{figure}[H]
\centering
\includegraphics[scale=0.8,clip,trim=6cm 1.5cm 8cm 2.5cm]{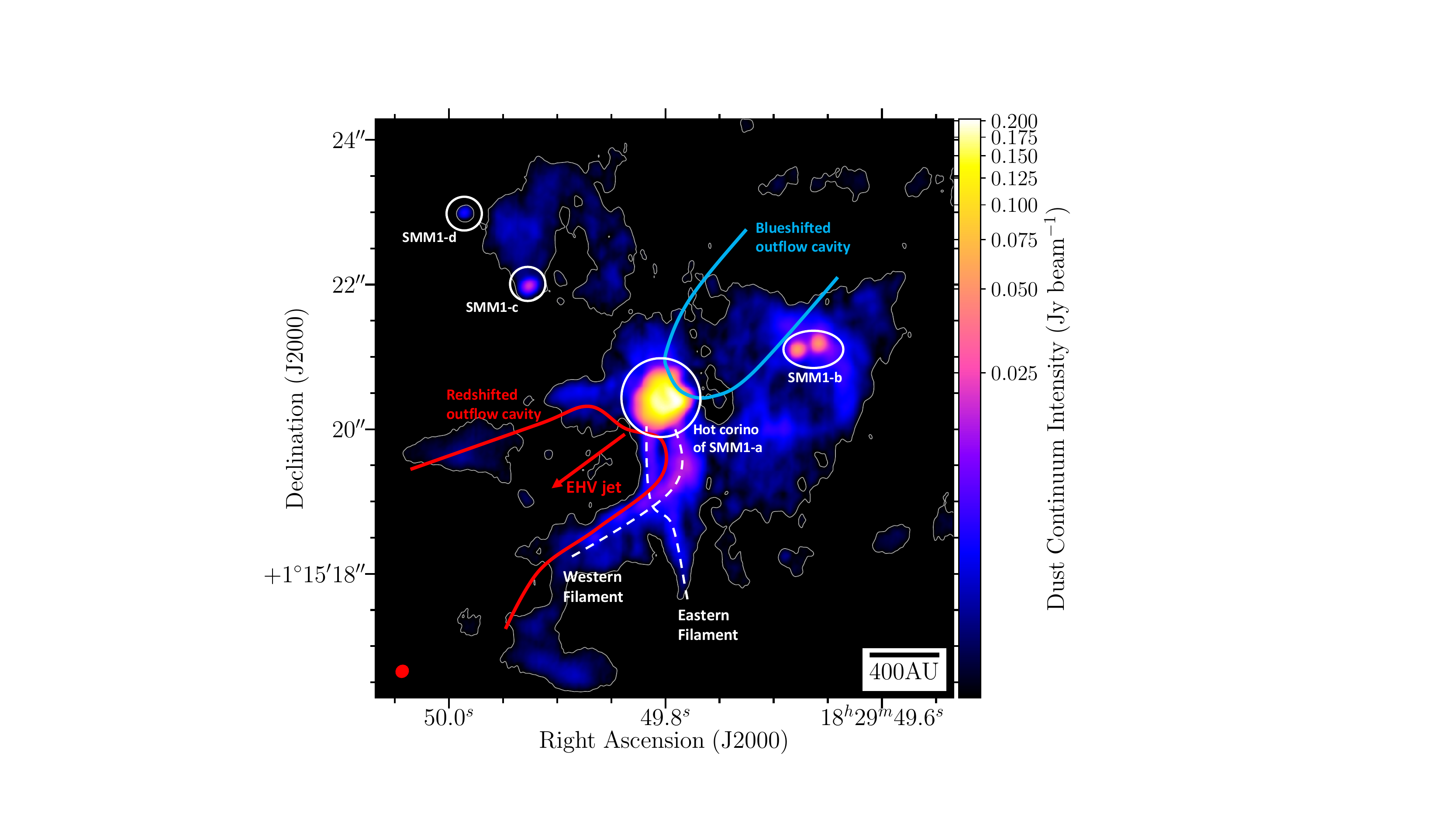}
\caption{\footnotesize Schematic view of Serpens SMM1. The color scale is the total intensity (Stokes $I$) thermal dust emission from Case-1, shown when $I\,>\,3\sigma_I$, from the combination of datasets A, B, and C; see Table \ref{t.imaging} Case-1. $\sigma_I$ = \rmsISMMcaseIUnits{}. The peak in the total intensity is \peakISMMcaseIUnits{}. The red ellipse in the lower-left corner represents the beam size, measuring \ie\xspace 0$\farcs$15 $\times$ 0$\farcs$14, with a position angle of --48.5$^\circ$.}
\vspace{0.3cm}
\end{figure}

\newpage

\subsection{\normalfont{\ref{sec:grad}. GRADIENT MAPS DERIVED FOR THE HISTOGRAMS IN SMM1 AND EMB 8(N)}}
\label{sec:grad}

The histograms presented in Figure \ref{fig:HRO} show the distributions of relative orientations between the magnetic field and the intensity gradients of both the continuum dust emission and the integrated \co moment 0 emission, within the Serpens SMM1-a and Emb 8(N) protostellar cores.

To produce the histograms, we select the magnetic field orientations in pixels with emission over the 3$\sigma_P$ threshold, where $\sigma_P$ is the rms noise value of the polarized intensity map. In the case of the SMM1-a core, the emission from its neighbors, \ie\xspace SMM1-b, c, and d, is removed by hand in order to only take into account the emission related to SMM1-a. We consider separately the emission emanating from the central inner core of SMM1-a (applying a threshold of 25$\sigma_I$, where $\sigma_I$ is the rms noise level of the Stokes $I$ map) and the emission emanating from the dust cavity walls and filaments (where the selected emission is between 7$\sigma_I$ and 25$\sigma_I$). As for Emb 8(N), we just select the value above 11$\sigma_I$ in the total intensity map, in order to focus on the central core and to avoid considering any emission from the large-scale dense, cold filament described in Section \ref{subsec:8N}. Concerning the blue- and redshifted moment 0 maps, we use the same ranges of integration as the those presented in Figures \ref{fig:emb8N_CO_pol} and \ref{fig:smm1_pol_CO}, applying a threshold of three times the rms noise level of the moment 0 maps.

\begin{figure}[H]
\centering
\begin{tabular}{cc}
\multicolumn{2}{c}{\subfigure{\includegraphics[scale=0.4,clip,trim=0cm 0cm 0.4cm 0cm]{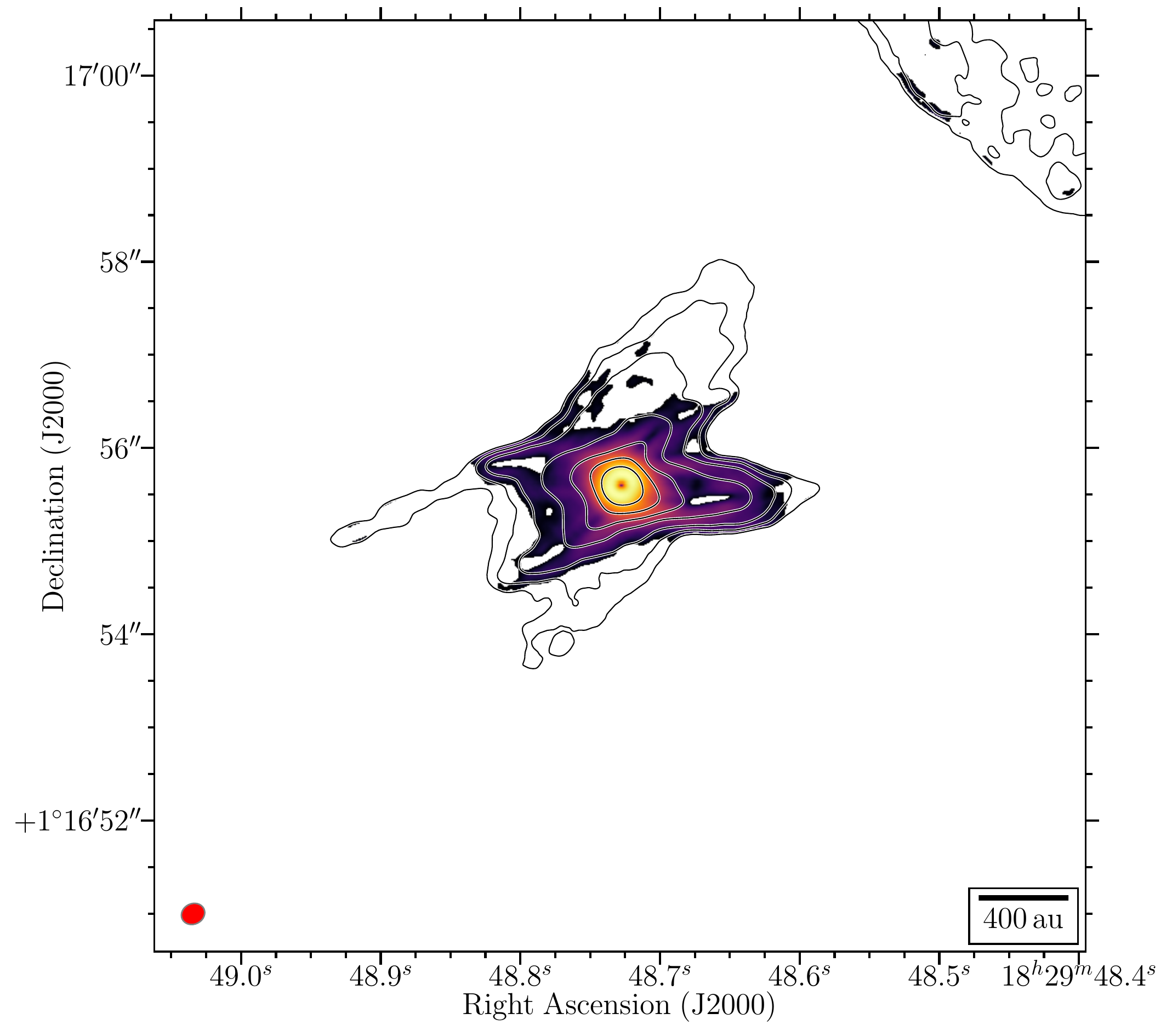}}}\\

\hspace{-4cm}
\subfigure{\includegraphics[scale=0.4,clip,trim=0cm 0cm 0.4cm 0cm]{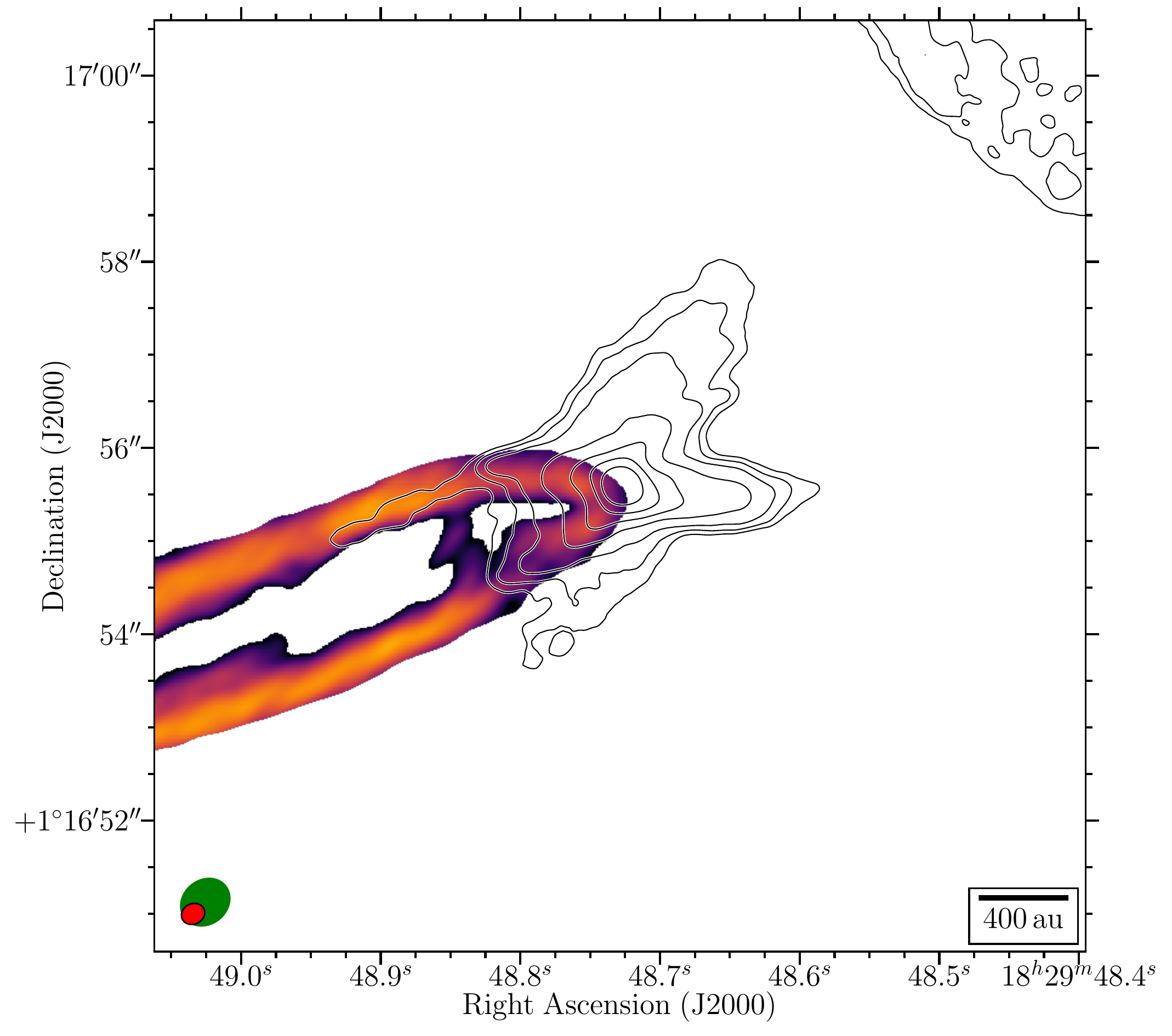}}
\subfigure{\includegraphics[scale=0.4,clip,trim=0cm 0cm 0.4cm 0cm]{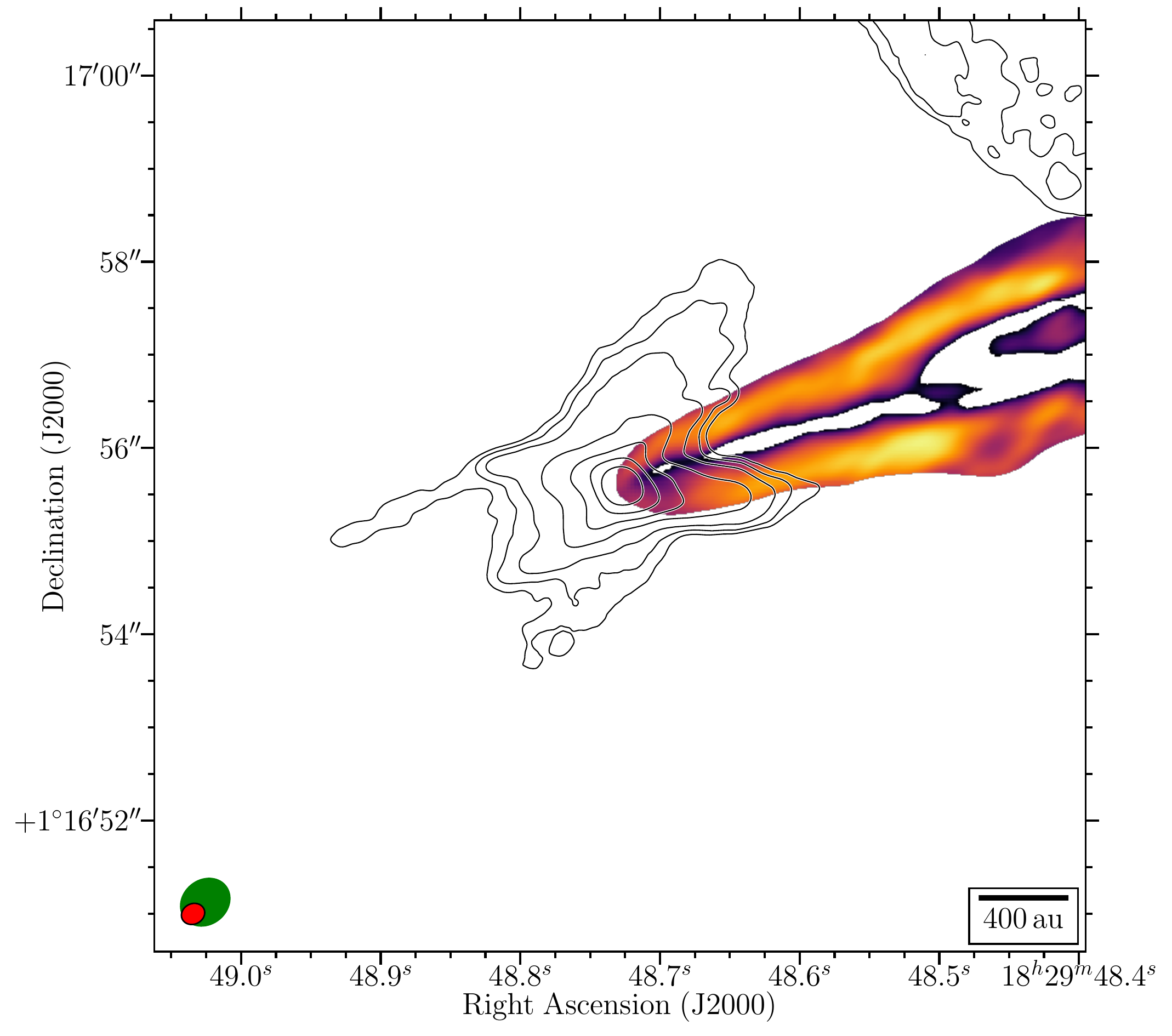}}
\end{tabular}
\caption{\footnotesize Gradient maps in Serpens Emb 8(N). The black contours trace the dust continuum from the Case-2 at 11, 16, 24, 44, 74, 128, 256 $\times$ $\sigma_{I}$ in the total intensity dust emission map, where $\sigma_{I}$ = \rmsIEmbEightNcaseIIUnits{}. \textit{Top left}: The color scale represents the gradient within the central zone of Emb 8(N), derived using the total intensity dust emission map. The gradients are computed where the emission is  >\,11 $\sigma_{I}$ in this central region. \textit{Top right}: gradient derived within the moment 0 map of the blueshifted CO emission integrated from --53 to 0 \kms{}. \textit{Bottom}: gradient derived within the moment 0 map of the redshifted CO emission integrated from 15 to 40 \kms{}. The noise value in both moment 0 maps is 0.1 \jybmkms{}; the gradients are calculated where the integrated CO emission is >\,3\,$\times$ this value.  All gradients are displayed where where the gradient is >\,1/100 $\times$ (peak -- rms) in the moment 0 maps, and >\,1/1000 $\times$ (peak -- rms) in the dust emission maps.}
\label{fig:grad_emb8N}
\vspace{0.3cm}
\end{figure}

\begin{figure}[H]
\centering
\subfigure{\includegraphics[scale=0.43,clip,trim=0cm 0cm 1cm 0cm]{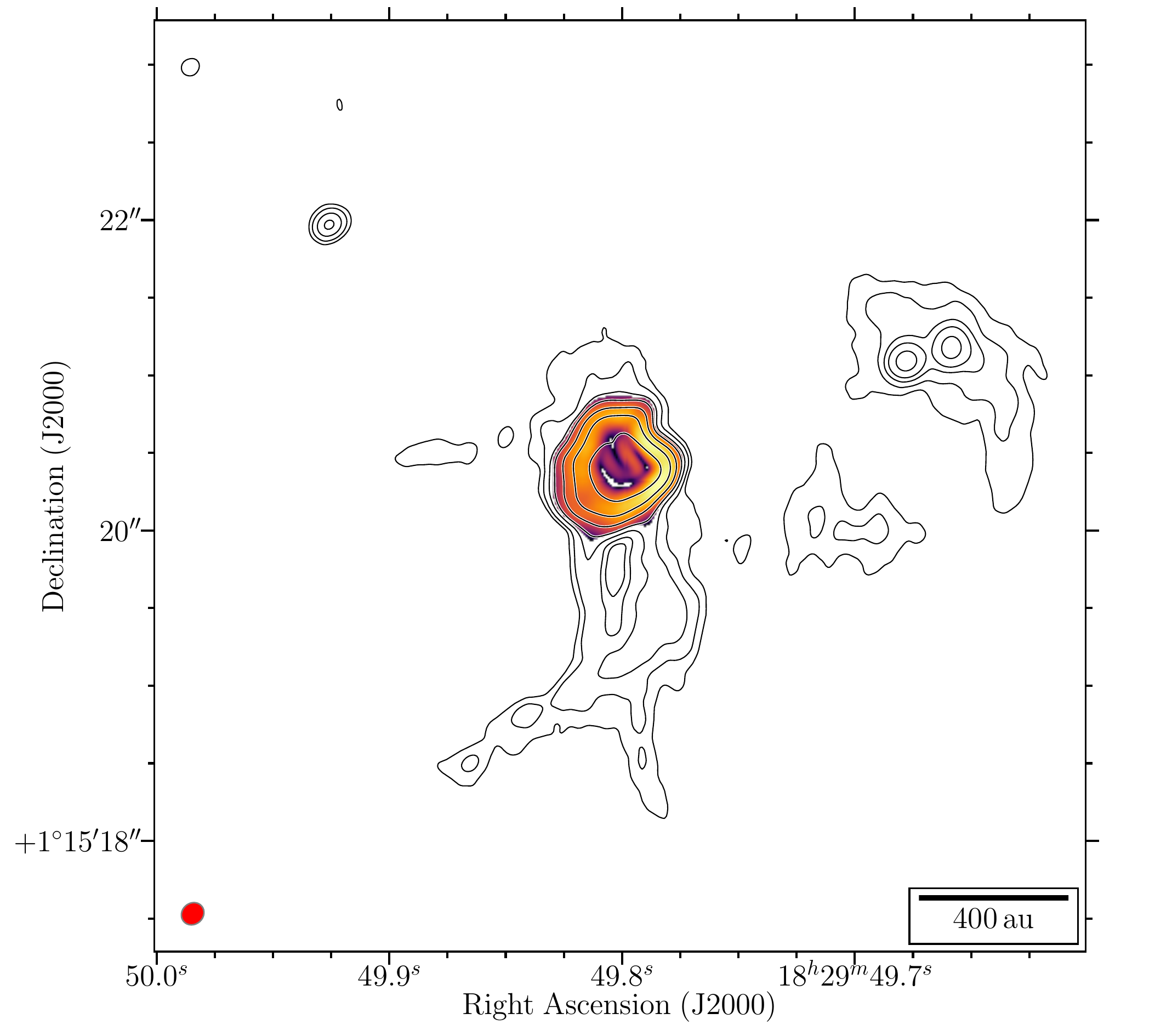}}
\subfigure{\includegraphics[scale=0.43,clip,trim=0cm 0cm 1cm 0cm]{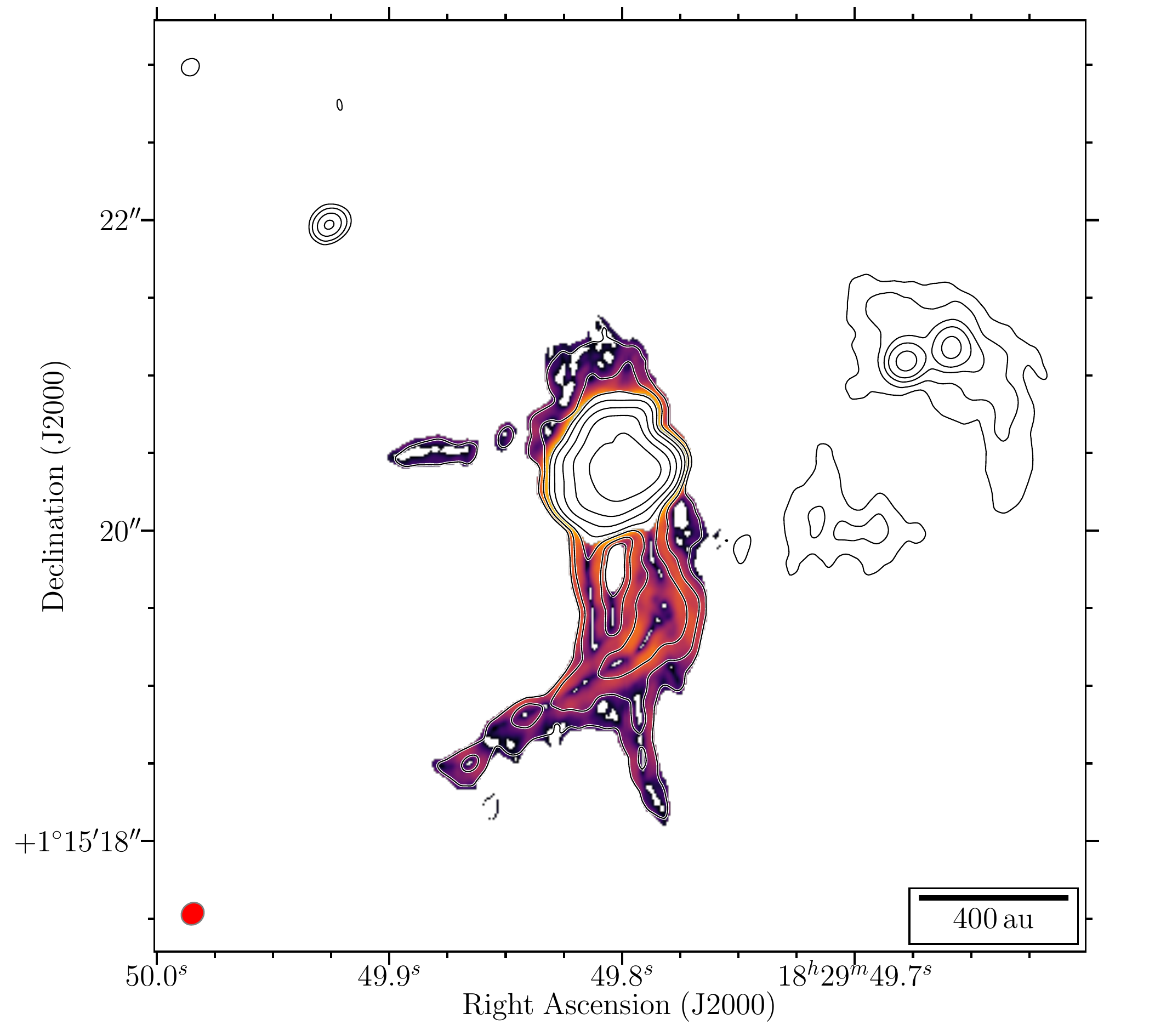}}
\subfigure{\includegraphics[scale=0.43,clip,trim=0cm 0cm 1cm 0cm]{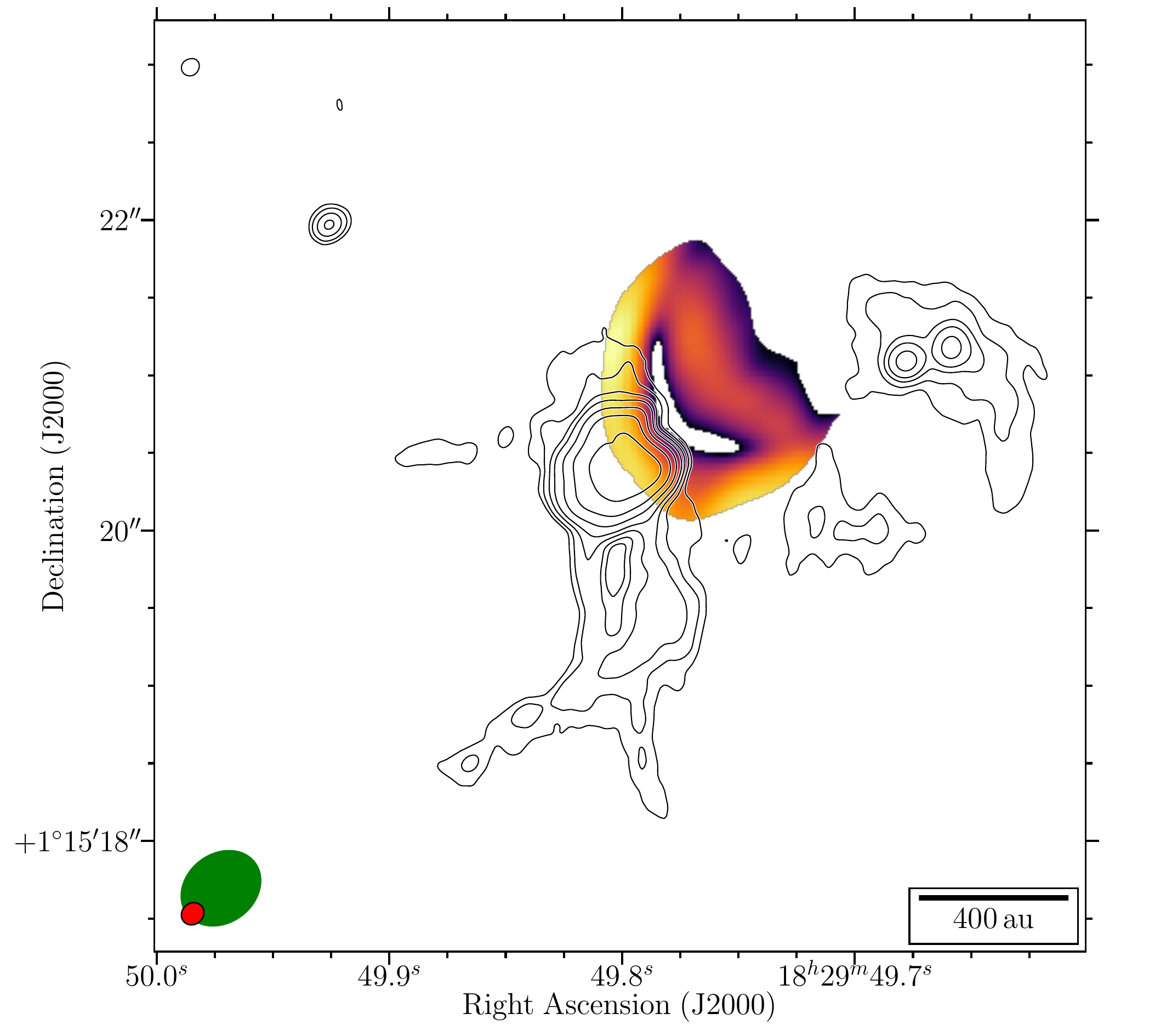}}
\subfigure{\includegraphics[scale=0.43,clip,trim=0cm 0cm 1cm 0cm]{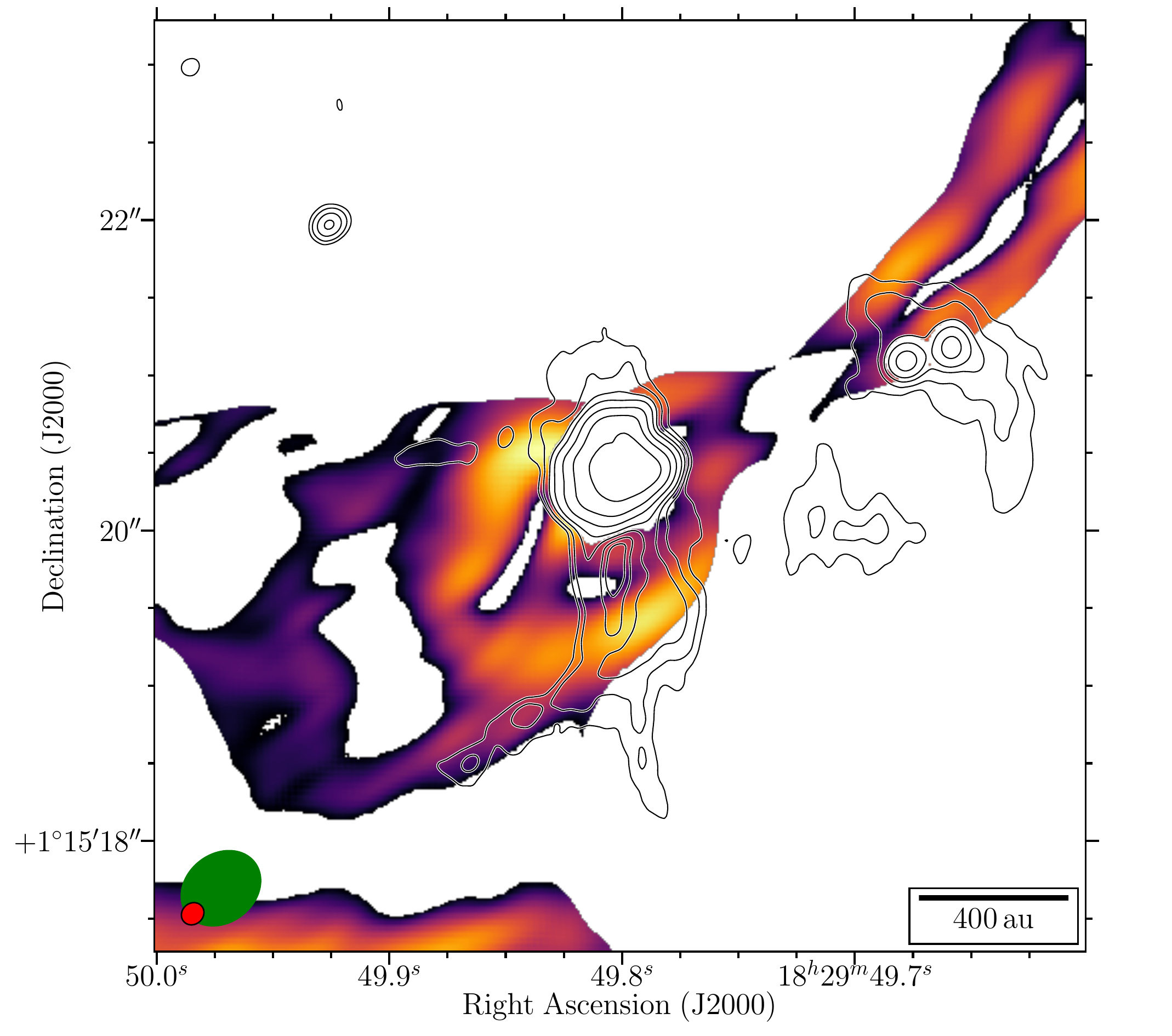}}
\caption{\footnotesize Gradient maps in Serpens SMM1-a. The black contours trace the dust continuum from the Case-1 at 8, 12, 20, 32, 64, 128, 256 $\times$ $\sigma_{I}$ in the total intensity dust emission map, where $\sigma_{I}$ = \rmsISMMcaseIUnits{}. \textit{Top left}: The color scale represents the gradient derived within the central zone of SMM1-a, derived using the total intensity dust emission map. The gradients are computed where the emission is >\,25 $\sigma_{I}$ in this central region. \textit{Top right}: gradient derived within the filaments and outflow cavities of SMM1-a, derived using the total intensity map. The gradients are computed where the emission is between the 7 $\sigma_{I}$ and the 25 $\sigma_{I}$ values; we also remove the emission from the other cores SMM1-b, c, and d. \textit{Bottom left}: gradient derived within the moment 0 map of the blueshifted CO emission integrated from --13 to 4 \kms{}. \textit{Bottom right}: gradient derived within the moment 0 map of the redshifted CO emission integrated from 10.5 to 30 \kms{}. 
The noise values in the blue- and redshifted CO moment 0 maps are 0.58 and 0.43 \jybmkms{}, respectively.  The gradients are calculated where the integrated CO emission is >\,3\,$\times$ these values.  All gradients are displayed where where the gradient is >\,1/100 $\times$ (peak -- rms) in the moment 0 maps, and >\,1/1000 $\times$ (peak -- rms) in the dust emission maps.  Emission in the central zone of SMM1-a and from SMM1-b have been removed in order to focus on the outflow cavities of SMM1-a.}
\label{fig:grad_smm1}
\vspace{0.3cm}
\end{figure}

After calculating the gradient, we apply a selection on the gradient values in order to focus on the relationship between the magnetic field and the emission gradients in regions of \textit{high} gradient. This selects, for example, the clear outflow cavity walls in the dust emission and moment 0 maps, the filamentary structures of SMM1-a, and the central inner cores of our protostars. In order to select regions of high gradient, we consider only gradients >\,1/100 $\times$ (peak -- rms) for the moment 0 maps,  and >\,1/1000 $\times$ (peak -- rms) for the dust continuum maps. Where both a gradient value and a magnetic field position angle are selected in the same location, a point is added to the distribution. Finally, we Nyquist sample the final distributions presented in the histograms, \ie using 4 points per synthesized beam (two points in RA and two in DEC) in both the dust or CO moment 0 maps. We present the gradient maps in Figures \ref{fig:grad_emb8N} and \ref{fig:grad_smm1}.

\end{document}